\documentclass[12pt,letterpaper,a4paper]{article}
\usepackage[includeheadfoot,
            marginratio={1:1,2:3},
            width=500pt,
            height=720pt,]{geometry}
\usepackage{amsmath}
\usepackage{amsfonts}
\usepackage{amssymb}
\usepackage{graphicx}
\usepackage{cancel}
\usepackage{empheq}
\usepackage{color}
\usepackage{hyperref}

\numberwithin{equation}{section}
\usepackage{float}
\restylefloat{table}
\restylefloat{figure}
\usepackage[utf8]{inputenc}
\usepackage{amsmath, amsthm, amssymb, amsfonts}
\usepackage[font=small, labelfont=bf]{caption}
\usepackage{amsmath, amsthm, amssymb, amsfonts}
\usepackage{multirow}
\usepackage{graphicx}
\usepackage{float}
\usepackage[numbers,sort&compress]{natbib}
\usepackage{enumerate}
\usepackage{amsmath}
\usepackage{amsfonts}
\usepackage{amssymb}
\usepackage{graphicx}
\usepackage{cancel}
\usepackage{empheq}
\usepackage{color}
\usepackage{hyperref}
\usepackage{float}
\restylefloat{table}
\restylefloat{figure}

\newcommand{\nc}{\newcommand}
\nc{\beq}{\begin{equation}}
\nc{\eeq}{\end{equation}}
\nc{\bea}{\begin{eqnarray}}
\nc{\eea}{\end{eqnarray}}

\def\ov{\overline}

\begin{document}
{\hfill
IFT-UAM/CSIC-18-038

\hfill 
arXiv:1805.05748}

\vspace{1.0cm}
\begin{center}
{\Large
On Missing Bianchi Identities in Cohomology Formulation}
\vspace{0.4cm}
\end{center}

\vspace{0.35cm}
\begin{center}
 Xin Gao$^\dag$\footnote{Email: xgao@roma2.infn.it}, Pramod Shukla$^{\ddag\Diamond}$\footnote{Email: pshukla@ictp.it}, Rui Sun$^\spadesuit$\footnote{Email: rsun@tsinghua.edu.cn}
\end{center}

\vspace{0.1cm}
\begin{center}
{
$^{\dag}$Dipartimento di Fisica, Universita di Roma ``Tor Vergata", and \\
I.N.F.N. Sezione di Roma ``Tor Vergata",\\
Via della Ricerca Scientifica, 00133 Roma, Italy \\
\vskip0.4cm
$^{\ddag}$Departamento de F\'{\i}sica Te\'orica and Instituto de F\'{\i}sica Te\'orica UAM/CSIC,\\
Universidad Aut\'onoma de Madrid, Cantoblanco, 28049 Madrid, Spain\\
\vskip0.4cm
$^{\Diamond}$Abdus Salam ICTP, Strada Costiera 11, Trieste 34151, Italy \\
\vskip0.4cm
$^{\spadesuit}$Yau Mathematical Sciences Center, Tsinghua University, Haidian District, \\Beijing 100084, China
}
\end{center}

\vspace{1cm}

\abstract{In this article, we perform a deep analysis of the Bianchi identities in the two known formulations developed for the four-dimensional effective type IIA supergravity theory with (non-)geometric fluxes. In what we call the `first formulation', fluxes are expressed in the real six-dimensional indices while in the `second formulation', fluxes are written in the cohomology form. We find that the set of flux constraints arising from these two known formulations are not equivalent, and there are missing identities in the cohomology version which need to be supplemented  to match with the first formulation. By analyzing two explicit examples, we conjecture a model independent form for (the most of) the missing identities. These identities have been mostly overlooked in the previous attempts of studying moduli stabilization, particularly for the models developed in the beyond toroidal frameworks, where they could play some important role.}

\clearpage

\tableofcontents

\section{Introduction}
\label{sec_intro}
In the context of Type II supergravity theories, a successive application of the T-duality on the NS-NS three-form flux $H_3$ results in a chain of geometric and non-geometric fluxes, which can be given as under \cite{Hellerman:2002ax,Dabholkar:2002sy,Hull:2004in,Derendinger:2004jn, Derendinger:2005ph, Shelton:2005cf, Wecht:2007wu},
\bea
\label{eq:Tdual}
& & H_{ijk} \longrightarrow \omega_{ij}{}^k  \longrightarrow Q_{i}{}^{jk}  \longrightarrow R^{ijk}\, .
\eea
%Generically, such fluxes can appear as parameters in the four-dimensional (4D) effective potential, and subsequently can help in developing a suitable scalar potential which could be useful for various model building purposes. Moreover, interesting connections among the toolkits of superstring flux-compactifications, the gauged supergravities and the Double Field Theory (DFT) via non-geometric fluxes have given the platform for approaching phenomenology based goals from these three directions  \cite{ Derendinger:2004jn, Derendinger:2005ph, Shelton:2005cf, Wecht:2007wu, Aldazabal:2006up, Dall'Agata:2009gv, Aldazabal:2011yz, Aldazabal:2011nj, Geissbuhler:2011mx, Grana:2012rr, Dibitetto:2012rk, Andriot:2013xca, Andriot:2014qla, Blair:2014zba, Andriot:2012an, Geissbuhler:2013uka}.
A consistent incorporation of the various possible fluxes makes the compactification background richer and more flexible for model building. In this regard, a continuous progress has been made since more than a decade towards moduli stabilization \cite{Aldazabal:2006up, Ihl:2006pp, Ihl:2007ah, Blumenhagen:2015kja}, in constructing de-Sitter vacua \cite{deCarlos:2009qm, Danielsson:2012by, Blaback:2013ht, Damian:2013dwa, Blumenhagen:2015xpa} and also in realizing the minimal aspects of inflationary cosmology \cite{Damian:2013dq, Hassler:2014mla, Blumenhagen:2014gta, Blumenhagen:2015qda}. 

One of the important aspects of model building in non-geometric flux compactification is to consistently satisfy all the quadratic flux constraints coming from the various Bianchi identities and the tadpole cancellation conditions. This can be very crucial as sometimes it can simplify the scalar potential to a great length by canceling many terms. In this regard, it is worth to mention that the 4D non-geometric scalar potentials arising from a concrete construction, very often consist of quite huge number of terms. For example, in the two concrete setups which we will consider in this article, we find that there are thousands of terms in the scalar potential. Subsequently, it is anticipated that it can get even hard to analytically solve the extremization conditions because the same would demand to solve very high degree polynomials. Unfortunately there is nothing like LARGE volume scenarios \cite{Balasubramanian:2005zx} in these non-geometric constructions, and therefore all the terms being at tree level are equally important and cannot be naturally hierarchical. The difficulty in dealing with the extremization conditions is so much involved that one has to look either for simplified ansatz by switching-off certain flux components at a time, or else one has to opt for an involved numerical analysis \cite{Aldazabal:2008zza,Font:2008vd,Guarino:2008ik,Danielsson:2012by,Damian:2013dq, Damian:2013dwa}.   

Moreover, it is still not fully known how many and which type of fluxes can be simultaneously turned-on on a given background. In this regard,  there are two main formulations of Bianchi identities which one utilizes for simplifying the type II effective potentials. One formulation involves fluxes denoted by the real six dimensional indices (e.g. $H_{ijk}$ etc.) \cite{Shelton:2005cf, Ihl:2007ah} while in the latter one, all flux components are written out using cohomology indices of the complex threefold $X_3$; e.g. $H_K, \omega_{a K}$ etc., where $K \in \{0, 1, ..,h^{2,1}(X_3)\}$ and $a \in \{1, 2, .., h^{1,1}_-(X_3) \}$ \cite{Gurrieri:2002wz, Grana:2005ny, Ihl:2007ah, Benmachiche:2006df, Grana:2006hr}. The first formulation has been always utilized for simplifying the scalar potential of the toroidal examples \cite{Aldazabal:2008zza,Font:2008vd,Guarino:2008ik,Danielsson:2012by,Damian:2013dq, Damian:2013dwa} while the recent interests beyond the toroidal setups have used the second formulation \cite{Blumenhagen:2015qda, Blumenhagen:2015kja,Blumenhagen:2015jva, Blumenhagen:2015xpa,  Li:2015taa}. However, it turns out that the known versions of these two formulations of Bianchi identities do not produce an equivalent set of constraints. To the best of our knowledge, this mismatch has been observed/emphasized only in \cite{Ihl:2007ah,Robbins:2007yv, Shukla:2016xdy}, which might play some important role in moduli stabilization and any subsequent phenomenological applications, e.g. such missing identities might be relevant in the recent interesting studies made in \cite{Blumenhagen:2015qda, Blumenhagen:2015kja,Blumenhagen:2015jva, Blumenhagen:2015xpa,  Li:2015taa}. Therefore, it is worth as well as timely to bring the attention of the model builders on this aspect. On these lines we have the following plans.
\begin{itemize}
\item{We carefully investigate the two formulations of the Bianchi identities in two concrete setups. This analysis is motivated by some observations made in \cite{Ihl:2007ah,Robbins:2007yv, Shukla:2016xdy}, in which it has been found that the two formulations in their currently known version do not result in an equivalent set of flux constraints. The first formulation has all the second formulation identities along with some additional ones, which we call as `missing' identities. In this article, we plan to investigate the $(1,1)$- and the $(2,1)$-cohomology structure in the missing identities in some detail.}
\item{Unlike the type IIB studies made in \cite{Robbins:2007yv, Shukla:2016xdy} along these motivations, we show that in type IIA orientifold setup it is easier to observe this mismatch for some simpler class of models. In  particular, the ones in which orientifold involution results in a trivial even (1,1)-cohomology. As we will explain later, this leads to the fact that we have just a single identity in the `second formulation' while the `first formulation' consists of five distinct classes of identities. %Note that this direct observation of the mismatch could not be made in \cite{Ihl:2007ah} which is based on the ${\mathbb T}^6/{\mathbb Z}_4$ orientifold for type IIA compactification as this setup has a non-trival (1,1)-cohomology.
}
\item{Recently in \cite{Gao:2017gxk}, we have presented a symplectic formulation of the 4D type IIA scalar potential with non-geometric fluxes. Being very compact, this formulation creates the possibility of studying the model independent moduli stabilization, i.e. for an arbitrary number of K\"ahler- and complex structure- moduli. In this regard, knowing the generic form of the missing Bianchi identities is a crucial step to take.}
\end{itemize}

\noindent
The main strategy which we follow in our approach is such that first we extract the second formulation identities from the set of constraints which arise from translating the identities of the first formulation into cohomological form. We take this step in great detail for both of the  explicit models. Subsequently we separate out one version of the `missing' identities which are not the part of the second formulation. This is a tricky step because there are non-unique ways of reshuffling the set of missing Bianchi identities such that one could create more than one equivalent sets of constraints which apparently possess distinct cohomological structure, and therefore it is hard to club them into a particular form for claiming their generality. For that purpose one would need as many explicit examples as possible to check the mutual consistency for any generic guess. 

However, it is not impossible to invoke some structure among the missing identities from the two concrete models. For example we show how the $(1,1)$-cohomology structure in the missing Bianchi identities can be encoded in the triple intersection numbers of the complex threefold while the $(2,1)$-cohomology structure has some insights from the complex structure moduli dependent prepotential. We compare these cohomology sectors for both the explicit examples to look for a model independent generalization which could produce them as particular cases, and this is what we mainly aim to achieve in this work.  

The article is organized as follows: Firstly, in section \ref{sec_basics} we provide the relevant details on the two formulation of the Bianchi identities, and subsequently in section \ref{sec_BIs} we perform a deep investigation of the Bianchi identities for the two concrete examples to illustrate that the two known formulations of Bianchi identities do not result in an equivalent set of flux constraints. In section \ref{sec_BIsGeneric}, we study the possibility of rewriting the missing identities in a model independent manner by investigating the $(1,1)$- and the $(2,1)$-cohomology sectors in the two explicit examples. Finally the important conclusions are presented in the section \ref{sec_conclusions} followed by three appendices. The first appendix \ref{sec_appendix} provides a derivation of Bianchi identities in the second formulation. The appendix \ref{sec_setups} presents the relevant details about the two concrete setups while the appendix \ref{sec_lengthyBIs} consists of the Bianchi identities which are too lengthy to be part of the main sections.

\section{Two formulations of the Bianchi identities}
\label{sec_basics}
%\subsection{Fixing the convention}
In this work, we consider type IIA superstring theory compactified on an orientifold of a Calabi Yau (CY) threefold $X_3$ with the presence of $O6$-planes. In this regard, the orientifold is constructed via modding out the CY with a
discrete symmetry ${\cal O}$ which includes the world-sheet parity $\Omega_p$ combined with the space-time fermion number in the left-moving sector $(-1)^{F_L}$. In addition
${\cal O}$ can act non-trivially on the CY manifold so that one has altogether,
\bea
\label{eq:orientifoldO}
& & {\cal O} = \Omega_p\, (-1)^{F_L}\, \sigma
\eea
where $\sigma$ is an involutive symmetry (i.e. $\sigma^2=1$) of the internal CY and acts trivially on the four flat dimensions.
The massless states in the four dimensional effective theory are in one-to-one correspondence with various involutively even/odd harmonic forms, and hence do generate the equivariant cohomology groups $H^{(p,q)}_\pm(X_3)$. Subsequently, the various field ingredients can be expanded in appropriate bases of the equivariant cohomologies. %For that purpose, let us first start with fixing the conventions following the notations of \cite{Ihl:2007ah, Robbins:2007yv} with some small symbolic changes. 
To begin with, let us fix our conventions by considering the following representations for the various involutively even and odd harmonic forms \cite{Grimm:2004ua},
\begin{table}[H]
\begin{center}
\begin{tabular}{|c|| c| c| c| c| c| c|} 
\hline
%&&&&&&\\
Cohomology group & $H^{(1,1)}_+$ & $H^{(1,1)}_-$ & $H^{(2,2)}_+$ & $H^{(2,2)}_-$ & $H^{(3)}_+$ & $H^{(3)}_-$ \\
%&&&&&&\\
\hline\hline
%&&&&&&\\
 Dimension & $h^{1,1}_+$ & $h^{1,1}_-$ & $h^{1,1}_-$ & $h^{1,1}_+$ & $h^{2,1}+ 1$ & $h^{2,1} + 1$ \\
 %&&&&&&\\
 Basis & $\mu_\alpha$ & $\nu_a$ & $\tilde\nu^a$ & $\tilde\mu^\alpha$ & $\alpha_I$ & $\beta^J$ \\
 %&&&&&&\\
 \hline
\end{tabular}
\end{center}
\caption{Representation of various forms and their counting}
\label{tab_1}
\end{table}
\noindent
Here $\mu_\alpha$ and $\nu_a$ denote the bases of even and odd real harmonic two-forms respectively, while $\tilde\mu^\alpha$ and $\tilde\nu^a$ denote the bases of  odd and even four-forms. Further, $\alpha_I$ and $\beta^J$ form the bases of even and odd real three-forms. In addition, the zero form {\bf 1} is even while there is an involutively odd six-form $\Phi_6$. Moreover, the triple intersection numbers and the normalization of the various forms are fixed as under,
\bea
\label{eq:intersectionBases}
& & \int_{X_3} \Phi_6 \equiv f = 1, \qquad \int_{X_3} \nu_a \wedge \nu_b \wedge \nu_c= \kappa_{abc} , \qquad \int_{X_3} \nu_a \wedge \mu_\alpha \wedge \mu_\beta = \hat{\kappa}_{a \alpha \beta},\\
& & \quad \int_{X_3} \nu_a \wedge \tilde{\nu}^b \equiv d_a{}^b = \delta_a{}^b, \qquad \int_{X_3} \mu_\alpha \wedge \tilde{\mu}^\beta \equiv \hat{d}_\alpha{}^\beta = {\delta}_\alpha{}^\beta , \qquad \int_{X_3} \alpha_I \wedge \beta^J=\delta_I{}^J\,. \nonumber
\eea
Our above convention slightly differs from the normalizations considered in \cite{Ihl:2007ah, Robbins:2007yv}.
%Note that our notations are a bit more generic and flexible for appropriate normalization of forms. Of course, for an appropriate choice of the bases of four-forms to be dual to the respective two-forms, one would have $d_a{}^b = \delta_a{}^b$ and $\hat{d}_\alpha{}^\beta = \delta_\alpha{}^\beta$.
The effective four-dimensional supergravity theory is governed by the dynamics of the complexified chiral variables $T^a$ and $N^K$ which are defined as under,
\bea
\label{eq:chiralvariables}
& & {\cal J}_c \equiv B_2 + i \, {\cal J} = b^a \, \nu_a + \, i\, t^a \, \nu_a = T^a\, \nu_a\,\\
& & \Omega_c \equiv C_3 + 2\, i\, e^{-D} \, {\rm Re}(\Omega_3) = \xi^K \, \alpha_K + 2 \, i \, e^{-D} \, {\cal X}^K \, \alpha_K = 2\, N^K\, \alpha_K\,,\nonumber
\eea
where the K\"ahler form ${\cal J}$ is expanded as ${\cal J} = t^a\, \nu_a$ where $t^a$'s are volume of the two-cycles, and the holomorphic three-form $\Omega_3$ is expanded in terms of the period vectors $\left({\cal X}^K, {\cal F}_K\right)$ as $\Omega_3 = {\cal X}^K \alpha_K - {\cal F}_K\, \beta^K$. Further, $B_2 = b^a \, \nu_a$ and $C_3 = \xi^K\, \alpha_K$ respectively denote the NS-NS two-form potential and the RR three-form potential expanded in their respective odd/even bases, and $D$ denotes the four-dimensional dilaton which is related to the ten-dimensional dilaton $\phi$ via $e^{-D} = e^{-\phi}\, \sqrt{\cal V}$ where ${\cal V}$ is the volume of the complex threefold.

For studying moduli stabilization and any subsequent phenomenology, a very crucial step to follow is to impose the constraints from various NS-NS Bianchi identities as well as RR tadpoles to get the {\it true} non-vanishing contribution to the effective four dimensional scalar potential. We have two formulations for representing the (NS-NS) Bianchi identities, and we emphasize here that both sets of Bianchi identities have their own advantages and limitations. The `first formulation' is in which all fluxes, moduli and fields are expressed using the real six-dimensional indices. e.g. $B_{lm}, H_{lmn}$, $\omega_{lm}{}^n, \, Q_l{}^{mn}$ and $R^{lmn}$ where $l,m,n$ are indices corresponding to  the real coordinates of the real sixfold. In the `second formulation', all the fluxes, moduli and fields are counted by cohomology indices. 

In the generic case, some naive counting suggests that there is an upper bound on the maximum number of flux components which can be non-trivial in a given setting, and they have to further satisfy several constraints arising from the orientifold action and the non-trivial Bianchi identities, and subsequently it is hard to present a model independent counting for the number of independent flux components. However, for the fluxes in both the formulations, we have presented the counting for the upper bound on the number of fluxes as mentioned in Table \ref{tab_BIscount}, where we have only assumed the anti-symmetry of the various flux components in the so-called `first formulation'. It is well anticipated that in a given orientifold construction, many of the flux components would be non-trivially coupled, and so would significantly reduce the number of `independent' flux components. Also note that in our current conventions, the flux components $H^K$, $R^K$, $\omega_a{}^K$, $Q^{aK}$, $\hat{\omega}_{\alpha K}$ and $\hat{Q}^\alpha{}_K$ are projected out, and so they do not appear in the flux counting presented in Table \ref{tab_BIscount}. As said before, let us reiterate that this simple counting corresponds to an upper bound for the maximum number of the flux components which can be further significantly constrained by the Bianchi identities, and hence can influence the moduli stabilization and any subsequent phenomenology. 
\begin{table}[h!]
 \centering
 \begin{tabular}{|c|c||c|c|}
\hline
 & & & \\
Flux type & Max. number of  & Flux type  & Max. number of \\
 & flux components &  & flux components \\
\hline
 & & & \\
$H_{ijk}$ \qquad & \qquad 20 \qquad & \qquad \, $H_{K}$ \qquad & \, $h^{2,1}+1$ \qquad\\
 & & & \\
$\omega_{ij}{}^k$  \qquad & \qquad 90 \qquad & \qquad $\omega_{aK}$ \qquad & $h^{1,1}_-\, (h^{2,1}+1)$ \qquad \\
  \qquad & \qquad \qquad & \qquad $\hat{\omega}_{\alpha}{}^K$ \qquad & $h^{1,1}_+\, (h^{2,1}+1)$ \qquad \\
 & & & \\
$Q_i{}^{jk}$    \qquad & \qquad 90  \qquad & \qquad $Q^a{}_K$ \qquad & $h^{1,1}_-\, (h^{2,1}+1)$ \qquad\\
 \qquad & \qquad  \qquad & \qquad $\hat{Q}^{\alpha K}$ \qquad & $h^{1,1}_+\, (h^{2,1}+1)$ \qquad\\
 & & & \\
$R^{ijk}$      \qquad & \qquad 20 \qquad & \qquad $R_K$ \qquad & $h^{2,1}+1$ \qquad \\
& & & \\
 \hline
& & & \\
 {\bf Total } \qquad & \qquad 220 \qquad & \qquad {\bf Total } \qquad & $2\, (h^{1,1} + 1)\, (h^{2,1}+1)$ \qquad  \\
 & & & \\
 \hline
 \end{tabular}
\caption{Maximum number of flux components in the two formulations.}
 \label{tab_BIscount}
\end{table}

Further, let us note that it is not necessary to have a bijection among the two sets of fluxes mentioned in Table \ref{tab_BIscount}, especially among the respective set of $\omega$-flux and the non-geometric $Q$-flux. Nevertheless in several examples, the bijection between the respective set of fluxes in the two formulation does hold; e.g. the orientifold setups built from the orbifolds ${\mathbb T}^6/{\Gamma}$, where $\Gamma$ corresponds to the crystallographic actions ${\mathbb Z}_2 \times {\mathbb Z}_2, {\mathbb Z}_3, {\mathbb Z}_3 \times {\mathbb Z}_3, {\mathbb Z}_4$ and ${\mathbb Z}_6$-I \cite{Villadoro:2005cu, Camara:2005dc, DeWolfe:2005uu, Ihl:2007ah, Blumenhagen:2013hva}.
However, there is always a bijection between the respective $H$-flux and $R$-flux components for which the `actual' counting follows from the cohomology formulation. 

\subsection{First formulation}
This formulation has five classes of Bianchi identities supplemented by an extra constraint as presented in Table \ref{tab_BIsFirstFormulation}. 
\begin{table}[h!]
  \centering
 \begin{tabular}{|c||c|c|}
\hline
& & \\
Class & Bianchi Identities of the & Maximum no. of \\
 & First formulation & identities \\
 \hline
 & & \\
{\bf (I)} & $H_{m[\underline{ij}} \, \omega_{\underline{kl}]}{}^{m}= 0$ & 15 \\
& & \\
{\bf (II)} & $ \omega_{[\underline{ij}}{}^{m} \, {\omega}_{\underline{k}]m}{}^{l} \, = \, {Q}_{[\underline{i}}{}^{lm} \, {H}_{\underline{jk}]m}$ & 120 \\
& & \\
{\bf (III)} & ${H}_{ijm} \, {R}^{mkl} + {\omega}_{{ij}}{}^{m}{} \, {Q}_{m}{}^{{kl}} =
     4\, {Q}_{[\underline{i}}{}^{m[\underline{k}} \, {\omega}_{\underline{j}]m}{}^{\underline{l}]}$ & 225 \\
&& \\
{\bf (IV)} & ${Q}_{m}{}^{[\underline{ij}}  \, {Q}_{l}{}^{\underline{k}]m} \, = \, {\omega}_{lm}{}^{[\underline{i}}{}\,\, {R}^{\underline{jk}]m}$ & 120 \\
& & \\
{\bf (V)} & $R^{m [\underline{ij}}\,  Q_{m}{}^{\underline{kl}]} \, =0$ & 15 \\
& & \\
\hline
{Extra} & & \\
{constraint} & $\frac{1}{6}\, H_{ijk} \, R^{ijk} + \frac{1}{2} \, \omega_{ij}{}^k \, Q_k{}^{ij} = 0$ & 1 \\
& & \\
\hline
\hline
& & \\
& {\bf Total} & 496 \\
& & \\
 \hline
  \end{tabular}
  \caption{Bianchi identities of the first formulation and their counting}
  \label{tab_BIsFirstFormulation}
 \end{table}
 \noindent
For our current interest, we consider the fluxes to be constant parameters, however for the non-constant fluxes and in the presence of sources, these Bianchi identities are modified \cite{Geissbuhler:2013uka, Aldazabal:2013sca, Blumenhagen:2013hva, Andriot:2014uda}. In addition, let us also note that the ``Extra constraint" is automatically satisfied in the orientifold setting since there are no scalars which are invariant under the orbifolding and odd under the involution.

There have been several ways of deriving these sets of constraints; for example see \cite{Ihl:2007ah, Shelton:2005cf, Aldazabal:2013sca, Blumenhagen:2013hva, Aldazabal:2006up, Blumenhagen:2012pc, Geissbuhler:2013uka, Andriot:2014uda}. We do not intend to provide the detailed derivation, however let us sketch a couple of routes to arrive at these constraints. One way to derive these identities is via the Jacobi identities of the following Lie brackets for the NS-NS fluxes \cite{Shelton:2005cf, Aldazabal:2006up, Ihl:2007ah},
\bea  
& & \left[Z_i, Z_j  \right] \, \, \, \, = \omega_{ij}{}^k\, Z_k - \, H_{ijk}\, X^k, \quad \\
& & \left[ Z_i, X^j \right] \, = Q_i{}^{jk} Z_k \, -\, \omega_{ik}{}^j\, X^k \, , \quad \nonumber\\
& & \left[ X^i, X^j\right] =  Q_k{}^{ij} X^k \, - R^{ijk} Z_k  \,, \nonumber
\eea
where $Z_i$ and $X^i$'s are generators of the gauge transformations corresponding to the two gauge groups consisting of two sets of $d$-dimensional vectors obtained, from the metric and the $B$-field respectively, via the reduction of  type II superstring theory on a $d$-dimensional torus.

Another route to derive these identities is via considering the nilpotency of a twisted differential operator ${\cal D}$ defined as under \cite{Shelton:2006fd},
\bea
%\label{eq:twistedD}
& & {\cal D} = d + H \wedge . - \omega \triangleleft . +Q \triangleright . - R \bullet . \, ,
\eea
where the action of various (non-)geometric fluxes via $\triangleleft$, $\triangleright$ and $\bullet$ on a $p$-from changes them into a $(p+1)$-form, a $(p-1)$-form and a $(p-3)$-form respectively. To be more specific, if we consider a generic $p$-form to be given as $X_p = \frac{1}{p!} X_{i_1 ....i_p} dx^1 \wedge dx^2 ....\wedge dx^{p}$, then the various flux-actions can be defined as under \cite{Robbins:2007yv,Shelton:2006fd},
\bea
\label{eq:action0}
& & \hskip-0.75cm (\omega \triangleleft X)_{i_1i_2...i_{p+1}} = \left(\begin{array}{c}p+1\\2\end{array}\right) \, \, \omega_{[\underline{i_1 i_2}}{}^{j} X_{j|\underline{i_3.....i_{p+1}}]} + \frac{1}{2} \left(\begin{array}{c}p+1\\1\end{array}\right) \, \, \omega_{[\underline{i_1}\, j}{}^{j} X_{\underline{i_2 i_3.....i_{p+1}}]} \, ,\\
& & \hskip-0.75cm (Q \triangleright X)_{i_1i_2...i_{p-1}} = \frac{1}{2}\left(\begin{array}{c}p-1\\1\end{array}\right) \, \, Q_{[\underline{i_1}}{}^{jk}{} X_{jk|\underline{i_2.....i_{p-1}}]} + \frac{1}{2} \left(\begin{array}{c}p-1\\0\end{array}\right) \, \, Q_{j}{}^{kj}{} X_{k |i_1 i_2.....i_{p-1}} \, , \nonumber\\
& & \hskip-0.75cm (R \bullet X)_{i_1i_2...i_{p-3}} = \frac{1}{3!}\left(\begin{array}{c}p-3\\0\end{array}\right) \, \, R^{jkl} X_{jkl \, i_1.....i_{p-3}} \, ,\nonumber
\eea 
where the underlined indices inside the brackets $[..]$ are anti-symmetrized. The set of Bianchi identities and the `extra constraint' given in Table \ref{tab_BIsFirstFormulation} can be derived from the nilpotency of twisted differential operator ${\cal D}$ via ${\cal D}^2 A_p = 0$. %For the first formulation Bianchi identities, let us make the following two useful remarks,

As a side remark, let us note from the table \ref{tab_BIsFirstFormulation} that the maximum number of flux-constraints in the first formulation is bounded by 496 which is quite a peculiar number in string theory, and it would be interesting to know if there is any fundamental reason behind this, or its just a matter of counting.

\subsubsection*{A weaker set of identities} Let us mention that demanding the vanishing of ${\cal D}^2$ on a generic $p$-form $X_p$ apparently also results in the following additional set of constraints \cite{Ihl:2007ah},
\bea
\label{eq:bianchids1additional}
&  H_{kl[\ov i} \,\, \, Q_{\ov j]}{}^{kl}{} -\frac{1}{2}\, Q_k{}^{kl}\,H_{lij}-\,\frac{1}{2}\, \omega_{kl}{}^k\, \omega_{ij}{}^l &=0 , \\
& H_{kli}\, R^{klj} - Q_i{}^{kl}{}\, \omega_{kl}{}^j  -\omega_{kl}{}^k\, Q_i{}^{lj}{}-\, Q_k{}^{kl}\,\omega_{li}{}^j &= 0 ,\nonumber\\
& \omega_{kl}{}^{[\ov i} \, R^{kl \ov j]}  +\frac{1}{2}\,\omega_{kl}{}^k\, R^{lij} + \frac{1}{2}\,Q_k{}^{kl}{}\,Q_l{}^{ij}{} &  = 0, \nonumber\\
&  2\, H_{klm}\, R^{klm} + 3\, \omega_{kl}{}^k\, Q_m{}^{ml} &= 0\,. \nonumber 
\eea
However, a closer look ensures that all these identities in eqn. (\ref{eq:bianchids1additional}) can be obtained by contracting more indices from their respective main identities given in Table \ref{tab_BIsFirstFormulation}. It is worth to note that the last constraint in eqn. (\ref{eq:bianchids1additional}) generically holds by the orientifold construction itself as there are generically no zero-forms (scalars) which are odd under involution. Thus, these apparently additional identities are effectively not the new ones to worry about. Nevertheless, we will explain their relevance in a different sense while we compare the two formulations in explicit examples later on.
\subsubsection*{Tracelessness condition}
In order for the $\omega$-flux and $Q$-flux to be individually $T$-dual to the $H$-flux, they must satisfy the following so-called tracelessness condition \cite{Shelton:2006fd, Wecht:2007wu},
\bea
\label{eq:tracelessQw}
& & \hskip-2cm \omega_{ij}{}^i = 0\, , \qquad \qquad \qquad  Q_i{}^{ij}  = 0\, ,
\eea
It might be worth to mention that imposing this condition (\ref{eq:tracelessQw}) has been quite common in the literature \cite{Shelton:2006fd, Wecht:2007wu}. Also, a Calabi Yau threefold does not have any homologically non-trivial one- as well as five-cycles, and hence for the Calabi Yau orientifold cases it would be well justified to demand that all flux components having effectively one (real six-dimensional) free-index are trivial. %However, in our generic construction, to begin with we will assume that the components $\omega_{ij}{}^i$ or $Q_i{}^{ij}$, if allowed by the orientifold projection, can generically receive non-trivial values. Later on, we will also discuss the need for switching-off such fluxes, and their subsequent effects on the identities.

\subsection{Second formulation}
Considering the relevant flux actions for the type IIA orientifold setup as given in eqn. (\ref{eq:fluxActions0}), and ensuring the nilpotency of the twisted differential ${\cal D}$ on the harmonic forms via ${\cal D}^2 = 0$ results in 10 Bianchi identities \cite{Robbins:2007yv} which can be further classified into the five classes as mentioned in the first formulation. These identities are collected in Table \ref{tab_BIsSecondFormulation}, and a proof of the same has also been presented in the appendix \ref{sec_appendix}. 
\begin{table}[h]
  \centering
 \begin{tabular}{|c||c|c||}
\hline
& & \\
Class & Bianchi Identities of the & Maximum no. of \\
 & Second formulation & identities \\
 \hline
 & & \\
{\bf (I)} & $H_K\, \hat{\omega}_\alpha{}^K = 0$ & $h^{1,1}_+$\\
& & \\
{\bf (II)} & $H_K\, \hat{Q}^{\alpha K} = 0$ & $h^{1,1}_+$ \\
 & $\omega_{aK}\, \hat{\omega}_\alpha{}^K = 0$ & $h^{1,1}_+ \, h^{1,1}_-$ \\
& & \\
{\bf (III)} & $ \omega_{aK}\, \hat{Q}^{\alpha K} =0$ & $h^{1,1}_+ \, h^{1,1}_-$ \\
& $\hat{\omega}_\alpha{}^{K} \,Q^a{}_K=0$ & $h^{1,1}_+ \, h^{1,1}_-$ \\
& $\hat{\omega}_\alpha{}^{[K}\, \hat{Q}^{\alpha J]} = 0$ & $\frac{1}{2} h^{2,1} (h^{2,1} +1)$ \\
& $H_{[K} \, R_{J]} = \omega_{a[K}\, Q^a{}_{J]}$ & $\frac{1}{2} h^{2,1} (h^{2,1} +1)$ \\
& & \\
{\bf (IV)} & $R_K \, \hat{\omega}_\alpha{}^K = 0$ & $h^{1,1}_+$ \\
& $Q^a{}_K \, \hat{Q}^{\alpha K} = 0$ & $h^{1,1}_+ \, h^{1,1}_-$ \\
& & \\
{\bf (V)} & $R_K\, \hat{Q}^{\alpha K} = 0$ & $h^{1,1}_+$ \\
& & \\
\hline
\hline
& & \\
& \qquad If $h^{1,1}_+ \neq 0$,  \qquad \qquad {\bf Total} =& $4\, h^{1,1}_+ (1 + h^{1,1}_-) + h^{2,1}(1+ h^{2,1})$ \\
& \qquad If $h^{1,1}_+ = 0$, \qquad \qquad {\bf Total} =& $\frac{1}{2} h^{2,1} (h^{2,1} +1)$ \\
& & \\
 \hline
  \end{tabular}
  \caption{Bianchi identities of the second formulation and their counting}
  \label{tab_BIsSecondFormulation}
 \end{table}
 
Before coming to the specific models, let us mention that if one uses an orientifold involution such that the even (1,1)-cohomology and its dual odd (2,2)-cohomology are trivial, i.e. $h^{1,1}_+(X_3/\sigma) = 0$, then all the `hatted' fluxes which are counted via $\alpha\in h^{1,1}_+$ are projected out. The choice of such involutions are quite common in Type IIA Calabi Yau orientifold compactification as these are simpler setups to study, and subsequently the second formulation tells us that the only Bianchi identities which could be non-trivial turns out to be the following one,
\bea
& & H_{[K} \, R_{J]} = \omega_{a[K}\, Q^a{}_{J]}\,.
\eea
This happens because all the other identities involve `hatted' fluxes which are projected out. Such a situation provides a strong constraint on the set of second formulation identities as it suggests that all the Bianchi identities of the class {\bf (I)}, {\bf (II)}, {\bf (IV)} and {\bf (V)} are identically trivial ! Moreover, even only one of the four Identities within the class {\bf (III)} is non-trivial. However, as we will see in the two explicit models, the identities in class {\bf (I)}, {\bf (II)}, {\bf (IV)} and {\bf (V)} indeed provide non-trivial flux constraints while being computed from the first formulation. This has been possible to check in the concrete toroidal models in which both formulations can be explicitly computed. We will exemplify these arguments in two concrete models:
\begin{itemize}
\item{{\bf Model A:} In this setup we will consider the orientifold of a ${\mathbb T}^6/({\mathbb Z}_2 \times {\mathbb Z}_2)$ orbifold, with an anti-holomorphic involution which results in $h^{1,1}_+(X_3/\sigma) = 0$, and hence no `hatted' fluxes being present in this construction.}
\item{{\bf Model B:} In this setup we will consider the orientifold of a ${\mathbb T}^6/{\mathbb Z}_4$ orbifold, with an anti-holomorphic involution which results in $h^{1,1}_+(X_3/\sigma) \neq 0$, and hence there would be non-trivial `hatted' fluxes being present in this construction.}
\end{itemize}
From the Table \ref{tab_BIsSecondFormulation} we see that given the topological data about the orientifold, the maximum number of the (non-trivial) Bianchi identities in the second formulation can be read-off in terms of some Hodge numbers. To summarize this part, we have five classes of Bianchi identities in the first formulation while the second formulation has in total 10 distinct types of constraints. However, all the identities are coupled in a complex manner and generically it is hard to find the set of inequivalent flux constraints. The maximum number of such constraints is bounded by 496 in the first formulation while in the second formulation this number is bounded by $4\, h^{1,1}_+ (1 + h^{1,1}_-) + h^{2,1}(1+ h^{2,1})$. However, there is generically no bijection between the set of constraints because there is no bijection at first place between the respective $\omega$-flux and the $Q$ flux representations in the two formulations.

It has been observed in \cite{Ihl:2007ah, Robbins:2007yv, Shukla:2016xdy} that these two formulations of Bianchi identities do not lead to equivalent set of constraints. In fact, the first formulation has some additional constraints which cannot be derived from the identities of the second formulation. As most of the non-geometric scalar potential studies are motivated from toroidal examples, such an observation is worth to explore more insights of this mismatch. To be specific in this regard, let us mention that the mismatch in the two formulations of Bianchi identities have been observed for type IIA case in \cite{Ihl:2007ah}, however without having much attention on the insights of the mismatch, for example so that one could promote the same to the case of beyond toroidal setups such as those using CY orientifold. Moreover, motivated by the interesting type IIB model building efforts as made in \cite{Blumenhagen:2015qda, Blumenhagen:2015kja,Blumenhagen:2015jva, Blumenhagen:2015xpa, Li:2015taa} which have used the second formulation identities only,  if one attempts to make similar efforts for type IIA model building, then it is very much anticipated that such models and any subsequently realized vacua should be heavily under-constarined as most of the Bianchi identities would not be captured in the second formulation. Such a clear manifestation of the mismatch between the Bianchi identities of the two formulations, which we see from the type IIA setup, cannot be observed from type IIB setups, and our aim in this article is to investigate more on this and invoke the possible structure which could be generalised in a model independent manner to some more generic (beyond-toroidal) setups.

\section{Bianchi identities in the cohomology formulation}
\label{sec_BIs}
In this section we will compute the Bianchi identities for two toroidal models using the two formulations we have described, and subsequently we will compare if the set of Bianchi identities are equivalent or not. The main idea is to translate the first formulation identities into cohomology version using some flux conversion relations for the two formulations, and subsequently to perform some reshuffling in the first formulation constraints to recover the second formulation, and then the rest is what we term as the `missing identities' which cannot be obtained from the known version of the second formulation, i.e. from the Table \ref{tab_BIsSecondFormulation}. However, let us mention at the outset that we will provide more than one equivalent set of the `missing' constraints as there are non-unique ways of rewriting or clubbing the identities for invoking some model independent insights out of a complicated collection of flux-squared relations.

\subsection{Missing identities in Model A}
In this section we will compute the Bianchi identities using the two formulations for our Model A, which corresponds to a type IIA setup with a ${\mathbb T}^6/{({\mathbb Z}_2 \times {\mathbb Z}_2)}$-orientifold. The various explicit details about this model can be found in \cite{Gao:2017gxk}, and the relevant ingredients have been also briefly collected in the appendix \ref{sec_setups}. For this setup, focussing only on the untwisted sector, we have $h^{1,1}_+ = 0, \, h^{1,1}_- = 3$ and $h^{2,1} = 3$, and therefore there are six second-formulation identities which are to be imposed on 32 flux components. For translating the first formulation identities into the cohomology form, we will need the following flux conversion relations,
\bea
\label{eq:fluxconversion1}
& H_K & = \left[{\begin{array}{cccc}
  -\, H_{135}\, , & H_{146}\, , & H_{236}\, , & H_{245} \\
  \end{array} } \right], \\
& & \nonumber\\
& \omega_{aK} & = \begin{bmatrix}
  -\,  \omega_{35}{}^{2} \, , & \omega_{46}{}^{2} \, , & -\, \omega_{36}{}^{1} \, , & - \, \omega_{45}{}^{1} \\
  -\,  \omega_{51}{}^{4} \, , & -\, \omega_{61}{}^{3} \, , & \omega_{62}{}^{4} \, , & -\, \omega_{52}{}^{3} \\
  -\,  \omega_{13}{}^{6} \, , & -\, \omega_{14}{}^{5} \, , & - \,\omega_{23}{}^{5} \, , & \omega_{24}{}^{6} \\
  \end{bmatrix} \, ,\nonumber\\
& & \nonumber\\
& Q^{a}_{K} & = \begin{bmatrix}
   -\, Q_1{}^{46} , & Q_1{}^{35} , & - \, Q_2{}^{45}  , & -\, Q_2{}^{36}  \\
  -\,  Q_3{}^{62} , & - \, Q_4{}^{52} , & Q_3{}^{51}  , & - \,Q_4{}^{61}  \\
-\, Q_5{}^{24} , & - \,Q_6{}^{23} , & - \, Q_6{}^{14}  , & Q_5{}^{13}  \\
  \end{bmatrix}, \nonumber\\
& & \nonumber\\
& R_K & = \left[{\begin{array}{cccc}
 -\,  R^{246}\, , & R^{235}\, , & R^{145}\, , & R^{136} \\
  \end{array} } \right] \,. \nonumber
\eea
Let us restate that the fluxes appearing with three indices always correspond to the first formulation, i.e. the non-cohomology case. Moreover, as all the indices are denoted by numbers, in order to avoid any possible confusion let us also mention that the flux components in the second formulation are given with the `ordering' as per denoted in $\{ \omega_{aK}, Q^a{}_K\}$ where the first index ``$a$'' is counted by $h^{1,1}_-$ while the second index ``$K$" is counted by $(1+h^{2,1})$. In this example, we have $a = \{1, 2, 3\}$ and $K = \{0, 1, 2, 3\}$. As we see from the eqn. (\ref{eq:fluxconversion1}), there are no flux components present which are of the type $\omega_{ij}{}^i$ and $Q_i{}^{ij}$, and so the tracelessness conditions given in eqn. (\ref{eq:tracelessQw}) are automatically satisfied after imposing the full orientifold requirements. In addition, the normalization of the various forms is appropriately fixed as mentioned in eqn. (\ref{eq:intersectionBases}).

\subsubsection*{Second formulation}
In this setup we have $h^{1,1}_+(X_3/\sigma) =0$, and so no `hatted' fluxes counted by $\alpha$ indices are present, and subsequently the ten identities mentioned in Table \ref{tab_BIsSecondFormulation} produce only one class of non-trivial constraints given as under,
\bea
\label{eq:secondModelA}
& & H_{[K}\, R_{J]} = \, \omega_{a[K} \, Q^a{}_{J]} \,.
\eea
This subsequently results in six identities as $a \in\{1, 2, 3\}$ and $J, K \in \{0, 1, 2, 3\}$. It is clear that the $(HQ+\omega^2)$-type and $(R\omega+Q^2)$-type identities, which would be obtained via translating the first formulation later on, cannot be obtained from the second formulation. This is one of the main message we want to convey and so it is worth emphasising.

\subsubsection*{Cohomology version of the first formulation}
Now the plan is to compute the five classes of Bianchi identities of the first formulation using Table \ref{tab_BIsFirstFormulation} and subsequently to translate the same into cohomology form via using the conversion relations in eqn. (\ref{eq:fluxconversion1}). First we note that we have $H_{ijk} R^{ijk} = 0 = \omega_{jk}{}^i \, Q_i{}^{jk}$, and therefore the `extra constraint' of Table \ref{tab_BIsFirstFormulation} is trivially satisfied. This is well anticipated by the choice of the orientifold action itself which in the present setup also guarantees the so-called tracelessness conditions for the fluxes denotes as $\omega_{ij}{}^i = 0 = Q_i{}^{ij}$. Further, it turns out that the Bianchi identities in the class {\bf (I)} and class {\bf (V)} of the first formulation as presented in Table \ref{tab_BIsFirstFormulation}, are trivially satisfied. Moreover, the remaining three classes of identities result in a total of 48 flux constraints in which $(HQ+\omega^2)$-type and $(R\omega+Q^2)$-type have 12 constraints each while the remaining 24 constraints correspond to $(HR+\omega Q)$-type. Using the conversion relations, these identities can be classified as we discuss below.
\vskip0.2cm
\noindent
{\bf (II). $(HQ+\omega \omega)$-type :} This identity results in 12 flux constraints which are explicitly given in the eqn. (\ref{eq:12HQww0}) of the appendix \ref{sec_lengthyBIs}. All these 12 identities can be equivalently expressed in terms of two simple relations given as under
\bea
\label{eq:12HQww}
& & \hskip-1cm H_{(\underline i} \, Q^a{}_{\underline 0)} = \omega_{b\, (\underline i} \, \omega_{c \, \underline 0)} \,, \quad \,\, a\neq b \neq c \, \, \& \, \, i \neq a \, ,\\
& & \hskip-1cm H_{(\underline i} \, Q^a{}_{\underline j)}  = \omega_{b\, (\underline i} \, \omega_{c \, \underline j)} \,, \quad \, \, a\neq b \neq c \, \, \& \, \, i \neq a = j \,, \nonumber
\eea
where the bracket $(..)$ denotes the symmetrization of the underlined indices, and we have $\{a, b, c\} \in \{1, 2, 3\}$ and  $\{i, j\} \in \{1, 2, 3\}$. %Moreover, the above two constraints can also be joined into a single relation given as under,
%\bea
%& & \hskip-1cm H_{(\underline I} \, Q^a{}_{\underline J)} = \omega_{b\, (\underline I} \, \omega_{c \, \underline J)}\, \quad \quad \forall \, \, a\neq b %\neq c, \quad 0 \neq I \neq a, \quad J \in \{0, \, \, a \}. 
%\eea
Now let us take a note on the presence of $a\neq b \neq c$ in the above identity along with the existence of the only non-trivial triple-intersection number being $\kappa_{123} =1$ for this setup. This can be considered as indicative of some insight in the $(1,1)$-cohomology sector. The same leads to the observation that the above identities can be also expressed by using the intersection number as under,
\bea
& & \hskip-2cm H_{(\underline i} \, Q^a{}_{\underline 0)} = \frac{1}{2} \, \kappa_{abc}^{-1} \, \omega_{b\, (\underline i} \, \omega_{c \, \underline 0)},\, \quad \quad \, \, i \neq a \, ,\\
& & \hskip-2cm H_{(\underline i} \, Q^a{}_{\underline j)} = \frac{1}{2} \, \kappa_{abc}^{-1} \, \omega_{b\, (\underline i} \, \omega_{c \, \underline j)},\, \quad \quad\, \, i \neq j = a \, ,\nonumber
\eea
%\bea
%& & \hskip-2cm H_{(\underline I} \, Q^a{}_{\underline J)} = \frac{1}{2} \, \kappa_{abc}^{-1} \, \omega_{b\, (\underline I} \, \omega_{c \, \underline J)}%\, \quad \quad \forall \, \, a, \quad 0 \neq I \neq a, \quad J \in \{0, \, \, a \} \, ,
%\eea
where we have defined $\kappa_{abc}^{-1} = 1/\kappa_{abc}$ for the all the fixed values of $a, b$ and $c$, whenever $\kappa_{abc}$ is non-zero. %This suggest that it might be possible that the missing identities could be generically determined from the topological numbers such as Hodge number and the intersection numbers.
\vskip0.2cm
\noindent
{\bf (III). $(HR +\omega Q)$-type :} This identity results in a total of 24 flux constraints which are explicitly given the eqn. (\ref{eq:24HRwQ0}) of the appendix \ref{sec_lengthyBIs}. However after some reshuffling, we realize that this number can be effectively reduced to 18 constraints, and the same can be collected in the following compact form,
\bea
\label{eq:24HRwQ}
& \hskip-1cm (i). & \quad H_{[K}\, R_{J]} = \, \omega_{a[K} \, Q^a{}_{J]}, \qquad \forall \, \, I, \, J \in \{0, 1, 2, 3\}\,, \\
& & \nonumber\\
& \hskip-1cm (ii). & \quad H_{(\underline i} \, R_{\underline 0)} = \omega_{a' \, (\underline i} \, Q^{a'}{}_{\underline 0)} \,,  \quad \quad \, \, i =a', \, \quad a' \, {\rm is \, \, not \, \, summed} \, ,\nonumber\\
& \hskip-1cm (iii). & \quad H_{(\underline i} \, R_{\underline j)} = \omega_{a' \, (\underline i} \, Q^{a'}{}_{\underline j)} \,,  \quad \quad \, \, i \neq j \neq a', \, \quad a' \, {\rm is \, \, not \, \, summed} \, , \nonumber\\
& \hskip-1cm (iv). & \quad \omega_{a' (\underline i} \, Q^{a'}{}_{\underline 0)} = \omega_{b' \, (\underline i} \, Q^{b'}{}_{\underline 0)} \,,  \quad \, \, a' \neq b' \neq i, \, \quad a' \,, b' \, {\rm are \, \, not \, \, summed} \, ,\nonumber\\
& \hskip-1cm (v). & \quad \omega_{a' (\underline i} \, Q^{a'}{}_{\underline j)} = \omega_{b' \,(\underline i} \, Q^{b'}{}_{\underline j)} \,,  \, \quad \, \, a' = i, \, j = b', \, i \neq j, \quad a' \,,  b' \, {\rm are \, \, not \, \, summed} \, , \nonumber
\eea
where the bracket $(..)$ is used to present the symmetrization of the underlined indices while the bracket $[..]$ is used for anti-symmetrization of two indices, and the primed indices $a'$ and $b'$ are not summed over.
%\bea
%\label{eq:24HRwQ}
%& \hskip-1cm (i) & H_{[K}\, R_{J]} = \, \omega_{a[K} \, Q^a{}_{J]}, \quad \forall \, \, a \in \{1, 2, 3\} \, \, {\rm and} \, \, \{I, J\} \in \{0, 1, 2, 3\} \,, \\
%& \hskip-1cm (ii) & H_i \, R_{0} + R_0 \, H_{i} = \omega_{a' i} \, Q^{a'}{}_{0} + \omega_{a' 0} \, Q^{a'}{}_{i} \,,  \quad i =a', \, \quad a' \, {\rm not \, \, summed} \, ,\nonumber\\
%& \hskip-1cm (iii) & H_i \, R_{j} + R_j \, H_{i} = \omega_{a' i} \, Q^{a'}{}_{j} + \omega_{a' j} \, Q^{a'}{}_{i} \,,  \quad i \neq j \neq a', \, \quad a' \, {\rm not \, \, summed} \, , \nonumber\\
%& \hskip-1cm (iv) & \omega_{a' i} \, Q^{a'}{}_{0} + \omega_{a' 0} \, Q^{a'}{}_{i}  = \omega_{b' i} \, Q^{b'}{}_{0} + \omega_{b' 0} \, Q^{b'}{}_{i} \,,  \quad a' \neq b' \neq i, \, \quad a' \, b' \, {\rm \, not \, \, summed} \, ,\nonumber\\
%& \hskip-1cm (v) & \omega_{a' i} \, Q^{a'}{}_{j} + \omega_{a' j} \, Q^{a'}{}_{i} = \omega_{b' i} \, Q^{b'}{}_{j} + \omega_{b' j} \, Q^{b'}{}_{i} \,,  \quad i \neq j \neq a' \neq b', \, \quad a' \, b' \, {\rm \, \, not \, \, summed} \, , \nonumber
%\eea
%where $a'$ and $b'$ are not summed. 
In this collection (\ref{eq:24HRwQ}), the first identity precisely corresponds to the six identities of the second formulation given in eqn. (\ref{eq:secondModelA}) while the remaining 12 identities cannot be derived from the known version of the second formulation. Also there are non-unique ways of reshuffling the set of missing identities, and it is hard to guess any $(1,1)$-cohomology sector insights in terms of intersection number $\kappa_{abc}$ as we have seen in the previous case. However after making some back and forth checks on the compatibility of this model with the model B (to be discussed later on), we propose the following version of the 12 missing identities in eqn. (\ref{eq:24HRwQ}),
\bea
& & H_{(\underline 0} \, R_{\underline k)} - \, \omega_{a \, (\underline 0} \, Q^{a}{}_{\underline k)} \, \delta_{a k} =0\,,\, \hskip3.75cm \forall \,\, k\,; \\
& & H_{(\underline 0} \, R_{\underline k)} - \omega_{a \, (\underline 0} \, Q^{a}{}_{\underline k)} + 2\, \omega_{a \, (\underline 0} \, Q^{a}{}_{\underline j)}\, \, \delta_{aj} = 0\,,\qquad \qquad \, \, j \neq k \,; \nonumber\\
& & H_{(\underline j} \, R_{\underline k)} - \omega_{a \, (\underline j} \, Q^{a}{}_{\underline k)} + 2\, \omega_{a \, (\underline j} \, Q^{a}{}_{\underline k)}\, \, \delta_{aj} = 0\,, \qquad \qquad  \, \, j \neq k \,;\nonumber\\
& & H_{(\underline j} \, R_{\underline k)} - \, \omega_{a \, (\underline j} \, Q^{a}{}_{\underline k)} \, \delta_{ai} =0\,, \hskip3.9cm \, \, i \neq j \neq k\,. \nonumber
\eea
\vskip0.2cm
\noindent
{\bf (IV). $(R\omega+Q Q)$-type :} This identity results in 12 flux constraints which are explicitly given the eqn. (\ref{eq:12RwQQ0}) of the appendix \ref{sec_lengthyBIs}. All these 12 identities can be equivalently expressed in terms of two simple relations given as under
\bea
\label{eq:12RwQQ}
& & \hskip-1cm R_{(\underline i} \, \omega_{a \underline 0)} = Q^{c}{}_{(\underline i} \, Q^{b}{}_{\underline 0)}\,, \quad  \, \, a\neq b \neq c \, \, \& \, \, i \neq a \,;\\
& & \hskip-1cm R_{(\underline i} \, \omega_{a \underline j)} = Q^{c}{}_{(\underline i} \, Q^{b}{}_{\underline j)}\,, \quad \,\, a\neq b \neq c \, \, \& \, \, i \neq a = j \,, \nonumber
\eea
where the bracket $(..)$ denotes the symmetrization of the underlined indices, and we have $\{a, b, c\} \in \{1, 2, 3\}$ and  $\{i, j\} \in \{1, 2, 3\}$. 
%Moreover, the above two constraints can also be joined into a single relation given as under,
%\bea
%& & \hskip-1cm R_{(\underline I} \, \omega_{a \underline J)} = Q^{c}{}_{(\underline I} \, Q^{b}{}_{\underline J)}\,, \quad {\rm where} \quad a\neq b %\neq c, \quad 0 \neq I \neq a, \quad J \in \{0, \, \, a \}\,. 
%\eea
Now let us again note that the presence of $a\neq b \neq c$ in the above identity and existence of the only non-trivial triple-intersection numbers being $\kappa_{abc} =1$ for this setup can be considered as indicative of some insight as one can rewrite the same as under,
\bea
& & \hskip-2cm R_{(\underline i} \, \omega_{a \underline 0)} = \frac{1}{2} \, \kappa_{abc} \, Q^{b}{}_{(\underline i} \, Q^{c}{}_{\underline 0)}\,, \quad \quad  \, \,  i \neq a\,; \\
& & \hskip-2cm R_{(\underline i} \, \omega_{a \underline j)} = \frac{1}{2} \, \kappa_{abc} \, Q^{b}{}_{(\underline i} \, Q^{c}{}_{\underline j)}\,, \quad \quad \, \,  i \neq j = a\,. \nonumber
\eea
%\bea
%& & \hskip-2cm R_{(\underline I} \, \omega_{a \underline J)} = \frac{1}{2} \, \kappa_{abc} \, Q^{b}{}_{(\underline I} \, Q^{c}{}_{\underline J)}\, \quad %\quad \forall \, \, a, \quad 0 \neq I \neq a, \quad J \in \{0, \, \, a \} \,.
%\eea
{\it This analysis and the subsequent observations in Model A suggest that it might be possible that the missing identities could be generically determined from the topological numbers such as Hodge number and the intersection numbers of the complex threefold background.} This has been our central aim to achieve.

\subsection{Missing identities in Model B}
In this section we will compute the Bianchi identities using the two formulations for our Model B, which corresponds to a type IIA setup with a ${\mathbb T}^6/{\mathbb Z}_4$-orientifold. Considering the untwisted sector, in this model we have $h^{1,1}_+ = 1, \, h^{1,1}_- = 4$ and $h^{2,1} = 1$ which results in $22$ second formulation identities for a total number of 24 flux components. 

Using the orbifold of ${\mathbb T}^6/{\mathbb Z}_4$ sixfold we consider {\it two explicit constructions} which differ in the choice of complexified coordinate of the ${\mathbb T}^6$ torus and their three-form bases. Despite these differences, as the global topological quantities in the untwisted sector such as the Hodge numbers and the triple intersection numbers are the same in these two constructions, so we naively expected some correlation of their respective sets of Bianchi identities in the cohomology formulations, and hence also among the missing Bianchi identities. This indeed turns out to be the case as we will elaborate in this section. We call these two constructions as Model B1 and Model B2 with the following details,
\begin{itemize}
\item{{\bf Model B1:} This construction was used for studying moduli stabilization using standard fluxes and without including any non-geometric fluxes \cite{Ihl:2006pp}. We have explicitly computed all the (non-)geometric flux components allowed in this setup, and the relevant details about the setup is presented in the appendix \ref{sec_setups}. However, let us present here the following flux conversion relations which we use for translating the first formulation identities to capture the missing identities,
\bea
\label{eq:fluxconversion2}
\hskip-4.5cm & & \hskip-0.5cm H_K = \left[{\begin{array}{cccc}
   H_{136}\, , & - \, H_{135} \\
  \end{array} } \right], \\
%& & \nonumber\\
& & \hskip-0.5cm  \omega_{aK} = \begin{bmatrix}
  \omega_{46}{}^{1} \, , & - \, \omega_{35}{}^{1} \\
  \omega_{62}{}^{3} \, , & \, - \, \omega_{51}{}^{3} \\
  -\, \omega_{13}{}^{5} \, , &  -\, \omega_{13}{}^{6} \\
\frac{1}{2}\left(\omega_{26}{}^{1}  + \omega_{36}{}^{3} \right)\, , & \quad -\frac{1}{2}\left(\omega_{15}{}^{1} - \omega_{45}{}^{3}\right)\,\\
  \end{bmatrix} \,, \nonumber\\
& & \nonumber\\
& & \hskip-0.5cm \hat\omega_{\alpha}{}^{K} = \begin{bmatrix}
    \frac{1}{2}\left(\omega_{15}{}^{1} +\omega_{45}{}^{3}\right)\, , &\, \, -\frac{1}{2}\left(\omega_{26}{}^{1}  - \omega_{36}{}^{3}\right) \,\\
  \end{bmatrix} \, ,\nonumber\\
& & \nonumber\\
& & \hskip-0.5cm Q^{a}{}_{K} = \begin{bmatrix}
    Q_1{}^{35} \, , & - \,Q_1{}^{36}  \\
     Q_3{}^{51} \, , & -\, Q_3{}^{61}  \\
    -\, Q_6{}^{13}\, , &  Q_5{}^{13}  \\
    \frac{1}{2}\left(Q_2{}^{51}  - Q_3{}^{35} \right) \, , & \, \,   -\frac{1}{2}\left(Q_1{}^{16} - Q_3{}^{46}\right)  \\
  \end{bmatrix}, \nonumber\\
& & \nonumber\\
& & \hskip-0.5cm \hat{Q}^{\alpha K} = \begin{bmatrix}
    \frac{1}{2}\left(Q_1{}^{16} + Q_3{}^{46}\right)\,, &  \, \,  -\frac{1}{2}\left(Q_2{}^{51} + Q_3{}^{35}\right) \\
  \end{bmatrix}\, , \nonumber\\
%& & \nonumber\\
& & \hskip-0.5cm R_K = \left[{\begin{array}{cccc}
 R^{135}\, , & R^{136} \\
  \end{array} } \right] \, .\nonumber
\eea
%Here the flux components are counted by $a =\{1, 2, 3, 4\}, \, \alpha = 1$ and $K =\{0, 1\}$, and the `ordering' in the indices are as per their symbolic notation.
}
\item{{\bf Model B2:} The second construction uses a different set of complexified coordinates $z^i$ on the ${\mathbb T^6}$ torus and also a different set of even/odd three-form bases. This construction was previously used for studying the supersymmetric moduli stabilization in \cite{Ihl:2007ah} and for a symplectic version of the scalar potential in \cite{Gao:2017gxk}. The relevant details about the setup is briefly presented in the appendix \ref{sec_setups}. However, here we present the following flux conversion relations which we use for translating the first formulation identities to capture the missing identities,
\bea
\label{eq:fluxconversion3}
& & \hskip-0.75cm H_K = \left[{\begin{array}{cccc}
   H_{135} - H_{136}\, , & - \, H_{136} \\
  \end{array} } \right], \\
%& & \nonumber\\
& & \hskip-0.75cm  \omega_{aK} = \begin{bmatrix}
  \omega_{36}{}^{1} \, , & - \, \omega_{46}{}^{1} \\
  \omega_{61}{}^{3} \, , & - \, \omega_{62}{}^{3} \\
  \omega_{13}{}^{5} +\omega_{13}{}^{6} \, , &  \omega_{13}{}^{5} \\
 \frac{1}{2}\left(\omega_{16}{}^{1} - \omega_{26}{}^{1}  - \omega_{36}{}^{3} - \omega_{46}{}^{3}\right)\, , & \qquad  \frac{1}{2}\left(- \omega_{16}{}^{1} - \omega_{26}{}^{1}  - \omega_{36}{}^{3} + \omega_{46}{}^{3}\right)\,\\
  \end{bmatrix} \,, \nonumber\\
& & \nonumber\\
& & \hskip-0.75cm \hat\omega_{\alpha}{}^{K} = \begin{bmatrix}
     \frac{1}{2}\left(\omega_{16}{}^{1} - \omega_{26}{}^{1}  + \omega_{36}{}^{3} + \omega_{46}{}^{3}\right)\, , &\quad  \frac{1}{2}\left(\omega_{16}{}^{1} + \omega_{26}{}^{1}  - \omega_{36}{}^{3} + \omega_{46}{}^{3}\right)\,\\
  \end{bmatrix} \, ,\nonumber\\
& & \nonumber
\eea
\bea
& & \hskip-0.75cm Q^{a}{}_{K} = \begin{bmatrix}
    - Q_1{}^{35} \, , & - Q_1{}^{35} - Q_1{}^{36}  \\
    - Q_3{}^{51} \, , & - Q_3{}^{51} - Q_3{}^{61}  \\
    - Q_5{}^{13} + Q_6{}^{13}\, , &  Q_6{}^{13}  \\
    - \frac{1}{2}\left(Q_1{}^{15} + Q_2{}^{51}  - Q_3{}^{35} + Q_4{}^{53} \right) \, , & \qquad   \frac{1}{2}\left(Q_1{}^{15} - Q_2{}^{51}  + Q_3{}^{35} + Q_4{}^{53}\right)  \\
  \end{bmatrix}, \nonumber\\
& & \nonumber\\
& & \hskip-0.75cm \hat{Q}^{\alpha K} = \begin{bmatrix}
     \frac{1}{2}\left(-Q_1{}^{15} - Q_2{}^{51}  - Q_3{}^{35} + Q_4{}^{53}\right)\,, &  \quad  \frac{1}{2}\left(-Q_1{}^{15} + Q_2{}^{51}  + Q_3{}^{35} + Q_4{}^{53}\right) \\
  \end{bmatrix}\, , \nonumber\\
& & \nonumber\\
& & \hskip-0.75cm R_K = \left[{\begin{array}{cccc}
  -\, R^{135} -\, R^{136} , & - R^{135} \\
  \end{array} } \right] \, .\nonumber
\eea
}
\end{itemize}
Unlike the conventions used in \cite{Gao:2017gxk}, here we fix our six-form to be integrated to unity over the Calabi Yau, and the four-forms being dual to the respective two-forms as mentioned in eqn. (\ref{eq:intersectionBases}). This helps in avoiding many normalizaton factors which otherwise would appear in the Bianchi identities and make them look more complicated, and hence more difficult to reshuffle and capture the missing identities. %However, let us also mention that the non-normalized conventions has helped us in invoking the structural insights in missing Bianchi identities, and after making the educated guess, we preferred to adopt the simpler conventions to illustrate the results with a cleaner presentation. 
The main motivation for considering the two models B1 and B2 which qualitatively look similar has been the fact that as they have different three-form bases, there are different intersection numbers on the mirror threefold, which facilitates some more freedom to make an educated guess for the $(2,1)$-cohomology sector of the Bianchi identities.

Considering all the non-zero flux components (for Model B1 and Model B2) as given in eqns. (\ref{eq:fluxcompB1}) and (\ref{eq:fluxcompB2}), it is evident that the generic tracelessness conditions (involving the summing over indices) as given in eqn. (\ref{eq:tracelessQw}) are automatically satisfied. However, from the flux conversion relations given in eqns. (\ref{eq:fluxconversion2}) and (\ref{eq:fluxconversion3}), one can observe a couple of non-zero flux components of the form $\omega_{i'j'}{}^{i'}$ or $Q_{i'}{}^{i'j'}$, where $i'$ is not summed over, being still allowed by the orientifold projection. This may create a curiosity/suspicion that may be after imposing the condition to make such flux components vanish, the mismatch goes away. To prove that it is not the case, one may consider our Model A itself in which the tracelessness condition (\ref{eq:tracelessQw}) is automatically satisfied along with having no flux components with a single free-index being allowed. For Model B also, we will show that even after imposing such a condition,  the `mismatch' between the two formulations does not go away. After imposing that each of the flux components with single free index vanish,
%\bea
%& & \omega_{15}{}^{1} = \omega_{25}{}^{2} = \omega_{36}{}^{3} = \omega_{46}{}^{4} =0, \qquad Q_1{}^{16} = Q_2{}^{26} = Q_3{}^{35} = Q_4{}^{45} =0\,.
%\eea
we find the following identifications that hold for both of the models B1 and B2,
\bea
\label{eq:tracelessQw1}
& & \hskip-1.5cm \omega_{40} = -\, \hat{\omega}_1{}^1\, , \quad \omega_{41} =\, \hat{\omega}_1{}^0 , \qquad \qquad Q^4{}_0 = -\, \hat{Q}^{11}, \quad Q^4{}_1 = \, \hat{Q}^{10}\,.
\eea
Thus we find that by imposing this conditions in eqn. (\ref{eq:tracelessQw1}), the effective flux components for $\omega$-flux as well as for the $Q$-flux get reduced from 10 to 8 in both the (cohomology and the non-cohomology) formulations, and hence the bijection between the respective flux counting in the two formulation remains intact.

\subsubsection*{Second formulation}
Unlike the previous model A, this setup has $h^{1,1}_+(X_3/\sigma) = 1$, and so the `hatted' fluxes counted by $\alpha$ indices are non-trivial. Subsequently, none of the second formulation identities mentioned in Table \ref{tab_BIsSecondFormulation} are identically trivial. Therefore, one might expect that this model would help us getting more insights of the Bianchi identities and the mismatch. The ten identities mentioned in Table \ref{tab_BIsSecondFormulation} produce  22 non-trivial constraints for the second formulation. All these constraints and their number can be explicitly read-off from the Table \ref{tab_BIsSecondFormulation} by considering $\alpha = 1, \, a\in \{1, 2, 3, 4 \}, \, K \in \{0, 1\}$ and the topological data given in eqn. (\ref{eq:intExB}). The explicit form of all these 22 second formulation constraints are listed in eqns. (\ref{eq:ModelBHwAndRQ}) and in eqns. (\ref{eq:ModelBHQwwExpand0}),  (\ref{eq:ModelBHRwQExpand0}) and (\ref{eq:ModelBRwQQExpand0}) of the appendix \ref{sec_lengthyBIs}.

\subsubsection*{Cohomology version of the first formulation}
Now we will compute the first formulation Bianchi identities given in Table \ref{tab_BIsFirstFormulation} and subsequently we will translate them into the cohomology formulation using the conversion relations given in eqn. (\ref{eq:fluxconversion2}) for Model B1, and in eqn. (\ref{eq:fluxconversion3}) for Model B2. Performing a similar analysis to what has been done for the Model A, we find that $H_{ijk} R^{ijk} = 0 = \omega_{jk}{}^i \, Q_i{}^{jk}$, and therefore the `extra constraint' of Table \ref{tab_BIsFirstFormulation} is trivially satisfied. As argued before, this should always be so due to the orientifold projection. Further, the first and the last Bianchi identities of the first formulation which corresponds to the class {\bf (I)} and class {\bf (V)} do not have any mismatch, although unlike the previous case in Model A, now these identities are generically non-trivial, and are explicitly given as,
\bea
\label{eq:ModelBHwAndRQ}
& & H_0 \, \hat\omega_1{}^0 + H_1 \, \hat\omega_1{}^1 = 0\, , \qquad \qquad \qquad R_0 \, {\hat Q}^{10} + R_1 \, {\hat Q}^{11} = 0 \,.
\eea
These are precisely the second formulation identities $H_K \, {\hat \omega}_{\alpha}{}^{K} = 0 = R_K \, {\hat Q}^{\alpha K}$ where $\alpha = 1$ and $K=\{0, 1\}$. The mismatch in the two formulation lies only in the class {\bf (II)}, {\bf (III)} and {\bf (IV)}. The explicit forms of all the identities belonging to these three classes are collected in the appendix \ref{sec_lengthyBIs}. Let us also mention that in each of the class {\bf (II)}, {\bf (III)} and {\bf (IV)} we do recover all the Bianchi identities of the second formulation, which are explicitly presented in eqns. (\ref{eq:ModelBHQwwExpand0}),  (\ref{eq:ModelBHRwQExpand0}) and (\ref{eq:ModelBRwQQExpand0}). %, and therefore from now onwards we will mainly focus on the missing identities only for the deeper analysis.

\subsubsection*{(II). \, $(HQ+\omega\omega)$-type : }
A simple conversion of all the identities of this type using the expressions in Table \ref{tab_BIsFirstFormulation} results in 16 identities after making some tedious reshuffling of the constraints. Subsequently we find that 5 of the 16 constraints are exactly the ones which belong to the second formulation and their explicit form is given in eqn. (\ref{eq:ModelBHQwwExpand0}). In addition, there are 11 constraints collected in eqns. (\ref{eq:ModelBHQwwExpand1}) and  (\ref{eq:ModelBHQwwExpand2}) in the appendix \ref{sec_lengthyBIs} which cannot be derived from the second formulation. Moreover, imposing the condition in eqn. (\ref{eq:tracelessQw1}) further reduces the number of missing identities to 10, which can be expressed in the following manner, 
\bea
\label{eq:ModelBHQww0}
& \hskip-1.0cm {\rm Model \,\, B1\, \& \, B2:}  & \quad H_{(\underline I} \, Q^a{}_{\underline J)} = \frac{1}{2}\, \kappa_{abc}^{-1} \, \omega_{b \, (\underline I} \, \omega_{c \, \underline J)} %+ \frac{1}{2} \, \hat{\kappa}_{a\alpha\beta}^{-1} \, \hat{\omega}_{\alpha}{}^{(\underline 0} \, \hat{\omega}_{\beta}{}^{\underline 1)} = 0
\,, \quad {\rm for} \, \, a \in \{1, 2, 4\}\, \, {\rm and} \, \, \forall \, I, \, J \,;\\
%& & \nonumber\\
& \hskip-1.5cm {\rm Model \,\, B1:} & \quad H_{(\underline 1} \, Q^a{}_{\underline 0)} - \frac{1}{2}\, \kappa_{abc}^{-1} \, \omega_{b \, (\underline 1} \, \omega_{c \, \underline 0)} - \frac{1}{2} \, \hat{\kappa}_{a\alpha\beta}^{-1} \, \hat{\omega}_{\alpha}{}^{(\underline 1} \, \hat{\omega}_{\beta}{}^{\underline 0)} = 0\,,  \, \, \, \, \, \quad {\rm for} \, \, a  = 3 \,;\nonumber \\
& \hskip-1.5cm {\rm Model \,\, B2:} & \quad H_{(\underline 0} \, Q^a{}_{\underline 0)} - \frac{1}{2}\, \kappa_{abc}^{-1} \, \omega_{b \, (\underline 0} \, \omega_{c \, \underline 0)} - \frac{1}{2} \, \hat{\kappa}_{a\alpha\beta}^{-1} \, \hat{\omega}_{\alpha}{}^{(\underline 0} \, \hat{\omega}_{\beta}{}^{\underline 0)} \, \nonumber\\
& & \hskip1cm = \, H_{(\underline 1} \, Q^a{}_{\underline 1)} - \frac{1}{2} \, \kappa_{abc}^{-1} \, \omega_{b \, (\underline 1} \, \omega_{c \, \underline 1)} - \frac{1}{2}\, \hat{\kappa}_{a\alpha\beta}^{-1} \, \hat{\omega}_{\alpha}{}^{(\underline 1} \, \hat{\omega}_{\beta}{}^{\underline 1)}\,,  \quad {\rm for} \, \, a  = 3 \,.\nonumber
%& & \nonumber\\
%& (iii). & \quad (\hat{d}^{-1})_\beta{}^\alpha \, H_{(\underline 0} \, \hat{Q}^{\beta \underline 1)} = \frac{1}{2} \, \hat{\kappa}_{a\alpha\beta}^{-1} \, {\omega}_{a (\underline 0} \, \hat{\omega}_{\beta}{}^{\underline 1)}\,, \qquad  \forall \, \, \{\alpha, \, \beta\}, \, \, {\rm i. \, e. \,} \, \,  \alpha = 1 = \beta \, . \nonumber
\eea
Here as before, the underlined indices within $(..)$ are symmetrized, and we have considered $\kappa_{abc}^{-1} = 1/\kappa_{abc}$ and $\hat{\kappa}_{a\alpha\beta}^{-1} = 1/\hat{\kappa}_{a\alpha\beta}$ for fixed values of $\{a, b, c\}$ and $\{\alpha, \beta\}$, and whenever these intersections are non-zero. These topological numbers are given in eqn. (\ref{eq:intExB}). 

\subsubsection*{(III). \, $(HR + \omega Q)$-type : }
This constitutes the most complicated part of the Bianchi identities, and a simple conversion of all the identities of this type using the expressions in Table \ref{tab_BIsFirstFormulation}  results in many complicated constraints heavily mixing all the four NS-NS fluxes, and a priori it does not reflect that one would precisely recover all the identities of the second formulation. However some rigorous reshuffling of various flux constraints results in 26 identities which indeed includes all the 10 identities of the second formulation as collected in eqn. (\ref{eq:ModelBHRwQExpand0}) of the appendix \ref{sec_lengthyBIs}. Further, the remaining 16 missing Bianchi identities are collected in eqns. (\ref{eq:ModelBHRwQExpand1}) and  (\ref{eq:ModelBHRwQExpand2}) of the appendix \ref{sec_lengthyBIs}. These are the ones which cannot be derived from the second formulation. Moreover, imposing the condition in eqn. (\ref{eq:tracelessQw1}) further reduces the number of missing identities into 13, which can be expressed in the following manner, 
\bea
\label{eq:ModelBHRwQ0}
& \hskip-1.1cm {\rm Model \,\, B1\, \& \, B2:}  & H_{(\underline I} \, R_{\underline J)} = \omega_{3\, (\underline I}\, Q^3{}_{\underline J)}, \quad \quad \forall \, \, \{I, \, J\} \,;\\
& \hskip-1.5cm & \kappa_{3ac} \, Q^c{}_{(\underline I} \, \omega_{b\, \underline J)}\,= \kappa_{3bc} \, Q^c{}_{(\underline I} \, \omega_{a\, \underline J)}, \quad \forall \, \, a, b \in \{1, \, 2, \, 4\}, \quad \forall \, \, \{I, \, J\}\,;\nonumber\\
%& \hskip-1.5cm (iii). & \hat{\kappa}_{3 \alpha\beta}\, \omega_{a (\underline I} \, \hat{Q}^{\beta \underline J)} = {\kappa}_{3ab} \, Q^b{}_{(\underline I} \, \hat{\omega}_\alpha{}^{\underline J)}, \quad \forall \, \, \alpha, \, \, \forall \, \, a \in \{1, \, 2, \, 4\}, \,\, \forall \, \, I \neq J, \nonumber\\
& \hskip-1.5cm {\rm Model \,\, B1:} & 3\, H_{(\underline 0} \, R_{\underline 1)} - \, \omega_{a (\underline 0} \, Q^a{}_{\underline 1)}  - \, \hat\omega_\alpha{}^{(\underline 0} \, \hat{Q}^{\alpha \underline 1)}= 0\,;\nonumber\\
& \hskip-1.5cm {\rm Model \,\, B2:} & 3\, H_0 \, R_0 - \, \omega_{a0} \, Q^a{}_{0}  - \, \hat\omega_\alpha{}^{0} \, \hat{Q}^{\alpha 0} - H_1 \, R_1 + \, \omega_{a 1} \, Q^a{}_{1}  + \, \hat\omega_\alpha{}^{1} \, \hat{Q}^{\alpha 1} =0\, . \nonumber
\eea

\subsubsection*{(IV). \, $(R\omega +Q Q)$-type : }
A simple conversion of all the identities of this type using the expressions in Table \ref{tab_BIsFirstFormulation} results in 16 constraints after making some tedious reshuffling of pieces. Subsequently we find that 5 of the 16 constraints are exactly the ones which belong to the second formulation and their explicit form is given in eqn. (\ref{eq:ModelBRwQQExpand0}). In addition, there are 11 constraints collected in eqns. (\ref{eq:ModelBRwQQExpand1}) and  (\ref{eq:ModelBRwQQExpand2}) in the appendix \ref{sec_lengthyBIs} which cannot be derived from the second formulation. Moreover, imposing the condition in eqn. (\ref{eq:tracelessQw1}) further reduces the number of missing identities to 10, which can be expressed in the following manner, 
\bea
\label{eq:ModelBRwQQ0}
& \hskip-0.50cm {\rm Model \,\, B1\, \& \, B2:} & \quad R_{(\underline I} \, \omega_{a \, \underline J)} = \frac{1}{2} \, \kappa_{abc} \, Q^{b}{}_{(\underline I} \, Q^{c}{}_{\underline J)} %+ \frac{1}{2} \, \hat{\kappa}_{a\alpha\beta} \, \hat{Q}^{\alpha \, (\underline 0} \, \hat{Q}^{\beta \, \underline 1)} = 0
\,, \qquad {\rm for} \, \, a \in \{1, 2, 4\}\, \, {\rm and} \, \, \forall \, I, \, J \, , \\
%& & \nonumber\\
& {\rm Model \, \, B1:} & \quad  R_{(\underline 1} \, \omega_{a \, \underline 0)} - \frac{1}{2} \, \kappa_{abc} \, Q^{b}{}_{(\underline 1} \, Q^{c}{}_{\underline 0)} - \frac{1}{2} \, \hat{\kappa}_{a\alpha\beta} \, \hat{Q}^{\alpha \, (\underline 1} \, \hat{Q}^{\beta \, \underline 0)} = 0\,, \, \, \, \quad \quad {\rm for} \, \, a  = 3 \,,\nonumber \\
& {\rm Model \, \, B2:} & \quad R_{(\underline 0} \, \omega_{a \, \underline 0)} - \frac{1}{2} \, \kappa_{abc} \, Q^{b}{}_{(\underline 0} \, Q^{c}{}_{\underline 0)} - \frac{1}{2} \, \hat{\kappa}_{a\alpha\beta} \, \hat{Q}^{\alpha \, (\underline 0} \, \hat{Q}^{\beta \, \underline 0)} \, \nonumber\\
& & \hskip1cm = R_{(\underline 1} \, \omega_{b \, \underline 1)} - \frac{1}{2} \, \kappa_{abc} \, Q^{b}{}_{(\underline 1} \, Q^{c}{}_{\underline 1)} - \frac{1}{2} \, \hat{\kappa}_{a\alpha\beta} \, \hat{Q}^{\alpha \, (\underline 1} \, \hat{Q}^{\beta \, \underline 1)}\,, \qquad {\rm for} \, \, a  = 3 \,.\nonumber
%& & \nonumber\\
%& (iii). & \quad f^{-1} \, R_{(\underline 0} \, \hat\omega_{\alpha}{}^{\underline 1)} = \frac{1}{2} \, \tilde{k}_{a\alpha\beta} \, {Q}^{a}{}_{(\underline 0} \, \hat{Q}^{\beta \, \underline 1)}\,, \qquad  \forall \, \, \{\alpha, \, \beta\}, \, \, {\rm i. \, e. \,} \, \,  \alpha = 1 = \beta \,, \nonumber
\eea

\section{On generic structure of the missing identities}
\label{sec_BIsGeneric}
In the previous section we have presented some educated guess for the cohomology structure in the K\"ahler moduli space, i.e. in the $(1,1)$-cohomology sector via intersection numbers $\kappa_{abc}$ and $\hat{\kappa}_{a\alpha\beta}$. Now we plan to investigate the $(2,1)$-cohomology structure on the side of the complex structure moduli space, via looking at the intersection numbers on the mirror threefold. 

\subsection{Insights for the $(1,1)$-cohomology sector}
First we collect the results regarding the $(1,1)$-cohomology sector by presenting all the missing identities at one place which are given as under,
\subsubsection*{Model A}
%There are a total of 36 missing identities in model A, and each of the three classes have 12 such identities collected as under,
\bea
\label{eq:modelAH11}
& \hskip-0cm  {\rm \bf (II):} & \quad H_{(\underline i} \, Q^a{}_{\underline 0)} = \frac{1}{2} \, \kappa_{abc}^{-1} \, \omega_{b\, (\underline i} \, \omega_{c \, \underline 0)},\, \hskip4cm i \neq a \,; \\
& & \hskip-0cm \quad H_{(\underline i} \, Q^a{}_{\underline j)} = \frac{1}{2} \, \kappa_{abc}^{-1} \, \omega_{b\, (\underline i} \, \omega_{c \, \underline j)},\, \hskip4cm i \neq j = a \,; \nonumber\\
& \hskip-0cm  {\rm \bf (III):} & \quad  H_{(\underline 0} \, R_{\underline k)} - \, \omega_{a \, (\underline 0} \, Q^{a}{}_{\underline k)} \, \delta_{a k} =0,\, \hskip3.75cm \forall \, \, k\,;\nonumber\\
& & \quad  H_{(\underline 0} \, R_{\underline k)} - \omega_{a \, (\underline 0} \, Q^{a}{}_{\underline k)} + 2\, \omega_{a \, (\underline 0} \, Q^{a}{}_{\underline j)}\, \, \delta_{aj} = 0\,, \qquad \qquad j \neq k \,;\nonumber\\
& & \quad  H_{(\underline j} \, R_{\underline k)} - \omega_{a \, (\underline j} \, Q^{a}{}_{\underline k)} + 2\, \omega_{a \, (\underline j} \, Q^{a}{}_{\underline k)}\, \, \delta_{aj} = 0\,, \qquad \qquad  j \neq k \,; \nonumber\\
& & \quad  H_{(\underline j} \, R_{\underline k)} - \, \omega_{a \, (\underline j} \, Q^{a}{}_{\underline k)} \, \delta_{ai} =0,\, \hskip3.8cm i \neq j \neq k\,; \nonumber\\
& \hskip-0cm  {\rm \bf (IV):} & \quad R_{(\underline i} \, \omega_{a \underline 0)} = \frac{1}{2} \, \kappa_{abc} \, Q^{b}{}_{(\underline i} \, Q^{c}{}_{\underline 0)},\, \hskip4.1cm \, \,  i \neq a\,;\nonumber\\
& \hskip-0cm & \quad R_{(\underline i} \, \omega_{a \underline j)} = \frac{1}{2} \, \kappa_{abc} \, Q^{b}{}_{(\underline i} \, Q^{c}{}_{\underline j)},\, \hskip4.1cm \, \,  i \neq j = a\,. \nonumber
\eea
\subsubsection*{Model B}
%There are a total of 38 missing identities in model B, in which the classes {\bf (II)} and {\bf (IV)} have 11 each while the class {\bf (III)} has 16 identities. These are collected as under,
\bea
\label{eq:modelBH11}
& {\rm \bf (II):}& \quad H_{(\underline I} \, Q^a{}_{\underline J)} = \frac{1}{2}\, \kappa_{abc}^{-1} \, \omega_{b \, (\underline I} \, \omega_{c \, \underline J)} %+ \frac{1}{2} \, \hat{\kappa}_{a\alpha\beta}^{-1} \, \hat{\omega}_{\alpha}{}^{(\underline 0} \, \hat{\omega}_{\beta}{}^{\underline 1)} = 0
\,, \quad {\rm for} \, \, a \in \{1, 2, 4\}\, \, {\rm and} \, \, \forall \, I, \, J \, , \\
%& & \nonumber\\
& \hskip-0.75cm {\rm Model \,\, B1:} & \quad H_{(\underline 1} \, Q^a{}_{\underline 0)} - \frac{1}{2}\, \kappa_{abc}^{-1} \, \omega_{b \, (\underline 1} \, \omega_{c \, \underline 0)} - \frac{1}{2} \, \hat{\kappa}_{a\alpha\beta}^{-1} \, \hat{\omega}_{\alpha}{}^{(\underline 1} \, \hat{\omega}_{\beta}{}^{\underline 0)} = 0\,,  \, \, \, \, \, \quad {\rm for} \, \, a  = 3 \,,\nonumber \\
& \hskip-0.75cm {\rm Model \,\, B2:} & \quad H_{(\underline 0} \, Q^a{}_{\underline 0)} - \frac{1}{2}\, \kappa_{abc}^{-1} \, \omega_{b \, (\underline 0} \, \omega_{c \, \underline 0)} - \frac{1}{2} \, \hat{\kappa}_{a\alpha\beta}^{-1} \, \hat{\omega}_{\alpha}{}^{(\underline 0} \, \hat{\omega}_{\beta}{}^{\underline 0)} \, \nonumber\\
& & \hskip1cm = \, H_{(\underline 1} \, Q^a{}_{\underline 1)} - \frac{1}{2} \, \kappa_{abc}^{-1} \, \omega_{b \, (\underline 1} \, \omega_{c \, \underline 1)} - \frac{1}{2}\, \hat{\kappa}_{a\alpha\beta}^{-1} \, \hat{\omega}_{\alpha}{}^{(\underline 1} \, \hat{\omega}_{\beta}{}^{\underline 1)}\,,  \quad {\rm for} \, \, a  = 3 \,.\nonumber\\
& & \nonumber\\
& {\rm \bf (III):}  & \quad H_{(\underline I} \, R_{\underline J)} = \omega_{3\, (\underline I}\, Q^3{}_{\underline J)}, \quad \quad \forall \, \, \{I, \, J\} \, ,\nonumber\\
& \hskip-0.75cm & \quad \kappa_{3ac} \, Q^c{}_{(\underline I} \, \omega_{b\, \underline J)}\,= \kappa_{3bc} \, Q^c{}_{(\underline I} \, \omega_{a\, \underline J)}, \quad \forall \, \, a, b \in \{1, \, 2, \, 4\}, \quad \forall \, \, \{I, \, J\}\,, \nonumber\\
%& \hskip-1.5cm (iii). & \hat{\kappa}_{3 \alpha\beta}\, \omega_{a (\underline I} \, \hat{Q}^{\beta \underline J)} = {\kappa}_{3ab} \, Q^b{}_{(\underline I} \, \hat{\omega}_\alpha{}^{\underline J)}, \quad \forall \, \, \alpha, \, \, \forall \, \, a \in \{1, \, 2, \, 4\}, \,\, \forall \, \, I \neq J, \nonumber\\
& \hskip-0.75cm {\rm Model \,\, B1:} & \quad 3\, H_{(\underline 0} \, R_{\underline 1)} - \, \omega_{a (\underline 0} \, Q^a{}_{\underline 1)}  - \, \hat\omega_\alpha{}^{(\underline 0} \, \hat{Q}^{\alpha \underline 1)}= 0\,,\nonumber\\
& \hskip-0.75cm {\rm Model \,\, B2:} & \quad 3\, H_0 \, R_0 - \, \omega_{a0} \, Q^a{}_{0}  - \, \hat\omega_\alpha{}^{0} \, \hat{Q}^{\alpha 0} - H_1 \, R_1 + \, \omega_{a 1} \, Q^a{}_{1}  + \, \hat\omega_\alpha{}^{1} \, \hat{Q}^{\alpha 1} =0\,,\nonumber\\
& & \nonumber\\
& {\rm \bf (IV):} & \quad R_{(\underline I} \, \omega_{a \, \underline J)} = \frac{1}{2} \, \kappa_{abc} \, Q^{b}{}_{(\underline I} \, Q^{c}{}_{\underline J)} %+ \frac{1}{2} \, \hat{\kappa}_{a\alpha\beta} \, \hat{Q}^{\alpha \, (\underline 0} \, \hat{Q}^{\beta \, \underline 1)} = 0
\,, \qquad {\rm for} \, \, a \in \{1, 2, 4\}\, \, {\rm and} \, \, \forall \, I, \, J \, , \nonumber\\
%& & \nonumber\\
& \hskip-0.75cm {\rm Model \, \, B1:} & \quad  R_{(\underline 1} \, \omega_{a \, \underline 0)} - \frac{1}{2} \, \kappa_{abc} \, Q^{b}{}_{(\underline 1} \, Q^{c}{}_{\underline 0)} - \frac{1}{2} \, \hat{\kappa}_{a\alpha\beta} \, \hat{Q}^{\alpha \, (\underline 1} \, \hat{Q}^{\beta \, \underline 0)} = 0\,, \, \, \, \quad \quad {\rm for} \, \, a  = 3 \,,\nonumber \\
& \hskip-0.75cm {\rm Model \, \, B2:} & \quad R_{(\underline 0} \, \omega_{a \, \underline 0)} - \frac{1}{2} \, \kappa_{abc} \, Q^{b}{}_{(\underline 0} \, Q^{c}{}_{\underline 0)} - \frac{1}{2} \, \hat{\kappa}_{a\alpha\beta} \, \hat{Q}^{\alpha \, (\underline 0} \, \hat{Q}^{\beta \, \underline 0)} \, \nonumber\\
& & \hskip1cm = R_{(\underline 1} \, \omega_{b \, \underline 1)} - \frac{1}{2} \, \kappa_{abc} \, Q^{b}{}_{(\underline 1} \, Q^{c}{}_{\underline 1)} - \frac{1}{2} \, \hat{\kappa}_{a\alpha\beta} \, \hat{Q}^{\alpha \, (\underline 1} \, \hat{Q}^{\beta \, \underline 1)}\,, \qquad {\rm for} \, \, a  = 3 \,.\nonumber
\eea
Let us also mention that using the collection of missing identities given in eqn. (\ref{eq:modelBH11}), one can check that the following relations also hold,
\bea
\label{eq:ModelBHQww1}
& \hskip-0.75cm {\rm Model \,\, B1:} & \quad H_{(\underline 1} \, Q^a{}_{\underline 0)} - \frac{1}{2}\, \kappa_{abc}^{-1} \, \omega_{b \, (\underline 1} \, \omega_{c \, \underline 0)} - \frac{1}{2} \, \hat{\kappa}_{a\alpha\beta}^{-1} \, \hat{\omega}_{\alpha}{}^{(\underline 1} \, \hat{\omega}_{\beta}{}^{\underline 0)} = 0\,, \quad \quad \quad \forall \, \, a\,, \\
& \hskip-0.75cm {\rm Model \,\, B2:} & \quad H_{(\underline 0} \, Q^a{}_{\underline 0)} - \frac{1}{2}\, \kappa_{abc}^{-1} \, \omega_{b \, (\underline 0} \, \omega_{c \, \underline 0)} - \frac{1}{2} \, \hat{\kappa}_{a\alpha\beta}^{-1} \, \hat{\omega}_{\alpha}{}^{(\underline 0} \, \hat{\omega}_{\beta}{}^{\underline 0)} \, \nonumber\\
& & \hskip1cm = \, H_{(\underline 1} \, Q^a{}_{\underline 1)} - \frac{1}{2} \, \kappa_{abc}^{-1} \, \omega_{b \, (\underline 1} \, \omega_{c \, \underline 1)} - \frac{1}{2}\, \hat{\kappa}_{a\alpha\beta}^{-1} \, \hat{\omega}_{\alpha}{}^{(\underline 1} \, \hat{\omega}_{\beta}{}^{\underline 1)}\,,  \qquad \forall \, \, a\,.\nonumber
\eea
and
\bea
\label{eq:ModelBRwQQ1}
& \hskip-0.5cm {\rm Model \, \, B1:} & \quad  R_{(\underline 1} \, \omega_{a \, \underline 0)} - \frac{1}{2} \, \kappa_{abc} \, Q^{b}{}_{(\underline 1} \, Q^{c}{}_{\underline 0)} - \frac{1}{2} \, \hat{\kappa}_{a\alpha\beta} \, \hat{Q}^{\alpha \, (\underline 1} \, \hat{Q}^{\beta \, \underline 0)} = 0\,, \, \, \, \quad \quad \forall \, \, a \,,\\
& \hskip-0.5cm {\rm Model \, \, B2:} & \quad R_{(\underline 0} \, \omega_{a \, \underline 0)} - \frac{1}{2} \, \kappa_{abc} \, Q^{b}{}_{(\underline 0} \, Q^{c}{}_{\underline 0)} - \frac{1}{2} \, \hat{\kappa}_{a\alpha\beta} \, \hat{Q}^{\alpha \, (\underline 0} \, \hat{Q}^{\beta \, \underline 0)} \, \nonumber\\
& & \hskip1cm = R_{(\underline 1} \, \omega_{b \, \underline 1)} - \frac{1}{2} \, \kappa_{abc} \, Q^{b}{}_{(\underline 1} \, Q^{c}{}_{\underline 1)} - \frac{1}{2} \, \hat{\kappa}_{a\alpha\beta} \, \hat{Q}^{\alpha \, (\underline 1} \, \hat{Q}^{\beta \, \underline 1)}\,, \qquad \forall \, \, a\,.\nonumber
\eea
These two sets of relations hold for all the values of $a$, i.e. $\forall a \in \{1, 2, 3, 4\}$, and hence they appear to represent some better insights for the $(1,1)$-cohomology sector, and their completions for the $(2,1)$-cohomology sector can be invoked by looking at the prepotential as we discuss in the next step.

\subsection{Insights for the $(2,1)$-cohomology sector}
The K\"ahler potential descending from the ${\cal N}=2$ quaternion sector is determined by a prepotential ${\cal F}$ via the following relation \cite{Grimm:2004ua, Grimm:2005fa},
\bea
& & K^{\rm Q} = - 2\, \ln\left(4\, i\, {\cal F}(n^{\prime I}) \right) \,,
\eea
where we have defined $n^{\prime I}$ as the imaginary part of the complexified chiral variable $N^I$ which is defined in eqn. (\ref{eq:chiralvariables}), i.e. $n^{\prime I} \equiv e^{-D} \, {\cal X}^I$ following the notations of \cite{Carta:2016ynn, Herraez:2018vae}. Moreover, splitting the index $I$ as $\{0, i\}$ where $i$ is counted via the Hodge number $h^{2,1}$ of the threefold, the prepotential ${\cal F}(n^{\prime I})$ which is a homogeneous function of degree two in variables $n^{\prime I}$ can also be written as,
\bea
& & {\cal F}(n^{\prime I}) = - \, i\, (n^{\prime 0})^2\, f\left(u^i\right), \quad 
\eea
where $f\left(u^i\right)$ is now a new function of the inhomogeneous variables $q^i(u^i)=\frac{n^{\prime i}}{n^{\prime 0}} = \frac{{\cal X}^i}{{\cal X}^0}$ which implicitly depends on the complex structure moduli $u^i$ as we will illustrate in our concrete examples. Such a function can be given as a cubic polynomial which can take the following form,
\bea
\label{eq:pref}
& & \hskip-1cm f\left(u^i\right) = \frac{1}{6} \, l_{ijk}\, {u}^i\, {u}^j \, {u}^k + \frac{1}{2} \, l_{0ij}\, {u}^i\, {u}^j + l_{00i}\, {u}^i + \frac{1}{2} \, l_{000}\, \, + {\rm non} \, \, {\rm pert.}\, ,
\eea
Here the quantities $l_{ijk}$ are the triple intersection numbers on the mirror threefold while $l_{0ij}, \, l_{00i}$ and $l_{000}$ can be determined from the other topological quantities \cite{Hosono:1994av}. Also we would be neglecting the non-perturbaive effects in the prepotential assuming the large complex structure limit. 

\subsubsection*{Model A}
The prepotential for Model A has been computed in the appendix \ref{sec_setups} in detail, and it is given by the eqn. (\ref{eq:prepotentialA}), using which we find that
\bea
& & {\cal F}(n^{\prime K}) = - \,\,i \, \sqrt{n^{\prime 0} \, n^{\prime 1} \, n^{\prime 2} \, n^{\prime 3}} = - \, i \, (n^{\prime 0})^2 \, u^1\, u^2\, u^3\,,
\eea 
where in the second equality we have used 
\bea
& & \hskip-1cm q^1 = \frac{n^{\prime 1}}{n^{\prime 0}} = \frac{{\cal X}^1}{{\cal X}^0}= u^2\, u^3, \quad q^2 = \frac{n^{\prime 2}}{n^{\prime 0}}= \frac{{\cal X}^2}{{\cal X}^0} = u^1\, u^3, \quad q^3 = \frac{n^{\prime 3}}{n^{\prime 0}} = \frac{{\cal X}^3}{{\cal X}^0} = u^1\, u^2 \,,
\eea
which follows from the definition of the complex structure moduli as given in eqn. (\ref{eq:TUModelA}) and the period vectors given in eqn. (\ref{eq:periodExA}). Therefore, for Model A, we have the following intersection numbers,
\bea
& & l_{123} = 1, \qquad l_{0ij} = 0, \qquad l_{00i} = 0, \qquad l_{000} = 0 \,.
\eea
With these ingredients, the missing identities for Model A can be written as,
\bea
& & H_{(\underline i}\, Q^a{}_{\underline 0)} - \frac{1}{2} \, \kappa_{abc}^{-1} \, \omega_{b (\underline i}\, \omega_{c \, \underline 0)} =0, \hskip4.3cm  \, \, i \neq a \,; \\
& & H_{(\underline 0} \, R_{\underline k)} - \, \omega_{a \, (\underline 0} \, Q^{a}{}_{\underline k)} \, \delta_{a k} =0,\, \hskip4.7cm \forall \, \, k\,;\nonumber\\
& & H_{(\underline 0} \, R_{\underline k)} - \omega_{a \, (\underline 0} \, Q^{a}{}_{\underline k)} + 2\, \omega_{a \, (\underline 0} \, Q^{a}{}_{\underline j)}\, \, \delta_{aj} = 0\,, \qquad \qquad \qquad \, \, j \neq k \,; \nonumber\\
& & R_{(\underline i}\, \omega_{a \underline 0)} - \frac{1}{2} \, \kappa_{abc} \, Q^{b}{}_{(\underline i}\, Q^{c}{}_{\underline 0)} =0, \hskip4.3cm\, \, i \neq a\,; \nonumber
\eea
and
\bea
& & \frac{1}{2} \, \, l_{ijk}^{-1} \, \biggl[H_{(\underline j}\, Q^a{}_{\underline k)} - \, \frac{1}{2} \, \kappa_{abc}^{-1} \, \omega_{b \,(\underline j}\, \omega_{c \,\underline k)} \biggr] = 0, \qquad \qquad \qquad \, \, i \neq a\,; \\
& & \frac{1}{2}\, l_{ijk}^{-1} \,\biggl[H_{(\underline j} \, R_{\underline k)} - \omega_{a \, (\underline j} \, Q^{a}{}_{\underline k)} + 2\, \omega_{a \, (\underline j} \, Q^{a}{}_{\underline k)}\, \, \delta_{aj}\biggr] = 0\,, \, \, \, \quad \quad \forall \, \, i \,; \nonumber\\
& & \frac{1}{2}\, l_{ijk}^{-1} \,\biggl[H_{(\underline j} \, R_{\underline k)} - \, \omega_{a \, (\underline j} \, Q^{a}{}_{\underline k)} \, \delta_{ai}\biggr] =0,\, \hskip3.3cm \forall \, \, i\,; \nonumber\\
& & \frac{1}{2} \, \, l_{ijk}^{-1} \biggl[ R_{(\underline j}\, \omega_{a \underline k)} - \frac{1}{2} \, \kappa_{abc} \, Q^{b}{}_{(\underline j}\, Q^{c}{}_{\underline k)} \biggr] =0, \qquad \qquad \qquad \quad i \neq a\,.\nonumber
\eea

\subsubsection*{Model B1}
The prepotential for Model B1 has been computed in the appendix \ref{sec_setups} in detail, and it is given by the eqn. (\ref{eq:prepotentialB1}), using which we find that
\bea
& & \hskip-2cm  {\cal F}(n^{\prime K}) = - \, i \, n^{\prime 0} \, n^{\prime 1} = - \, i \, (n^{\prime 0})^2 \, u\,,
\eea 
where we have used $q(u) = \left(\frac{n^{\prime 1}}{n^{\prime 0}}\right) = \frac{{\cal X}^1}{{\cal X}^0}= U =\, u$ from eqn. (\ref{eq:periodExB1}), and therefore we have the following intersection numbers,
\bea
& & l_{ijk} = 0, \qquad l_{0ij} = 0, \qquad l_{001} = 1, \qquad l_{000} = 0\,.
\eea
With these ingredients, (most of) the missing identities for Model B1 can be written as,
\bea
& & \hskip-1cm l_{00i}^{-1} \, \biggl[H_{(\underline i}\, Q^a{}_{\underline 0)} - \, \frac{1}{2} \, \kappa_{abc}^{-1} \, \omega_{b \,(\underline i}\, \omega_{c \,\underline 0)} \biggr] = l_{00i} \biggl[\frac{1}{2} \, \hat\kappa_{a\alpha\beta}^{-1} \, \, \, \hat\omega_{\alpha}{}^{(\underline i}\, \hat\omega_{\beta}{}^{\underline 0)} \biggr], \, \, \qquad \qquad \, \, \forall \, a\,; \\
& & \hskip-1cm l_{00i}^{-1} \,\biggl[H_{(\underline 0} \, R_{\underline i)} - \omega_{a \, (\underline 0} \, Q^{a}{}_{\underline i)} + 2\, \omega_{a \, (\underline 0} \, Q^{a}{}_{\underline i)}\, \, \delta_{ai}\biggr] = l_{00i} \,\hat\omega_{\alpha}{}^{(\underline 0} \, \hat{Q}^{\alpha\, \underline i)}\,, \qquad \quad \quad \,\,\,\nonumber\\
& & \hskip-1cm l_{00i}^{-1} \biggl[ R_{(\underline i}\, \omega_{a \underline 0)} - \frac{1}{2} \, \kappa_{abc} \, Q^{b}{}_{(\underline i}\, Q^{c}{}_{\underline 0)} \biggr] = l_{00i}\, \biggl[\frac{1}{2} \, \hat\kappa_{a\alpha\beta} \, \hat{Q}^{\alpha\,(\underline i}\, \hat{Q}^{\beta \, \underline 0)} \biggr], \qquad \qquad \, \, \, \forall \,\, a\,;\nonumber
\eea
and
\bea
& & \hskip-1cm l_{00i}^{-1} \, \biggl[H_{(\underline 0}\, Q^a{}_{\underline 0)} - \, \frac{1}{2} \, \kappa_{abc}^{-1} \, \omega_{b \,(\underline 0}\, \omega_{c \,\underline 0)} \biggr] =0, \,\, \quad \qquad \qquad \qquad \forall \, i, \,\, \forall \, a: \hat\kappa_{a\alpha\beta} = 0\,; \\
& & \hskip-1cm l_{00i}^{-1} \,\biggl[H_{(\underline 0} \, R_{\underline 0)} - \omega_{a \, (\underline 0} \, Q^{a}{}_{\underline 0)} + 2\, \omega_{a \, (\underline 0} \, Q^{a}{}_{\underline 0)}\, \, \delta_{ai}\biggr] =0,\, \, \qquad \quad \forall \, i \, ; \nonumber\\
& & \hskip-1cm l_{00i}^{-1} \biggl[ R_{(\underline 0}\, \omega_{a \underline 0)} - \frac{1}{2} \, \kappa_{abc} \, Q^{b}{}_{(\underline 0}\, Q^{c}{}_{\underline 0)} \biggr] =0, \qquad \qquad \qquad \qquad \forall \, i, \,\, \forall \, a: \hat\kappa_{a\alpha\beta} = 0\,.\nonumber
\eea

\subsubsection*{Model B2}
The prepotential for Model B2 has been computed in the appendix \ref{sec_setups} in detail, and it is given by the eqn. (\ref{eq:prepotentialB2}), using which we find that
\bea
& & \hskip-2cm  {\cal F}(n^{\prime K}) = - \, \frac{i}{2} \, (n^{\prime 0})^2 \biggl[1- \, \left(\frac{n^{\prime 1}}{n^{\prime 0}}\right)^2 \biggr] = - \, \frac{i}{2} \, (n^{\prime 0})^2 \biggl[1- \, u^2 \biggr]\,,
\eea 
where we have used $q(u) = \left(\frac{n^{\prime 1}}{n^{\prime 0}}\right) = \frac{{\cal X}^1}{{\cal X}^0} =\, u$, and therefore we have the following intersection numbers,
\bea
& & l_{ijk} = 0, \qquad l_{011} = -1, \qquad l_{00i} = 0, \qquad l_{000} = 1 \,.
\eea
With these ingredients, the missing identities for Model B2 can be written as,
\bea
& & \hskip-1cm l_{0IJ}^{-1} \, \biggl[H_{(\underline I}\, Q^a{}_{\underline J)} - \, \frac{1}{2} \, \kappa_{abc}^{-1} \, \omega_{b \,(\underline I}\, \omega_{c \,\underline J)} \biggr] = l_{0IJ} \biggl[\frac{1}{2} \, \hat\kappa_{a\alpha\beta}^{-1} \, \, \, \hat\omega_{\alpha}{}^{(\underline I}\, \hat\omega_{\beta}{}^{\underline J)} \biggr], \, \, \qquad \qquad \, \, \forall \, a\,; \\
& & \hskip-1cm l_{0IJ}^{-1} \,\biggl[H_{(\underline I} \, R_{\underline J)} - \omega_{a \, (\underline I} \, Q^{a}{}_{\underline J)} + 2\, \omega_{a \, (\underline I} \, Q^{a}{}_{\underline J)}\, \, \delta_{ai}\biggr] = l_{0IJ} \,\hat\omega_{\alpha}{}^{(\underline I} \, \hat{Q}^{\alpha\, \underline J)}\,, \qquad \quad \quad \,\,\,\nonumber\\
& & \hskip-1cm l_{0IJ}^{-1}\,\biggl[3\, H_{(\underline I} \, R_{\underline J)} - \, \omega_{a (\underline I} \, Q^a{}_{\underline J)}\biggr] = l_{0IJ}\,\,\, \hat\omega_\alpha{}^{(\underline I} \, \hat{Q}^{\alpha \underline J)}, \nonumber\\
& & \hskip-1cm l_{0IJ}^{-1} \biggl[ R_{(\underline I}\, \omega_{a \underline J)} - \frac{1}{2} \, \kappa_{abc} \, Q^{b}{}_{(\underline I}\, Q^{c}{}_{\underline J)} \biggr] = l_{0IJ}\, \biggl[\frac{1}{2} \, \hat\kappa_{a\alpha\beta} \, \hat{Q}^{\alpha\,(\underline I}\, \hat{Q}^{\beta \, \underline J)} \biggr], \qquad \qquad \, \, \, \forall \,\, a\,;\nonumber
\eea
and
\bea
& & \hskip-1cm l_{0IJ}^{-1} \, \biggl[H_{(\underline 0}\, Q^a{}_{\underline J)} - \, \frac{1}{2} \, \kappa_{abc}^{-1} \, \omega_{b \,(\underline 0}\, \omega_{c \,\underline J)} \biggr] =0, \,\, \quad \quad \qquad \forall \, I \in \{0, i\}, \,\, \forall \, a: \hat\kappa_{a\alpha\beta} = 0\,; \\
& & \hskip-1cm l_{0IJ}^{-1} \,\biggl[H_{(\underline 0} \, R_{\underline J)} - \omega_{a \, (\underline 0} \, Q^{a}{}_{\underline J)} + 2\, \omega_{a \, (\underline 0} \, Q^{a}{}_{\underline J)}\, \, \delta_{ai}\biggr] =0\, ; \nonumber\\
& & \hskip-1cm l_{0IJ}^{-1} \biggl[ R_{(\underline 0}\, \omega_{a \underline J)} - \frac{1}{2} \, \kappa_{abc} \, Q^{b}{}_{(\underline 0}\, Q^{c}{}_{\underline J)} \biggr] =0, \quad \qquad \qquad \forall \, I \in \{0, i\}, \,\, \forall \, a: \hat\kappa_{a\alpha\beta} = 0\,.\nonumber
\eea
As a side remark, let us mention that we have also attempted to look for some identities in the first formulation which could directly translate into the second formulation. A couple of such constraints are presented in Table \ref{tab_BIsNew}.

\section{Conclusions and discussions}
\label{sec_conclusions}
This article has been focussed on investigating the two formulations of Bianchi identities in type IIA supergravity with the (non-)geometric fluxes. In what we call the `first formulation', all the fluxes are written using real six-dimensional indices ($H_{ijk}, \omega_{ij}^k$ etc.) while in the `second formulation', the fluxes are expressed using cohomology indices such as $H_K, \omega_{aK}$ etc., where $a$ index components are counted to be $h^{1,1}_-(X_3/\sigma)$ in number while the $K$ index components are counted to be $(1+h^{2,1}(X_3))$. Assuming the appropriate normalizations of forms, the two formulations of the Bianchi identities are summarized as in the table \ref{tab_BIs}. We have performed a deep analysis in search of the missing identities in the cohomology formulation, and subsequently we have conjectured a model independent form for (the most of) these identities which are collected in table \ref{tab_missingBIs}.
\begin{table}[h!]
  \centering
 \begin{tabular}{|c||c|c||}
\hline
& & \\
 BIs & First formulation & Second formulation  \\
 & & \\
 \hline
 & & \\
{\bf (I)} & $H_{m[\underline{ij}} \, \omega_{\underline{kl}]}{}^{m}= 0$ & $H_K\, \hat{\omega}_\alpha{}^K = 0$ \\
& & \\
{\bf (II)} & $ \omega_{[\underline{ij}}{}^{m} \, {\omega}_{\underline{k}]m}{}^{l} \, = \, {Q}_{[\underline{i}}{}^{lm} \, {H}_{\underline{jk}]m}$ & $H_K\, \hat{Q}^{\alpha K} = 0, \, \,  \omega_{aK}\, \hat{\omega}_\alpha{}^K = 0$ \\
& & \\
{\bf (III)} & ${H}_{ijm} {R}^{mkl} + {\omega}_{{ij}}{}^{m}{Q}_{m}{}^{{kl}} = 4{Q}_{[\underline{i}}{}^{m[\underline{k}} {\omega}_{\underline{j}]m}{}^{\underline{l}]}$ & $\omega_{aK} \hat{Q}^{\alpha K} =0, \, \, \hat{\omega}_\alpha{}^{K} \,Q^a{}_K=0,$ \\
& & $\hat{\omega}_\alpha{}^{[K}\, \hat{Q}^{\alpha J]} = 0, H_{[K} \, R_{J]} = \omega_{a[K}\, Q^a{}_{J]}$\\
&& \\
{\bf (IV)} & ${Q}_{m}{}^{[\underline{ij}}  \, {Q}_{l}{}^{\underline{k}]m} \, = \, {\omega}_{lm}{}^{[\underline{i}}{}\,\, {R}^{\underline{jk}]m}$ & $R_K \, \hat{\omega}_\alpha{}^K = 0, \, \, Q^a{}_K \, \hat{Q}^{\alpha K} = 0$ \\
& & \\
{\bf (V)} & $R^{m [\underline{ij}}\,  Q_{m}{}^{\underline{kl}]} \, =0$ & $R_K\, \hat{Q}^{\alpha K} = 0$ \\
& & \\
  \hline
  \end{tabular}
  \caption{Two formulations of the type IIA Bianchi identities.}
  \label{tab_BIs}
 \end{table}
\begin{table}[H]
  \centering
 \begin{tabular}{|c||c|c||}
\hline
&  \\
 BIs & Missing identities  \\
%& \\
 \hline
% &  \\
%{\bf (I)} & Nill \\
& \\
{\bf (II)} & \hskip-1cm $l_{0IJ}^{-1} \,\bigl[H_{(\underline I} \, Q^a{}_{\underline J)} - \frac{1}{2}\, \kappa_{abc}^{-1} \, \omega_{b \, (\underline I} \, \omega_{c \, \underline J)}\bigr] = l_{0IJ}\, \bigl[\frac{1}{2} \,\hat{\kappa}_{a\alpha\beta}^{-1} \, \hat{\omega}_{\alpha}{}^{(\underline I} \, \hat{\omega}_{\beta}{}^{\underline J)} \bigr], \, \quad \forall \, a\,;$\\
& \\
& $l_{IJ0}^{-1} \,\bigl[H_{(\underline J} \, Q^a{}_{\underline 0)} - \frac{1}{2}\, \kappa_{abc}^{-1} \, \omega_{b \, (\underline J} \, \omega_{c \, \underline 0)}\bigr] = 0, \, \quad \forall \,\, I \in \{0, i\}$ \& $\forall \, \, a: \hat{\kappa}_{a\alpha\beta} = 0\,;$\\
& \\
& \hskip-1cm  $l_{ijk}^{-1} \,\bigl[H_{(\underline j} \, Q^a{}_{\underline k)} - \frac{1}{2}\, \kappa_{abc}^{-1} \, \omega_{b \, (\underline j} \, \omega_{c \, \underline k)}\bigr] = l_{ijk}\, \bigl[\frac{1}{2} \,\hat{\kappa}_{a\alpha\beta}^{-1} \, \hat{\omega}_{\alpha}{}^{(\underline j} \, \hat{\omega}_{\beta}{}^{\underline k)} \bigr], \, \quad i \neq a\,;$\\
& \\
{\bf (III)} & \hskip-4.3cm  $l_{0IJ}^{-1}\,\bigl[3\, H_{(\underline I} \, R_{\underline J)} - \, \omega_{a (\underline I} \, Q^a{}_{\underline J)}\bigr] = l_{0IJ}\, \, \, \hat\omega_\alpha{}^{(\underline I} \, \hat{Q}^{\alpha \underline J)}$\,; \\
& \\
& \hskip-1.7cm  $l_{IJ0}^{-1} \,\bigl[H_{(\underline J} \, R_{\underline 0)} - \omega_{a \, (\underline J} \, Q^{a}{}_{\underline 0)} + 2\, \omega_{a \, (\underline J} \, Q^{a}{}_{\underline 0)}\, \, \delta_{ai}\bigr] = l_{IJ0} \,\hat\omega_{\alpha}{}^{(\underline J} \, \hat{Q}^{\alpha\, \underline 0)}$\,; \\
& \\
& \hskip-2.1cm $l_{ijk}^{-1} \,\bigl[H_{(\underline j} \, R_{\underline k)} - \omega_{a \, (\underline j} \, Q^{a}{}_{\underline k)} + 2\, \omega_{a \, (\underline j} \, Q^{a}{}_{\underline k)}\, \, \delta_{aj}\bigr] = 0\,, \, \, \, \quad \quad \forall \, \, i \,$; \\
&\\
& \hskip-2.4cm  $l_{ijk}^{-1} \,\bigl[H_{(\underline j} \, R_{\underline k)} - \, \omega_{a \, (\underline j} \, Q^{a}{}_{\underline k)} \, \delta_{ai}\bigr] =0,\, \hskip3.3cm \forall \, \, i$\,; \\
&\\
{\bf (IV)} & \hskip-0.85cm  $l_{0IJ}^{-1} \,\bigl[R_{(\underline I} \, \omega_{a \, \underline J)} - \frac{1}{2} \, \kappa_{abc} \, Q^{b}{}_{(\underline I} \, Q^{c}{}_{\underline J)} \bigr] = l_{0IJ}\, \bigl[\frac{1}{2} \, \hat{\kappa}_{a\alpha\beta} \, \hat{Q}^{\alpha \, (\underline I} \, \hat{Q}^{\beta \, \underline J)} \bigr], \, \quad \forall \, a$\, ;\\
& \\
& $l_{IJ0}^{-1} \,\bigl[R_{(\underline J} \, \omega_{a \, \underline 0)} - \frac{1}{2} \, \kappa_{abc} \, Q^{b}{}_{(\underline J} \, Q^{c}{}_{\underline 0)} \bigr] = 0, \, \quad \forall \,\, I \in \{0, i\}$ \& $\forall \, \, a: \hat{\kappa}_{a\alpha\beta} = 0$\,;\\
& \\
& \hskip-0.85cm  $l_{ijk}^{-1} \,\bigl[R_{(\underline j} \, \omega_{a \, \underline j)} - \frac{1}{2} \, \kappa_{abc} \, Q^{b}{}_{(\underline j} \, Q^{c}{}_{\underline k)} \bigr] = l_{ijk}\, \bigl[\frac{1}{2} \, \hat{\kappa}_{a\alpha\beta} \, \hat{Q}^{\alpha \, (\underline j} \, \hat{Q}^{\beta \, \underline k)} \bigr], \, \quad i \neq a$\,.\\
& \\
  \hline
  \end{tabular}
  \caption{A conjectural form for (some of) the missing identities. Here we have considered $\kappa_{abc}^{-1} = 1/\kappa_{abc}$ and $\hat{\kappa}_{a\alpha\beta}^{-1} = 1/\hat{\kappa}_{a\alpha\beta}$ for fixed values of $\{a, b, c\}$ and $\{\alpha, \beta\}$, whenever these intersections are non-zero, and similarly for the triple intersections on the mirror side.}
  \label{tab_missingBIs}
 \end{table}%The main motivation of this study has been the analysis and observations of the \cite{Ihl:2007ah, Robbins:2007yv, Shukla:2016xdy} where it has been found that the flux constraints generated from these two formulations of Bianchi identities are not equivalent. 
The main findings and observations from our detailed analysis can be summarized in the following points,
\begin{itemize}
\item{All the identities of the second formulations can be obtained via reshuffling the identities of the first formulation.}
\item{There are certainly several flux constraints in the first formulation which cannot be obtained from the known version of the second formulation.}
\item{In our type IIA orientifold construction, it is easier to generically see the mismatch in the two formulation, in particular for the choice of involution leading to no `hatted' fluxes, which are counted by the even (1,1)-cohomology index $\alpha$. Such fluxes are absent for $h^{1,1}_+(X_3/\sigma) = 0$ and subsequently one can observe that 9 of the 10 second formulation identities as collected in Table \ref{tab_BIs} are identically and generically trivial. Our Model A demonstrates the explicit insights behind these arguments.}
\item{There is no mismatch between the first ($H\omega$-type) and the last ($RQ$-type) of the five classes of the constraints presented in Table \ref{tab_BIs}. In Model A both of these classes, namely {\bf (I)} and {\bf (V)} are trivial while in Model B, they are non-trivial but identical in the two formulations. So the mismatch is present only in the {\bf (II)}, {\bf (III)} and {\bf (IV)} type of the constraints of Table \ref{tab_BIs}.}
\item{We have managed to (partially) express the set of missing Bianchi identities in a model independent manner by using the topological quantities of the complex threefold such as the triple-intersection numbers as defined in eqns. (\ref{eq:intersectionBases}). These are given in table \ref{tab_missingBIs}. %However by merely considering two simple toroidal examples which have the simplest kinds of intersection numbers, it is hard to convincingly claim to extrapolate these observations for generic models.
}
\item{From table \ref{tab_BIsFirstFormulation} we observe that in the first formulation the maximum number of Bianchi identities is 496 while from the second formulation as listed in table \ref{tab_BIsSecondFormulation}, we find that the maximum number of identities depend on the hodge number $h^{1,1}_\pm$ and $h^{2,1}$. Therefore there will be certainly some redundancy in the second formulation, especially for the orientifold settings having large hodge numbers, so that to make it consistent with the counting in the first formulation. However, it is hard to find/claim that there will be a perfect bijection in terms of the number of ``independent" flux constraints.
}
%\item{In our Model B, we find that imposing the tracelessness condition $\omega_{ij}{}^i = 0 = Q_i{}^{ij}$ results in relating all the `hatted' fluxes (which can appear through the $D$-terms only \cite{Robbins:2007yv}) being related to the `unhatted' fluxes (which appear only through the $F$-terms). It would be interesting to check if this could be generically true in different models.}
%\item{We find that the collective set of Bianchi identities in each of the (first as well as second) formulation remain to hold true even after replacing the $\{H, \omega, Q, R \}$ with their respective generalized versions $\{{\mathbb H}, {\mathbb \mho}, {\mathbb Q}, {\mathbb R} \}$.}
\end{itemize}
From the table \ref{tab_missingBIs}, we observe that the index structure for the $h^{1,1}$ and $h^{2,1}$ indices in the missing Bianchi identities are completely different from those of the second formulation. For example, after looking at the index structure in the generic identities of the cohomology formulation as presented in the table \ref{tab_BIsSecondFormulation}, one can convince that none of the missing identities, namely the ones given in eqns. (\ref{eq:12HQww0}), (\ref{eq:24HRwQ1}), (\ref{eq:12RwQQ0}) for model A and those given in eqns. (\ref{eq:ModelBHQwwExpand1}), (\ref{eq:ModelBHQwwExpand2}), (\ref{eq:ModelBHRwQExpand1}), (\ref{eq:ModelBHRwQExpand2}), (\ref{eq:ModelBRwQQExpand1}), (\ref{eq:ModelBRwQQExpand2}) for the model B, can be produced from the identities of table \ref{tab_BIsSecondFormulation}.

As it has been very standard thing to follow, we have investigated the two toroidal models by considering ingredients (e.g. fluxes and moduli) only in the untwisted sector, and therefore one might speculate/suspect that may be after including the twisted sector fluxes, the mismatch between the two sets of Bianchi identities goes away. However, this cannot happen because of the simple reason stated regarding the distinct index structures appearing in the missing Bianchi identities and the ones presented in the cohomology formulation. For the later case, the generic expressions are given in table \ref{tab_BIsSecondFormulation}, and therefore in order to include the twisted sector one has to simply change the range of the $h^{1,1}_\pm$ and $h^{2,1}_\pm$ indices; for example the ${\mathbb T}^6/({\mathbb Z}_2 \times {\mathbb Z}_2)$ setup will have 48 twisted moduli and hence one would need to change $a$ from $a =\{1, 2, 3\}$ to $a =\{1, 2, ..., 51\}$ subject to the appropriate choice of the involution. This would surely append/modify the set of identities with additional constraints but those would never fall in line with the index structure of the missing identities, e.g. in the sense of contraction of indices, symmetrization of $h^{2,1}$ indieces etc. However it would be interesting to investigate on these lines by performing some explicit computations including the twisted sector.

At least one reason for the mismatch between the Bianchi identities of the two formulations could be considered to be the fact that the first formulation is derived by imposing the nilpotency of the twisted differential ${\cal D}$ on a generic $p$-forms $A_p$, while the second formulation can be derived by imposing the nilpotency only on the harmonic forms. By finding some fundamental derivation of all these missing identities of the second formulation, it would be interesting to check/verify if the conjectured form of (some of) the missing identities proposed in table \ref{tab_missingBIs} generically holds or not. 

From our explicit examples, we have observed that the second formulation produces only around 15\% of the total number of Bianchi identities in Model A, and that of around 35\% in Model B. Therefore, one would expect the scalar potential to have some strong restrictions imposed from the missing identities which can further nullify several terms of the potential making it better or worse for a given model, depending on the outcome. For example, they can kill many terms upto the extent that the no-scale structure could win against some of the terms responsible for the stabilization of (some of) the moduli, and hence this could be risky for an already working model. However, these additional identities could make significant simplifications such that one could even think of studying moduli stabilization analytically, and possibly in a model independent manner, which appears to be extremely challenging task in concrete non-geometric setups. To conclude, we would like to make a cautionary remark that these identities might play some crucial role, particularly in the scenarios where one uses only the second formulation for building the phenomenologically motivated non-geometric models beyond the toroidal orientifolds. %In this regard, non-supersymmetric moduli stabilization in such non-geometric type IIA setups would be interesting to study and we leave this for future work.

\section*{Acknowledgments}
We are very thankful to Wieland Staessens for several useful discussions. We also thank Michael Fuchs and Fernando Marchesano for useful discussion. We would like to thank Ralph Blumenhagen for his encouraging comments on an earlier version of this work. PS is grateful to Luis Ib\'a\~nez for his kind support and encouragements throughout. The work of PS has been supported in part by the ERC Advanced Grant ``String Phenomenology in the LHC Era" (SPLE) under contract ERC-2012-ADG-20120216-320421.

\clearpage

\appendix

\setcounter{equation}{0}
\renewcommand{\theequation}{A.\arabic{equation}}

\section{Derivation of the second formulation of Bianchi identities}
\label{sec_appendix}
The twisted differential operator ${\cal D}$ is defined as under,
\bea
& & {\cal D} = d + H \wedge . - \omega \triangleleft . +Q \triangleright . - R \bullet . \nonumber
\eea
Here the action of various fluxes appearing in ${\cal D}$ is such that for an arbitrary $p$-form $A_p$, the pieces $H\wedge A_p$, $\omega \triangleleft A_p$, $Q \triangleright A_p$ and $R \bullet A_p$ denote a $(p+3)$-form, a $(p+1)$-form, a $(p-1)$-form and a $(p-3)$-form respectively. More specifically, there are the following flux actions on various harmonic-forms \cite{Ihl:2007ah},
\bea
\label{eq:fluxActions0}
& & \hskip-0.7cm H \wedge {\bf 1} = H_K\, \beta^K, \qquad \, \, H \wedge \alpha_K = - H_K \,  \Phi_6, \qquad H\wedge \beta^K = 0, \\
& &  \hskip-0.7cm \omega \triangleleft \nu_a = \omega_{a K}\, \beta^K, \qquad \omega \triangleleft \mu_\alpha = \hat{\omega}_{\alpha}^K\, \alpha_K, \qquad \omega \triangleleft \alpha_K = \, \omega_{a K} \, \tilde\nu^a, \quad \omega \triangleleft \beta^K = -\, \hat\omega_{\alpha}^K \, \tilde\mu^\alpha \,,\nonumber\\
& & \hskip-0.7cm Q \triangleright \tilde\nu^a = Q^{a}_{K}\, \beta^K, \qquad Q \triangleright  \tilde\mu^\alpha = \hat{Q}^{\alpha K}\, \alpha_K, \quad Q \triangleright \alpha_K = - Q^{a}_{K} \, \nu_a, \quad Q \triangleright \beta^K =\, \hat{Q}^{\alpha K} \, \mu_\alpha\,, \nonumber\\
& & \hskip-0.7cm R \bullet {\Phi_6} = R_K\, \beta^K, \qquad R \bullet \alpha_K = R_K \,  {\bf 1}, \qquad R \bullet \beta^K =0\,. \nonumber
\eea
Now as can be seen from these flux actions, the operations $\triangleleft, \triangleright$ and $\bullet$ changes a $p$-form into a $(p+1)$-form, a $(p-1)$-form and a $(p-3)$-form respectively, and we have 
\bea
&&{\cal D}\, A_p = d \, A_p + H \wedge\, A_p  - \omega \triangleleft \, A_p + Q \triangleright \, A_p - R \bullet \, A_p \,.
\eea
Subsequently, we find that $({\cal D}^2 \, A_p)$ has seven types of pieces written as $(p+i)$-forms where $i \in \{ 6, 4, 2 ,0, -2 , -4 , -6\}$. These are collected as under,
\bea
\label{eq:D2forms}
& (i). & H \wedge (H \wedge A_p) \\
& (ii). & d (H\wedge A_p) + H \wedge (d A_p) - H \wedge (\omega \triangleleft A_p) - \omega \triangleleft(H \wedge A_p) \nonumber\\
& (iii). &d^2 \, A_p - d \,(\omega \triangleleft A_p) - \omega \triangleleft  (d\, A_p) + H \wedge (Q \triangleright A_p) + Q \triangleright(H \wedge A_p) + \omega \triangleleft(\omega \triangleleft A_p)\nonumber\\
& (iv). & d\, (Q \triangleright A_p)+Q \triangleright(d\, A_p) - Q \triangleright(\omega \triangleleft A_p) -\omega \triangleleft(Q \triangleright A_p) - H \wedge (R\bullet A_p) - R \bullet(H \wedge A_p)\nonumber\\
& (v). & -d \, (R \bullet A_p) - R\bullet (d A_p) + \omega \triangleleft (R\bullet A_p) + Q \triangleright (Q \triangleright A_p) + R \bullet (\omega \triangleleft A_p) \nonumber\\
& (vi). & -Q \triangleright (R \bullet A_p) - R \bullet (Q \triangleright A_p) \nonumber\\
& (vii).& R\bullet (R\bullet A_p) \nonumber
\eea
For ensuring the identity ${\cal D}^2 \, A_p = 0$, each of these seven pieces has to vanish individually. Now given that the internal background is a real six-dimensional manifold,  one can observe from collection in eqn. (\ref{eq:D2forms}) that the first $(i)$ and the last $(vii)$ expressions are relevant only for $A_p$ being zero-form ${\bf 1}$ and six-form $\Phi_6$ respectively. However, the same leads to trivial constraints as,
\bea
& & H \wedge (H \wedge {\bf 1}) = 0, \quad R\bullet (R\bullet \Phi_6) = R_K (R\bullet \beta^K) = 0\,,
\eea
where we have used R-flux actions given in eqn. (\ref{eq:fluxActions0}). Now, for further simplifying the remaining five type of identities in eqn. (\ref{eq:D2forms}), we will assume that all fluxes are constant parameters which is considered due to the subsequent simpler phenomenological relevance. Moreover, one observation is very straight that mixing of fluxes in the remaining five constraints are of $H\omega, (\omega^2+HQ), (HR+Q\omega), (Q^2+ \omega R)$ and ($QR$) types which is in obvious connections with the first formulation of Bianchi identities as given in Table \ref{tab_BIsFirstFormulation}.  However, for the second formulation our aim is to compute Bianchi identities with fluxes written in various cohomology bases and not in the real six-dimensional indices. Let us take each constraint one-by-one via considering the flux actions in eqn. (\ref{eq:fluxActions0}). 
\begin{itemize}
\item{Using the fact that $H_3$ is a three-form constant flux, we find that $d (H\wedge A_p) + H \wedge (d A_p) = 0$, and so nullification of terms in the class $(ii)$ simplifies into $H \wedge (\omega \triangleleft A_p) + \omega \triangleleft(H \wedge A_p) = 0$. The relevant $A_p$-forms for expecting non-trivial relations correspond to $p= \{0, 2\}$, and both of these choices result into a single Bianchi identity as under, 
\bea 
H_K\, \hat{\omega}_\alpha{}^K = 0. 
\eea
}
\item{Using $d^2 = 0$ and constancy of fluxes, the nullification of terms in class $(iii)$ are reduced into satisfying: $H \wedge (Q \triangleright A_p) + Q \triangleright(H \wedge A_p) + \omega \triangleleft(\omega \triangleleft A_p) = 0$. This results into the following two types of Bianchi identities via considering $A_p = \{{\bf 1}, \alpha_K, \nu_a\}$, 
\bea
H_K\, \hat{Q}^{\alpha K} = 0, \qquad \omega_{aK}\, \hat{\omega}_\alpha{}^K = 0. 
\eea
}
\item{Demanding the nullification of terms in collection $(iv)$ results in the following %four classes of Bianchi identities:
\bea
& & \hskip-1cm \omega_{aK}\, \hat{Q}^{\alpha K} =0, \quad \hat{\omega}_\alpha{}^{K} \,Q^a{}_K=0, \quad \hat{\omega}_\alpha{}^{[K}\, \hat{Q}^{\alpha J]} = 0, \quad H_{[K} \, R_{J]} = \omega_{a[K}\, Q^a{}_{J]}, 
 \eea
where the first two identities follow from the two-forms: $A_p = \{\nu_a, \mu_\alpha, \tilde{\nu}^a, \tilde{\mu}^\alpha \}$ while the third and fourth identities follow from the choice $A_p = \beta^K$ and $A_p = \alpha_K$ respectively. In addition, the bracket $[..]$ denotes anti-symmetrization of $J$ and $K$ indices.
}
\item{Demanding the nullification of collection $(v)$, we get $\omega \triangleleft (R\bullet A_p) + Q \triangleright (Q \triangleright A_p) + R \bullet (\omega \triangleleft A_p) =0$, which results in the following two Bianchi identities,
\bea
& & R_K \, \hat{\omega}_\alpha{}^K = 0, \quad Q^a{}_K \, \hat{Q}^{\alpha K} = 0\,,
\eea
where the first one follows from $A_p = \{\Phi_6, \mu_\alpha\}$ while the second one from $A_p = \tilde{\nu}^a$.
}
\item{Finally, the nullification of collection $(vi)$ which is $Q \triangleright (R \bullet A_p) + R \bullet (Q \triangleright A_p) = 0$, gives another Bianchi identity for the relevant $p$-forms being $A_p = \{ \Phi_6, \tilde{\mu}^\alpha \}$,
\bea
& & R_K\, \hat{Q}^{\alpha K} = 0\,.
\eea}
\end{itemize}
In summary, we have the following set of Bianchi identities in the second formulation,
\bea
& & H_K\, \hat{\omega}_\alpha{}^K = 0, \quad H_K\, \hat{Q}^{\alpha K} = 0, \quad R_K\, \hat{Q}^{\alpha K} = 0, \quad R_K \, \hat{\omega}_\alpha{}^K = 0, \nonumber\\
& & \omega_{aK}\, \hat{\omega}_\alpha{}^K = 0, \quad \omega_{aK}\, \hat{Q}^{\alpha K} =0, \quad Q^a{}_K \, \hat{Q}^{\alpha K} = 0, \quad \hat{\omega}_\alpha{}^{K} \,Q^a{}_K=0,\\
& & \hat{\omega}_\alpha{}^{[K}\, \hat{Q}^{\alpha J]} = 0, \quad H_{[K} \, R_{J]} = \omega_{a[K}\, Q^a{}_{J]}\,. \nonumber
\eea

\section{Relevant details on the two orientifold setups}
\label{sec_setups}

\setcounter{equation}{0}
\renewcommand{\theequation}{B.\arabic{equation}}

\subsection{Type IIA on a ${\mathbb T}^6/({\mathbb Z}_2 \times {\mathbb Z}_2)$-orientifold}
Let us briefly review the first model which is constructed in the framework of the type IIA compactification on the orientifold of a ${\mathbb T}^6/({\mathbb Z}_2\times {\mathbb Z}_2)$ orbifold. This is the very often studied type IIA setup, and the orientifold related details can also be found in \cite{Camara:2005dc,Villadoro:2005cu, Aldazabal:2006up, Blumenhagen:2013hva}. However to establish the consistency with our current notations, we will briefly present the relevant ingredients about this setup. We consider the complexified coordinates on the torus ${\mathbb T}^6$ to be defined as under,
\bea
& & \hskip-0.5cm z^1 = R^1\, x^1 + i\, R^2\, x^2, \qquad z^2 = R^3\, x^3 + i\, R^4\, x^4, \qquad z^3 = R^5 \, x^5 + i\, R^6 \, x^6 \,,
\eea
where $0 \leq x^i \leq 1$ and $R^i$ denote the circumference of the $i$-th circle. Further, the two ${\mathbb Z}_2$ orbifold actions are defined as:
\bea
& & \theta: \, \left(z^1, \, z^2, \, z^3 \right) \quad \to \quad  \left( -\, z^1, \, -\, z^2, \, z^3 \right)\,, \\
& & \ov\theta: \, \left(z^1, \, z^2, \, z^3 \right) \quad \to \quad  \left( \, z^1, \, - \, z^2, \, - z^3 \right). \nonumber
\eea
In addition an anti-holomorphic involution $\sigma$ is defined by the following action:
\bea
& & \sigma: \, \left(z^1, \, z^2, \, z^3 \right) \quad \to \quad  \left( -\ov z^1, \, -\ov z^2, \, -\ov z^3 \right).
\eea
Note that the six $R^i$'s defining the complex coordinates $z^i$'s determine the three complex structure moduli $u^i$ and three K\"ahler moduli $t^i$ which can be given as,
\bea
\label{eq:TUModelA}
& & \hskip-1cm t^1 = R^1 R^2, \quad t^2 = R^3 R^4, \quad t^3 = R^5 R^6, \quad u^1 = \frac{R^1}{R^2}, \quad u^2 = \frac{R^3}{R^4}, \quad u^3 = \frac{R^5}{R^6}\,.
\eea 
The bases of various non-trivial forms are summarized as in table \ref{tab_baseModelA}.
\begin{table}[H]
\begin{center}
\begin{tabular}{|c||c|} 
\hline
&\\
Basis of & Shorthand notation $m \wedge n = dx^m \wedge dx^n$ etc. is used. \\
Forms &\\
\hline\hline
 &\\
$\nu_a$ & $\nu_1 = 1 \wedge 2, \quad \nu_2 = 3 \wedge 4, \quad \nu_3 = 5 \wedge 6$\\
 &\\
$\alpha_K$ & $\alpha_0 = 2 \wedge 4 \wedge 6, \, \,  \alpha_1 = - 2 \wedge 3 \wedge 5, \, \, \alpha_2 = - 1 \wedge 4 \wedge 5, \, \, \alpha_3 = - 1 \wedge 3 \wedge 6$ \\
 &\\
$\beta^K$ & $ \beta^0 = 1 \wedge 3 \wedge 5, \quad \beta^1 = - 1 \wedge 4 \wedge 6, \quad \beta^2 = - 2 \wedge 3 \wedge 6, \quad \beta^3 = - 2 \wedge 4 \wedge 5$ \\
 &\\
$\tilde\nu^a$& $\tilde{\nu}^1 = \nu_2 \wedge \nu_3, \quad  \tilde{\nu}^2 = \nu_3 \wedge \nu_1, \qquad \tilde{\nu}^3 = \nu_1 \wedge \nu_2 $\\
 &\\
$\Phi_6$ & $\Phi_6 \equiv \nu_1 \wedge \nu_2 \wedge \nu_3 = 1 \wedge 2 \wedge 3 \wedge 4 \wedge 5 \wedge 6$\\
 &\\
 \hline
\end{tabular}
\end{center}
\caption{Bases of the various non-trivial forms for Model A. }
\label{tab_baseModelA}
\end{table}
\noindent
As chosen in eqn. (\ref{eq:intersectionBases}) these basis elements are normalized accordingly as $\int_{X_3} \, \alpha_J \, \wedge \beta^K = \delta_J{}^K$, and we find that $d^a{}_b = \delta^a{}_b$ and $f=1$, i.e. the six-form is normalized to unity. Further, there is only one triple intersection number of the type $\kappa_{abc}$ which is non-zero, namely $\kappa_{123} =1$ while all the other intersection numbers including $\hat{\kappa}_{a\alpha\beta}$ are zero. Using the various even/odd forms in Table \ref{tab_baseModelA} and the flux actions given in eqn. (\ref{eq:fluxActions0}), one finds the following non-zero components for the various $H, \omega, Q$ and $R$ fluxes,
\bea
\label{eq:fluxcompA1}
& { H_{ijk}:} & H_{135}, \quad H_{146}, \quad H_{236}, \quad H_{245}\, , \\
& {\omega_{ij}{}^k:} & \omega_{13}{}^6, \,\, \omega_{14}{}^5, \,\, \omega_{23}{}^5,\,\, \omega_{24}{}^6,\,\, \omega_{35}{}^2, \,\,  \omega_{36}{}^1,\,\, \omega_{45}{}^1, \,\, \omega_{46}{}^2,\,\, \omega_{51}{}^4, \,\, \omega_{52}{}^3, \,\, \omega_{61}{}^3, \,\, \omega_{62}{}^4, \nonumber\\
& {Q_{i}{}^{jk}:} & Q_1{}^{46}, \,\,  Q_1{}^{35}, \,\, Q_2{}^{36}, \,\, Q_2{}^{45}, \,\, Q_3{}^{62}, \,\, Q_3{}^{51}, \,\, Q_4{}^{52}, \,\, Q_4{}^{61}, \,\, Q_5{}^{24}, \,\, Q_5{}^{13}, \,\, Q_6{}^{14}, \,\, Q_6{}^{23}, \nonumber\\
& {R^{ijk}:} & R_{246}, \quad R_{235}, \quad R_{145}, \quad R_{136}\,. \nonumber
\eea
Note that 16 components out of 20 for each of the $H_{ijk}$ and $R^{ijk}$ flux, while 78 components out of 90 for each of the $\omega_{ij}{}^k$ and $Q_i{}^{jk}$ flux, are identically zero under this orientifold construction. In the cohomology version, these flux components can be rewritten as in eqn. (\ref{eq:fluxconversion1}). Also, from the collection of the various non-zero flux components in eqn. (\ref{eq:fluxcompA1}) it is obvious to observe that the tracelessness condition (\ref{eq:tracelessQw}) is trivially satisfied.

Now the holomorphic $(3, 0)$ form $\Omega_3$ can be determined by the choice of the coordinates $z^i$'s up to an overall constant factor. The phase is automatically fixed by our choice of anti-holomorphic involution $\sigma$ via $\sigma^\ast (\Omega_3) = \ov \Omega_3$ which suggests to consider the following form for the holomorphic three-form $\Omega_3$,
\bea
& & \Omega_3 \equiv {\cal X}^K \, \alpha_K - {\cal F}_K\, \beta^K = \frac{i}{\sqrt{2}} \, dz^1 \wedge dz^2 \wedge dz^3\,.
\eea
Subsequently, the period vectors $\left({\cal X}^I, {\cal F}_J\right)$ are given as,
\bea
\label{eq:periodExA}
& & \hskip-1.5cm  {\cal X}^0 %= \frac{1}{2 \, \sqrt{2}} \, \sqrt{\frac{R^2 \, R^4 \, R^6}{R^1 \, R^3 \, R^5}} 
=  \gamma \, \sqrt{\frac{1}{u^1 \, u^2 \, u^3}}\, , %\\ & & \hskip-1cm 
\quad {\cal X}^1 %= \,\frac{1}{2 \, \sqrt{2}} \, \sqrt{\frac{R^2 \, R^3 \, R^5}{R^1 \, R^4 \, R^6}} 
= \gamma \, \sqrt{\frac{u^2 \, u^3}{u^1}}, %\nonumber\\ & & \hskip-1cm 
\quad {\cal X}^2 %= \, \frac{1}{2 \, \sqrt{2}} \, \sqrt{\frac{R^1 \, R^4 \, R^5}{R^2 \, R^3 \, R^6}} 
= \gamma \, \sqrt{\frac{u^1 \, u^3}{u^2}}, %\nonumber\\ & & \hskip-1cm  
\quad {\cal X}^3 %= \, \frac{1}{2 \, \sqrt{2}} \, \sqrt{\frac{R^1 \, R^3 \, R^6}{R^2 \, R^4 \, R^5}} 
= \gamma \, \sqrt{\frac{u^1 \, u^2}{u^3}}, 
\eea
where $\gamma=\frac{1}{2}$ and ${\cal F}_I = -\, i /(8 \, {\cal X}^I)$ for each $I \in \{0, 1, 2, 3\}$. One can observe that ${\cal X}^K$'s are real while ${\cal F}_K$'s are pure imaginary functions of the complex structure moduli.  Moreover, the overall scale factor has been normalized via $\int_{X_3} i \, \Omega_3 \wedge \ov\Omega_3 = 1$, which is subsequently equivalent to ${\cal X}^I \, {\cal F}_I = -i/2$ as could be easily verified. Now using the definitions of chiral variable given in eqn. (\ref{eq:chiralvariables}), we define ${n^\prime}^K \equiv {\rm Im}N^K = e^{-D}\, {\cal X}^K$, and observe that the following relation holds,
\bea
& & {\cal X}^0\, {\cal X}^1\, {\cal X}^2\, {\cal X}^3\,  = \frac{1}{16} \quad \Longrightarrow  \quad 16\,\,\, e^{4\, D} \, {n^\prime}^0 \, {n^\prime}^1\, {n^\prime}^2\,{n^\prime}^3 = 1\,,
\eea
which implies that the K\"ahler potential in the quaternion sector can be written as under,
\bea
& & K_Q \equiv 4\, D = -\, \ln\left(16\,\, {n^\prime}^0 \, {n^\prime}^1\, {n^\prime}^2\, {n^\prime}^3 \right).
\eea
This K\"ahler potential $K_{Q}$ can also be written as $K_Q = - 2\, \ln\left(4\, i\, {\cal F}({n^\prime}^I)\right)$ \cite{Grimm:2004ua}, and therefore it can be determined by a prepotential of the following form,
\bea
\label{eq:prepotentialA}
& & {\cal F}({n^\prime}^I) = - i\, \sqrt{{n^\prime}^0 \, {n^\prime}^1\, {n^\prime}^2\, {n^\prime}^3}\,,
\eea 
which is a homogeneous function of degree 2 in the ${n^\prime}^I$ variables.

\subsection{Type IIA on a ${\mathbb T}^6/{\mathbb Z}_4$-orientifold}
Now let us consider the type IIA compactification on the orientifold of a ${\mathbb T}^6/{\mathbb Z}_4$ orbifold. This type IIA orientifold setup has been considered for a couple of times for different purposes, e.g. regarding (supersymmetric) moduli stabilization in \cite{Ihl:2007ah, Ihl:2006pp}. Here we consider two constructions for this sixfold and will briefly present the relevant necessary ingredients.

\subsubsection*{Model B1}
This model has been considered for the standard moduli stabilization without using the non-geometric flux in \cite{Ihl:2006pp}. The complexified coordinates on the torus ${\mathbb T}^6$ are defined as 
\bea
& & \hskip-1.5cm z^1 = x^1 + i\, x^2, \qquad z^2 = x^3 + i\, x^4, \qquad z^3 = x^5 + i\, U \, x^6\,,
\eea
where there is a single complex structure modulus $U$. Further, the ${\mathbb Z}_4$ action $\Theta$ and the anti-holomorphic involution $\sigma$  acting on the various coordinates are defined as:
\bea
\label{eq:ThetasigmaModelB}
& & \hskip-1.5cm \Theta: \, \left(z^1, \, z^2, \, z^3 \right) \to \left( i\, z^1, \, i \, z^2, \, - z^3 \right)\,, \qquad  \sigma: \, \left(z^1, \, z^2, \, z^3 \right) \to  \left( \ov z^1, \, i \, \ov z^2, \, \ov z^3 \right)\,.
\eea
%Let us note here that $\Theta \, . \, \sigma = \sigma \, . \, \Theta^3$ and so the full orientifold action is isomorphic to the dihedral group ${\cal D}_4$ \cite{Blumenhagen:2002gw, Ihl:2006pp}. 
The bases of various even/odd forms are defined in the table \ref{tab_baseModelB}.
\begin{table}[H]
\begin{center}
\begin{tabular}{|c||c|} 
\hline
% & \\
Basis of & Shorthand notation $m \wedge n = dx^m \wedge dx^n$ etc. is used. \\
Forms &\\
\hline\hline
% &\\
$\nu_a$ & $\nu_1 = 1 \wedge 2, \quad \nu_2 = 3 \wedge 4, \quad \nu_3 = 5 \wedge 6$\\
 & $\nu_4 = \frac{1}{2}\left(1 \wedge 3 - 1 \wedge 4 + 2 \wedge 3+ 2 \wedge 4\right)$ \\
  &\\
$\mu_\alpha$ & $\mu_1 = \frac{1}{2}\left(1 \wedge 3 + 1 \wedge 4 - 2 \wedge 3+ 2 \wedge 4 \right)$\\
 &\\
$\alpha_K$ & $\alpha_0 = 1 \wedge 3 \wedge 5 - 2 \wedge 4 \wedge 5 + 1 \wedge 4 \wedge 5 + 2 \wedge 3 \wedge 5$ \\
 & $\alpha_1 = 1 \wedge 3 \wedge 6 - 2 \wedge 4 \wedge 6 - 1 \wedge 4 \wedge 6 - 2 \wedge 3 \wedge 6$ \\
 &\\
$\beta^K$ & $\beta^0 = 1 \wedge 3 \wedge 6 - 2 \wedge 4 \wedge 6 + 1 \wedge 4 \wedge 6 + 2 \wedge 3 \wedge 6$ \\
 & $\beta^1 = - 1 \wedge 3 \wedge 5 + 2 \wedge 4 \wedge 5 + 1 \wedge 4 \wedge 5 + 2 \wedge 3 \wedge 5$ \\
 &\\
$\tilde\nu^a$& $\tilde{\nu}^1 = \nu_2 \wedge \nu_3, \quad  \tilde{\nu}^2 = \nu_3 \wedge \nu_1, \quad \tilde{\nu}^3 = \nu_1 \wedge \nu_2, \quad \tilde{\nu}^4 = -\,\nu_3 \wedge \nu_4$\\
 &\\
$\tilde\mu^\alpha$ &  $\tilde{\mu}^1 = \nu_3 \wedge \mu_1$ \,\\
 &\\
$\Phi_6$ & $\Phi_6 \equiv \nu_1 \wedge \nu_2 \wedge \nu_3 = - \, \nu_3 \wedge \nu_4 \wedge \nu_4 = 1 \wedge 2 \wedge 3 \wedge 4 \wedge 5 \wedge 6$\\
% &\\
 \hline
\end{tabular}
\end{center}
\caption{Bases of the various non-trivial forms for Model B. }
\label{tab_baseModelB}
\end{table}
\noindent
The triple intersection numbers surviving under the orientifold action, and the other normalization factors for the integral overs forms are fixed as under:
\bea
\label{eq:intExB}
& & \hskip-1.0cm f = 1, \quad \qquad \qquad d_a{}^b = \delta_a{}^b, \quad \qquad \qquad \hat{d}_\alpha{}^\beta = \delta_{\alpha}{}^\beta, \\ %\quad \kappa_{123} = 1, \quad \kappa_{344} =  -1, \quad \hat{\kappa}_{311} = -1\,,
& & \hskip-1.5cm \bigl\{\kappa_{abc}: \qquad \kappa_{123} = 1, \quad \kappa_{344} =  -1\bigr\}, \qquad \qquad \bigl\{\hat{\kappa}_{a\alpha\beta}: \quad \hat{\kappa}_{311} = -1\bigr\}, \nonumber 
\eea
which slightly differs from the notations of \cite{Ihl:2007ah, Ihl:2006pp}. Now we consider the holomorphic three-form $\Omega_3$ to be of the following form,
\bea
& & \hskip-1cm \Omega_3 \equiv {\cal X}^K \, \alpha_K - {\cal F}_K\, \beta^K = \frac{1 - i}{2 \, \sqrt{U}} \, dz^1 \wedge dz^2 \wedge dz^3 \\
& & = \frac{1}{2\, \sqrt{U}} \biggl[\alpha_0 + U \, \alpha_1 + i\,U \, \beta^0 + i \, \beta^1\biggr] \,. \nonumber
\eea
From this holomorphic three-form $\Omega_3$, one reads the period vectors to be given as under,
\bea
\label{eq:periodExB1}
& & {\cal X}^0 = \frac{1}{2\, \sqrt{U}} = i \, {\cal F}_1, \qquad {\cal X}^1 = \frac{\sqrt{U}}{2} = \, i \, {\cal F}_0\,.
\eea
Now using the definitions of chiral variable given as ${n^\prime}^K \equiv {\rm Im}N^K = e^{-D}\, {\cal X}^K$, one observes that the following relation holds,
\bea
& & {\cal X}^0\, {\cal X}^1\,  = \frac{1}{4} \quad \Longrightarrow  \quad 4\, e^{2\, D} \, {n^\prime}^0 \, {n^\prime}^1\, = 1\,,
\eea
which implies that the K\"ahler potential in the quaternion sector can be written as under,
\bea
& & K_Q \equiv 4\, D = - 2\, \ln\left(4\,{n^\prime}^0 \, {n^\prime}^1\right).
\eea
Comparing this K\"ahler potential $K_{Q}$ with the relation $K_Q = - 2\, \ln\left(4\, i\, {\cal F}({n^\prime}^I)\right)$ \cite{Grimm:2004ua}, one finds that $K_Q$ is determined by a prepotential of the following form,
\bea
\label{eq:prepotentialB1}
& & {\cal F}({n^\prime}^I) = - i \,{n^\prime}^0 \, {n^\prime}^1\,,
\eea 
which is a homogeneous function of degree 2 in the ${n^\prime}^I$ variables.

Using the various even/odd forms in Table \ref{tab_baseModelB} and the flux actions given in eqn. (\ref{eq:fluxActions0}), one finds the following non-zero components for the various $H, \, \omega, \, Q$ and $R$ fluxes,
\begin{align}
\label{eq:fluxcompB1}
%& {H_{ijk} \, \, {\rm flux}:}  & {R^{ijk}\, \, {\rm flux}:} \hskip4.2cm \\
&  H_{136} = - \, H_{246} = \, H_{146} = \, H_{236}\, , & R^{136} =- \, R^{246} = - \, R^{146} = - \, R^{236}\, , \nonumber\\
&  H_{135} =- \, H_{245} = - \, H_{145} = - \, H_{235}\,, & R^{135} = \,- \, R^{245} = \, R^{145} = \, R^{235},\, \, \,\quad \nonumber\\
&& \\
%& {\omega_{ij}{}^k \, \, {\rm flux}:}  & {Q_{i}{}^{jk} \, \, {\rm flux}:} \hskip4.3cm \\
& \omega_{13}{}^5 = \omega_{14}{}^5 = \omega_{23}{}^5 = - \, \omega_{24}{}^5 , & Q_1{}^{35} = -\,  Q_1{}^{45} = -\,  Q_2{}^{35} = -\,  Q_2{}^{45}, \, \, \nonumber\\
& \omega_{13}{}^6 = -\, \omega_{14}{}^6 = -\, \omega_{23}{}^6 = -\, \omega_{24}{}^6 , & Q_1{}^{36} = \,  Q_1{}^{46} = \,  Q_2{}^{36} = -\,  Q_2{}^{46}, \, \, \, \, \, \quad \nonumber\\
& \omega_{15}{}^1 = -\, \omega_{25}{}^2 , & Q_6{}^{13} = \,  Q_6{}^{14} = \,  Q_6{}^{23} = -\,  Q_6{}^{24}, \,\, \, \, \, \quad \nonumber\\
& \omega_{35}{}^1 = \omega_{35}{}^2 = -\, \omega_{45}{}^2 = \omega_{45}{}^1 , & Q_1{}^{16} = -\,  Q_2{}^{26}, \hskip3.4cm \nonumber\\
& \omega_{46}{}^1 = -\, \omega_{36}{}^1 = \omega_{36}{}^2 = \omega_{46}{}^2 ,& Q_3{}^{51} = -\,  Q_3{}^{52} = -\,  Q_4{}^{52} = -\,  Q_4{}^{51}, \,\nonumber\\
& \omega_{36}{}^3 = -\, \omega_{46}{}^4 , & Q_2{}^{51} = -\,  Q_1{}^{25}, \, \hskip3.3cm \nonumber\\
& \omega_{51}{}^3 = \omega_{51}{}^4 = \omega_{52}{}^3 = -\, \omega_{52}{}^4 , & Q_3{}^{35} = -\,  Q_4{}^{45}, \hskip3.4cm \nonumber\\
& \omega_{45}{}^3 = -\, \omega_{53}{}^4 , & Q_3{}^{61} = \,  Q_4{}^{61} = \,  Q_3{}^{62}  = - \,  Q_4{}^{62}, \, \, \, \, \quad \nonumber\\
& \omega_{26}{}^1 = -\, \omega_{61}{}^2 , & Q_3{}^{46} = -\,  Q_4{}^{63}, \hskip3.4cm \nonumber\\
& \omega_{62}{}^3 = -\, \omega_{61}{}^3 = \omega_{61}{}^4 = \omega_{62}{}^4 ,& Q_5{}^{13} = -\,  Q_5{}^{14} = -\,  Q_5{}^{23} = -\,  Q_5{}^{24}\, . \nonumber
\end{align}
\noindent
Note that 12 components out of 20 for each of the $H_{ijk}$-flux and the $R^{ijk}$-flux, while 58 components out of 90 for each of the $\omega_{ij}{}^k$-flux and $Q_i{}^{jk}$-flux, are identically zero under this orientifold construction. Moreover, all the non-zero components are not independent, and we find that there are 6 constraints among the 8 non-zero components for each of the $H_{ijk}$-flux and the $R^{ijk}$-flux while there are 22 constraints among the 32 non-zero components for each of the $\omega_{ij}{}^k$-flux and $Q_i{}^{jk}$-flux. Subsequently it turns out that there are only two independent flux components for each of the $H_{ijk}$ and $R^{ijk}$ flux while there are only 10 independent flux components for each of the $\omega_{ij}{}^k$ and $Q_i{}^{jk}$ fluxes. These details are collected in eqn. (\ref{eq:fluxcompB1}) while the cohomology version of the flux components are presented the eqn. (\ref{eq:fluxconversion2}). Also, from the collection of the various non-zero flux components in eqn. (\ref{eq:fluxcompB1}) it is obvious to observe that the tracelessness condition (\ref{eq:tracelessQw}) is indeed satisfied.

\subsubsection*{Model B2}
Now we consider a different construction for the same orbifold ${\mathbb T}^6/{\mathbb Z}_4$, which has been used for studying supersymmetric moduli stabilization with the non-geometric fluxes in \cite{Ihl:2007ah}. In this case, the complexified coordinates on the ${\mathbb T}^6$ torus is defined with a shift in the real component of $z^3$ as compared to the previous case. This is given as under,
\bea
& & \hskip-1.5cm z^1 = x^1 + i\, x^2, \qquad z^2 = x^3 + i\, x^4, \qquad z^3 = x^5 + \left(\frac{1}{2}+ i\, U \right) \, x^6\,.
\eea
While the orbifold action ${\mathbb Z}_4$, the anti-holomorphic involution $\sigma$, the two-form bases ($\mu_\alpha, \nu_a$), the four-form bases $(\tilde\mu^\alpha, \tilde\nu^a)$ and the six-form ($\Phi_6$) remain the same as before, the new three-form basis is taken as under,
\bea
& & \hskip-1.0cm \alpha_0 = 1 \wedge 3 \wedge 5 + 1 \wedge 3 \wedge 6+ 1 \wedge 4 \wedge 5 +2 \wedge 3 \wedge 5  - 2 \wedge 4 \wedge 5  -2 \wedge 4 \wedge 6 \,,\\
& & \hskip-1.0cm \alpha_1 = 1 \wedge 3 \wedge 5 + 1 \wedge 4 \wedge 5 + 1 \wedge 4 \wedge 6 +2 \wedge 3 \wedge 5  + 2 \wedge 3 \wedge 6  -2 \wedge 4 \wedge 5 \,,\nonumber\\
& & \hskip-1.0cm \beta^0 = -1 \wedge 3 \wedge 5 + 1 \wedge 4 \wedge 5 + 1 \wedge 4 \wedge 6 +2 \wedge 3 \wedge 5  + 2 \wedge 3 \wedge 6  + 2 \wedge 4 \wedge 5 \,,\nonumber\\
& & \hskip-1.0cm \beta^1 = 1 \wedge 3 \wedge 5 + 1 \wedge 3 \wedge 6 - 1 \wedge 4 \wedge 5  - 2 \wedge 3 \wedge 5  - 2 \wedge 4 \wedge 5  -2 \wedge 4 \wedge 6 \,, \nonumber
\eea
where as before $1\wedge 3 \wedge 5 = dx^1 \wedge dx^3 \wedge dx^5$ etc., and one can easily check that $\alpha_K$'s are even under involution while $\beta^K$'s are odd under the involution. We consider the holomorphic three-form $\Omega_3$ to take the following form,
\bea
& & \hskip-0.5cm \Omega_3 \equiv {\cal X}^K \, \alpha_K - {\cal F}_K\, \beta^K = \frac{1 - i}{2 \, \sqrt{U}} \, dz^1 \wedge dz^2 \wedge dz^3 \\
& & \hskip-0.5cm = \frac{1}{2\, \sqrt{U}} \biggl[ \left(\frac{1}{2} + U \right) \, \alpha_0 + \left(\frac{1}{2} - U \right) \, \alpha_1 + i\, \left(\frac{1}{2} + U \right) \, \beta^0 - i \, \left(\frac{1}{2} - U \right) \, \beta^1\biggr] \,. \nonumber
\eea
From these relations, the period vectors can be read-off as under,
\bea
\label{eq:periodExB2}
& & {\cal X}^0 = \frac{1}{2\, \sqrt{U}} \left(\frac{1}{2} + U \right) = i \, {\cal F}_0, \qquad {\cal X}^1 = \frac{1}{2\, \sqrt{U}} \left(\frac{1}{2} - U \right) =  - \, i \, {\cal F}_1\,.
\eea
Now using the definitions of chiral variable given as ${n^\prime}^K \equiv {\rm Im}N^K = e^{-D}\, {\cal X}^K$, one observes that the following relation holds,
\bea
& & ({\cal X}^0)^2\, - \, ({\cal X}^1)^2 \,  = \frac{1}{2} \quad \Longrightarrow  \quad ({n^\prime}^0)^2\,- \, ({n^\prime}^1)^2\, = \frac{1}{2}\, e^{-2\, D}\,,
\eea
which implies that the K\"ahler potential in the quaternion sector can be written as under,
\bea
& & K_Q \equiv 4\, D = - 2\, \ln\left(2\,({n^\prime}^0)^2\,- 2\, ({n^\prime}^1)^2\right).
\eea
Comparing this K\"ahler potential $K_{Q}$ with the relation $K_Q = - 2\, \ln\left(4\, i\, {\cal F}({n^\prime}^I)\right)$ \cite{Grimm:2004ua}, one finds that $K_Q$ is determined by a prepotential of the following form,
\bea
\label{eq:prepotentialB2}
& & {\cal F}({n^\prime}^I) = - \frac{i}{2} \,\biggl[({n^\prime}^0)^2\,- \, ({n^\prime}^1)^2\biggr]\,,
\eea 
which is again a homogeneous function of degree 2 in the ${n^\prime}^I$ variables as expected.

Using the various even/odd forms in Table \ref{tab_baseModelB} and the flux actions given in eqn. (\ref{eq:fluxActions0}), one finds the following non-zero components for the various $H, \omega, Q$ and $R$ fluxes \cite{Ihl:2007ah},
\begin{align}
\label{eq:fluxcompB2}
%& {H_{ijk} \, \, {\rm flux}:}  & {R^{ijk}\, \, {\rm flux}:} \hskip4.2cm \\
&  H_{136} = - \, H_{246}, & R^{135} = \,- \, R^{245}, \hskip3.4cm \nonumber\\ 
&  H_{135} =- \, H_{245} = - \, H_{145} = - \, H_{235}\,, & R^{136} = - \, R^{246} = - \, R^{146} = - \, R^{236}, \, \,  \nonumber\\
& H_{146} = \, H_{236} = -\,H_{135} +H_{136} & R^{145} = \, R^{235} = R^{135} + R^{136} \hskip1.3cm \nonumber\\
& & \\
%& {\omega_{ij}{}^k \, \, {\rm flux}:}  & {Q_{i}{}^{jk} \, \, {\rm flux}:} \hskip4.3cm \\
& \omega_{13}{}^5 = - \, \omega_{24}{}^5 , & Q_1{}^{35} = -\,  Q_2{}^{45}, \hskip3.4cm \nonumber\\
& \omega_{13}{}^6 = -\, \omega_{14}{}^6 = -\, \omega_{23}{}^6 = -\, \omega_{24}{}^6 , & Q_1{}^{36} = \,  Q_1{}^{46} = \,  Q_2{}^{36} = -\,  Q_2{}^{46}, \, \, \, \, \, \quad \nonumber\\
& 2\, \omega_{16}{}^1 = \, \omega_{15}{}^1 = -\, \omega_{25}{}^2= - 2\, \omega_{26}{}^2 , & 2\, Q_1{}^{15} = - Q_1{}^{16}  = - 2\, Q_2{}^{25} = \,  Q_2{}^{26}, \nonumber\\
& \omega_{26}{}^1 = - \omega_{61}{}^2, & 2\, Q_4{}^{35} = - Q_4{}^{36}  = 2\, Q_3{}^{45} = -\,  Q_3{}^{46}, \nonumber\\
& \omega_{36}{}^1 = -\, \omega_{46}{}^2 ,& Q_3{}^{51} = -\,  Q_4{}^{52}, \hskip3.4cm \nonumber\\
& \omega_{46}{}^1 = \, \omega_{36}{}^2 , & Q_2{}^{51} = -\,  Q_1{}^{25}, \, \hskip3.3cm \nonumber\\
& \omega_{16}{}^3 = \omega_{62}{}^4  , & Q_3{}^{35} = -\,  Q_4{}^{45}, \hskip3.4cm \nonumber\\
& \omega_{26}{}^3 = \, \omega_{16}{}^4 , & Q_3{}^{61} = \,  Q_4{}^{61} = \,  Q_3{}^{62}  = - \,  Q_4{}^{62}, \, \, \, \, \quad \nonumber\\
& \omega_{36}{}^3 = -\, \omega_{46}{}^4 , & Q_6{}^{13} = -\,  Q_6{}^{24}, \hskip3.3cm \, \nonumber\\
& 2\, \omega_{46}{}^3 = \, \omega_{45}{}^3 = \omega_{35}{}^4 = 2\, \omega_{36}{}^4 ,& Q_5{}^{13} = -\,  Q_5{}^{14} = -\,  Q_5{}^{23} = -\,  Q_5{}^{24}, \nonumber\\
& & \nonumber\\
& \hskip-0.5cm {\rm More \, \, constraints:} & \nonumber\\
& \quad \omega_{35}{}^1 = \omega_{45}{}^1 = \omega_{35}{}^2 = -\, \omega_{45}{}^2 = \omega_{36}{}^1 + \omega_{46}{}^1, & Q_6{}^{14} = Q_6{}^{23} = - Q_5{}^{13}+ Q_6{}^{13}, \nonumber \\
& \quad \omega_{15}{}^3 = \omega_{25}{}^3 = \omega_{15}{}^4 = -\,\omega_{25}{}^4 = \omega_{16}{}^3 + \omega_{26}{}^3, & Q_4{}^{15} = Q_3{}^{25} = - Q_3{}^{15}- Q_3{}^{16}, \nonumber \\
& \quad \omega_{14}{}^5 = \omega_{23}{}^5 = \omega_{13}{}^5 + \omega_{13}{}^6, & Q_2{}^{35} = Q_1{}^{45} = - Q_1{}^{35} - Q_1{}^{36}. \nonumber
\end{align}
\noindent
Note that 12 components out of 20 for each of the $H_{ijk}$-flux and the $R^{ijk}$-flux, while 54 components out of 90 for each of the $\omega_{ij}{}^k$-flux and $Q_i{}^{jk}$-flux, are identically zero under this orientifold construction. Moreover, all the non-zero components are not independent, and we find that there are 6 constraints among the 8 non-zero components for each of the $H_{ijk}$-flux and the $R^{ijk}$-flux while there are 26 constraints among the 36 non-zero components for each of the $\omega_{ij}{}^k$-flux and $Q_i{}^{jk}$-flux. Subsequently it turns out that there are only two independent flux components for each of the $H_{ijk}$ and $R^{ijk}$ flux while there are only 10 independent flux components for each of the $\omega_{ij}{}^k$ and $Q_i{}^{jk}$ fluxes. These details are collected in eqn. (\ref{eq:fluxcompB2}) while the cohomology version of the flux components are presented the eqn. (\ref{eq:fluxconversion3}). Also, from the collection of the various non-zero flux components in eqn. (\ref{eq:fluxcompB2}) it is obvious to observe that the tracelessness condition (\ref{eq:tracelessQw}) is trivially satisfied.

\section{Cohomology version of the first formulation}
\label{sec_lengthyBIs}

\setcounter{equation}{0}
\renewcommand{\theequation}{C.\arabic{equation}}

\subsection*{Model A}
In this section, we present a set of lengthy Bianchi identities for Model A. 
\subsubsection*{Class (II) identities:}
The explicit form of the 12 constraints translated from the first formulation identities using the conversion relations in eqn. (\ref{eq:fluxconversion1}) can be collectively given as under,
\bea
\label{eq:12HQww0}
& & \hskip-1cm \omega_{22}\, \omega_{30} + \omega_{20}\, \omega_{32} = H_2 Q^1{}_0 + H_0 Q^1{}_2, \quad \omega_{23}\, \omega_{30} + \omega_{20}\, \omega_{33} = H_3 Q^1{}_0 + H_0 Q^1{}_3 \,, \nonumber\\
& & \hskip-1cm \omega_{11}\, \omega_{30} + \omega_{10}\, \omega_{31} = H_1 Q^2{}_0 + H_0 Q^2{}_1, \quad \omega_{13}\, \omega_{30} + \omega_{10}\, \omega_{33} = H_3 Q^2{}_0 + H_0 Q^2{}_3 \,, \nonumber\\
& & \hskip-1cm \omega_{11}\, \omega_{20} + \omega_{10}\, \omega_{21} = H_1 Q^3{}_0 + H_0 Q^3{}_1 , \quad \omega_{12}\, \omega_{20} + \omega_{10}\, \omega_{22} = H_2 Q^3{}_0 + H_0 Q^3{}_2 \,, \nonumber\\
& & \\
& & \hskip-1cm \omega_{22}\, \omega_{31} + \omega_{21}\, \omega_{32} = H_2 Q^1{}_1 + H_1 Q^1{}_2 , \quad 
\omega_{23}\, \omega_{31} + \omega_{21}\, \omega_{33} = H_3 Q^1{}_1 + H_1 Q^1{}_3\,, \nonumber\\
& & \hskip-1cm \omega_{13}\, \omega_{32} + \omega_{12}\, \omega_{33} = H_3 Q^2{}_2 + H_2 Q^2{}_3, \quad 
\omega_{12}\, \omega_{31} + \omega_{11}\, \omega_{32} = H_2 Q^2{}_1 + H_1 Q^2{}_2 \,, \nonumber\\
& & \hskip-1cm \omega_{13}\, \omega_{22} + \omega_{12}\, \omega_{23} = H_3 Q^3{}_2 + H_2 Q^3{}_3, \quad 
\omega_{13}\, \omega_{21} + \omega_{11}\, \omega_{23} = H_3 Q^3{}_1 + H_1 Q^3{}_3 \,.\nonumber
\eea
\subsubsection*{Class (III) identities:}
The explicit expressions of the 24 constraints directly coming from translating the first formulation identities using the conversion relations in eqn. (\ref{eq:fluxconversion1}) can be collectively given as under,
\bea
\label{eq:24HRwQ0}
& & \hskip-1cm  H_0 R_1+\omega_{31} Q^3{}_0=\omega_{10} Q^1{}_1+\omega_{20} Q^2{}_1\, , \quad  H_1 R_0+\omega_{30} Q^3{}_1=\omega_{11} Q^1{}_0+\omega_{21} Q^2{}_0 \, , \nonumber\\
& & \hskip-1cm H_0 R_2+\omega_{12} Q^1{}_0=\omega_{20} Q^2{}_2+\omega_{30} Q^3{}_2 \, , \quad  H_2 R_0+\omega_{30} Q^3{}_2=\omega_{12} Q^1{}_0+\omega_{22} Q^2{}_0  \, , \nonumber\\
& & \hskip-1cm H_0 R_3+\omega_{23} Q^2{}_0=\omega_{10} Q^1{}_3+\omega_{30} Q^3{}_3\, , \quad  H_3 R_0+\omega_{10} Q^1{}_3=\omega_{23} Q^2{}_0+\omega_{33} Q^3{}_0 \, , \nonumber\\
& & \hskip-1cm H_1 R_0+\omega_{20} Q^2{}_1=\omega_{11} Q^1{}_0+\omega_{31} Q^3{}_0\, , \quad H_0 R_1+\omega_{21} Q^2{}_0=\omega_{10} Q^1{}_1+\omega_{30} Q^3{}_1 \, , \nonumber\\
& & \hskip-1cm H_2 R_0+\omega_{10} Q^1{}_2=\omega_{22} Q^2{}_0+\omega_{32} Q^3{}_0 \, , \quad H_0 R_2+\omega_{32} Q^3{}_0=\omega_{10} Q^1{}_2+\omega_{20} Q^2{}_2 \, , \nonumber\\
& & \hskip-1cm H_3 R_0+\omega_{20} Q^2{}_3=\omega_{13} Q^1{}_0+\omega_{33} Q^3{}_0 \, , \quad H_0 R_3+\omega_{13} Q^1{}_0=\omega_{20} Q^2{}_3+\omega_{30} Q^3{}_3  \, , \nonumber\\
& & \\
& & \hskip-1cm H_2 R_1+\omega_{21} Q^2{}_2=\omega_{12} Q^1{}_1+\omega_{32} Q^3{}_1\, , \quad  H_2 R_1+\omega_{11} Q^1{}_2=\omega_{22} Q^2{}_1+\omega_{32} Q^3{}_1\, , \nonumber\\
& & \hskip-1cm H_1 R_2+\omega_{12} Q^1{}_1=\omega_{21} Q^2{}_2+\omega_{31} Q^3{}_2\, , \quad H_1 R_2+\omega_{22} Q^2{}_1=\omega_{11} Q^1{}_2+\omega_{31} Q^3{}_2 \, , \nonumber\\
& & \hskip-1cm H_2 R_3+\omega_{23} Q^2{}_2=\omega_{12} Q^1{}_3+\omega_{32} Q^3{}_3 \, , \quad  H_2 R_3+\omega_{33} Q^3{}_2=\omega_{12} Q^1{}_3+\omega_{22} Q^2{}_3 \, ,\nonumber\\
& & \hskip-1cm H_3 R_2+\omega_{22} Q^2{}_3=\omega_{13} Q^1{}_2+\omega_{33} Q^3{}_2 \, , \quad  H_3 R_2+\omega_{32} Q^3{}_3=\omega_{13} Q^1{}_2+\omega_{23} Q^2{}_2 \, , \nonumber\\
& & \hskip-1cm H_1 R_3+\omega_{13} Q^1{}_1=\omega_{21} Q^2{}_3+\omega_{31} Q^3{}_3\, , \quad  H_1 R_3+\omega_{33} Q^3{}_1=\omega_{11} Q^1{}_3+\omega_{21} Q^2{}_3 \, , \nonumber\\
& & \hskip-1cm H_3 R_1+\omega_{31} Q^3{}_3=\omega_{13} Q^1{}_1+\omega_{23} Q^2{}_1 \, , \quad H_3 R_1+\omega_{11} Q^1{}_3=\omega_{23} Q^2{}_1+\omega_{33} Q^3{}_1\, . \nonumber
\eea
However after some reshuffling, these 24 constraints are reduced into 18 constraints which consists of the following six constraints directly coming from the second formulation,
\bea
& & H_{[K}\, R_{J]} = \omega_{a[K}\, Q^a{}_{J]}, \qquad \forall \, \, I, \, J \in \{0, 1, 2, 3\}\,.
\eea
In addition, there are 12 missing constraints collected as under,
\bea
\label{eq:24HRwQ1}
& & \hskip-1cm H_0 \, R_1 + H_1 \, R_0 = \omega_{11}\, Q^1{}_0 + \omega_{10}\, Q^1{}_1, \quad \omega_{21}\, Q^2{}_0 + \omega_{20}\, Q^2{}_1 = \omega_{31}\, Q^3{}_0 + \omega_{30}\, Q^3{}_1, \nonumber\\
& & \hskip-1cmH_0 \, R_2 + H_2 \, R_0 = \omega_{22}\, Q^2{}_0 + \omega_{20}\, Q^2{}_2, \quad \omega_{12}\, Q^1{}_0 + \omega_{10}\, Q^1{}_2 = \omega_{32}\, Q^3{}_0 + \omega_{30}\, Q^3{}_2, \nonumber\\
& & \hskip-1cm H_0 \, R_3 + H_3 \, R_0 = \omega_{33}\, Q^3{}_0 + \omega_{30}\, Q^3{}_3, \quad \omega_{13}\, Q^1{}_0 + \omega_{10}\, Q^1{}_3 = \omega_{23}\, Q^2{}_0 + \omega_{20}\, Q^2{}_3, \nonumber\\
& & \\
& & \hskip-1cm H_1 \, R_2 + H_2 \, R_1 = \omega_{32}\, Q^3{}_1 + \omega_{31}\, Q^3{}_2, \quad \omega_{12}\, Q^1{}_1 + \omega_{11}\, Q^1{}_2 = \omega_{22}\, Q^2{}_1 + \omega_{21}\, Q^2{}_2, \nonumber\\
& & \hskip-1cm H_2 \, R_3 + H_3 \, R_2 = \omega_{13}\, Q^1{}_2 + \omega_{12}\, Q^1{}_3, \quad \omega_{23}\, Q^2{}_2 + \omega_{22}\, Q^2{}_3 = \omega_{33}\, Q^3{}_2 + \omega_{32}\, Q^3{}_3, \nonumber\\
& & \hskip-1cm H_3 \, R_1 + H_1 \, R_3 = \omega_{23}\, Q^2{}_1 + \omega_{21}\, Q^2{}_3, \quad \omega_{13}\, Q^1{}_1 + \omega_{11}\, Q^1{}_3 = \omega_{33}\, Q^3{}_1 + \omega_{31}\, Q^3{}_3\,. \nonumber
\eea

\subsubsection*{Class (IV) identities:}
The explicit form of the 12 constraints translated from the first formulation identities using the conversion relations in eqn. (\ref{eq:fluxconversion1}) can be collectively given as under,
\bea
\label{eq:12RwQQ0}
& & \hskip-1cm R_2 \omega_{30} + R_0 \omega_{32} = Q^1{}_2 Q^2{}_0 + Q^1{}_0 Q^2{}_2, \quad R_1 \omega_{30} + R_0 \omega_{31} = Q^1{}_1 Q^2{}_0 + Q^1{}_0 Q^2{}_1 \,,\nonumber\\
& & \hskip-1cm R_2 \omega_{10} + R_0 \omega_{12} = Q^2{}_2 Q^3{}_0 + Q^2{}_0 Q^3{}_2, \quad R_3 \omega_{10} + R_0 \omega_{13} = Q^2{}_3 Q^3{}_0 + Q^2{}_0 Q^3{}_3 \,,\nonumber\\
& & \hskip-1cm R_1 \omega_{20} + R_0 \omega_{21} = Q^1{}_1 Q^3{}_0 + Q^1{}_0 Q^3{}_1, \quad R_3 \omega_{20} + R_0 \omega_{23} = Q^1{}_3 Q^3{}_0 + Q^1{}_0 Q^3{}_3 \,,\nonumber\\
& & \\
& & \hskip-1cm R_3 \omega_{32} + R_2 \omega_{33} = Q^1{}_3 Q^2{}_2 + Q^1{}_2 Q^2{}_3, \quad R_3 \omega_{31} + R_1 \omega_{33} = Q^1{}_3 Q^2{}_1 + Q^1{}_1 Q^2{}_3\,,\nonumber\\
& & \hskip-1cm R_2 \omega_{11} + R_1 \omega_{12} = Q^2{}_2 Q^3{}_1 + Q^2{}_1 Q^3{}_2, \quad R_3 \omega_{11} + R_1 \omega_{13} = Q^2{}_3 Q^3{}_1 + Q^2{}_1 Q^3{}_3\,,\nonumber\\
& & \hskip-1cm R_2 \omega_{21} + R_1 \omega_{22} = Q^1{}_2 Q^3{}_1 + Q^1{}_1 Q^3{}_2, \quad R_3 \omega_{22} + R_2 \omega_{23} = Q^1{}_3 Q^3{}_2 + Q^1{}_2 Q^3{}_3\,.\nonumber
\eea
Let us give some remarks on the four identities mentioned in eqn. (\ref{eq:bianchids1additional}). These are a set of weaker constraints which can be derived from the first formulation by contracting some more six-dimensional indices. We try to investigate if this collection is somehow directly related with some of the second formulation identities. It turns out that the first, the third and the fourth identities of eqn. (\ref{eq:bianchids1additional}) are trivially satisfied for Model A while the second identity results in the following set of constraints,
\bea
\label{eq:bianchids1additional2}
& & \hskip-1cm H_2 R_1 + H_1 R_2=\omega_{32} Q^3{}_1 + \omega_{31} Q^3{}_2 \, , \qquad H_1 R_0 + H_0 R_1=\omega_{11} Q^1{}_0+\omega_{10} Q^1{}_1 \, , \nonumber\\
& & \hskip-1cm  H_3 R_2 + H_2 R_3=\omega_{13} Q^1{}_2 + \omega_{12} Q^1{}_3 \, , \qquad H_2 R_0 + H_0 R_2=\omega_{22} Q^2{}_0 + \omega_{20} Q^2{}_2 \, , \\
& & \hskip-1cm  H_3 R_1+H_1 R_3=\omega_{23} Q^2{}_1+\omega_{21} Q^2{}_3 \, , \qquad H_3 R_0 + H_0 R_3 = \omega_{33} Q^3{}_0 + \omega_{30} Q^3{}_3 \,. \nonumber
\eea
Notice that all of these identities are indeed contained in the generic constraints of class {\bf (III)} as collected in eqn. (\ref{eq:24HRwQ1}) or in the compact collection of eqn. (\ref{eq:24HRwQ}). In particular, the second and third relations in eqn. (\ref{eq:24HRwQ}) are precisely these identities.

\subsection*{Model B}
In this section, we present a set of lengthy Bianchi identities for Model B. Unlike the case for Model A, we do not present the results which directly come from translating the first formulation identities using the conversion relations in eqn. (\ref{eq:fluxconversion2}) and eqn. (\ref{eq:fluxconversion3}) as the same are not only too lengthy to be presented but also are not very illuminating. However after reshuffling such relations, one can separate out the constraints which come from the second formulation, and in addition there are some which cannot be obtained from the second formulation. We present such identities in the following collections.
\subsubsection*{Class (II) identities:} In this case, there are 5 constraints which correspond to the second formulation, and the same are given as under,
\bea
\label{eq:ModelBHQwwExpand0}
& & H_0 \, \hat{Q}^{10} + H_1 \, \hat{Q}^{11} = 0, \quad \omega_{10}\, \hat{\omega}_{1}{}^0 + \omega_{11}\, \hat{\omega}_{1}{}^1=0, \\
& & \omega_{20}\, \hat{\omega}_{1}{}^0 + \omega_{21}\, \hat{\omega}_{1}{}^1=0, \quad \omega_{30}\, \hat{\omega}_{1}{}^0 + \omega_{31}\, \hat{\omega}_{1}{}^1=0, \quad \omega_{40}\, \hat{\omega}_{1}{}^0 + \omega_{41}\, \hat{\omega}_{1}{}^1=0 \,, \nonumber
\eea
while there are 11 constraints which are not the part of the second formulation. Nine of these are the followings which appears in both the Models B1 and B2,
\bea
\label{eq:ModelBHQwwExpand1}
& & \hskip-0.0cm H_0 \, Q^1{}_0 = \omega_{20} \, \omega_{30}, \quad H_1 \, Q^1{}_1= \omega_{21} \, \omega_{31},  \quad H_1 \, Q^1{}_0 + H_0 \, Q^1{}_1 = \omega_{21} \, \omega_{30} + \omega_{20} \, \omega_{31} \, ,\\
& & \hskip-0.0cm H_0 \, Q^2{}_0 = \omega_{10} \, \omega_{30}, \quad  H_1 \, Q^2{}_1 = \omega_{11} \, \omega_{31}, \quad H_1 \, Q^2{}_0 + H_0 \, Q^2{}_1 = \omega_{11} \, \omega_{30} + \omega_{10} \, \omega_{31} \,, \nonumber\\
& & \hskip-0.0cm H_0 \, Q^4{}_0 + \omega_{40} \, \omega_{30}=0, \quad H_1 \, Q^4{}_1 + \omega_{41} \, \omega_{31}=0, \quad  H_1 \, Q^4{}_0 + H_0 \, Q^4{}_1 + \omega_{31} \, \omega_{40} + \omega_{30} \, \omega_{41}=0, \nonumber
\eea
while there are two more identities which apparently differ in the two constructions, and these are given as under,
\bea
\label{eq:ModelBHQwwExpand2}
& & \hskip-0.5cm {\rm Model \, \, B1:} \quad H_1 \, \hat{Q}^{11} - H_0 \, \hat{Q}^{10} + \omega_{31} \, \hat\omega_{1}{}^1 - \omega_{30} \, \hat\omega_{1}{}^0 = 0\, , \nonumber\\
& & \hskip1.5cm H_1 Q^3{}_0 + H_0 Q^3{}_1 - \omega_{10} \, \omega_{21} - \omega_{11} \, \omega_{20} + \, \omega_{40}\,  \omega_{41} + \hat{\omega}_1{}^0 \, \hat{\omega}_1{}^1 = 0\, . \nonumber\\
& & \\
& & \hskip-0.5cm {\rm Model \, \, B2:} \quad H_1 \, \hat{Q}^{10} + H_0 \, \hat{Q}^{11} + \omega_{31} \, \hat\omega_{1}{}^0 + \omega_{30} \, \hat\omega_{1}{}^1 =0 \, , \nonumber\\
& & \hskip1.5cm H_0 Q^3{}_0 - \omega_{10} \, \omega_{20} + \frac{1}{2} \, \omega_{40}^2 + \frac{1}{2} \, (\hat{\omega}_1{}^0)^2  = H_1 Q^3{}_1 - \omega_{11} \, \omega_{21} + \frac{1}{2} \, \omega_{41}^2  + \frac{1}{2} \, (\hat{\omega}_1{}^1)^2 \, . \nonumber
\eea

\subsubsection*{Class (III) identities:}
In this case, there are 10 constraints which correspond to the second formulation, and the same are given as under,
\bea
\label{eq:ModelBHRwQExpand0}
& & \hskip-1cm \hat\omega_1{}^0 \, Q^1{}_0 + \hat\omega_1{}^1 \, Q^1{}_1 = 0, \quad \hat\omega_1{}^0 \, Q^2{}_0 + \hat\omega_1{}^1 \, Q^2{}_1 = 0, \quad \hat\omega_1{}^0 \, Q^3{}_0 + \hat\omega_1{}^1 \, Q^3{}_1 = 0, \\
& & \hskip-1cm \hat\omega_1{}^0 \, Q^4{}_0 + \hat\omega_1{}^1 \, Q^4{}_1 = 0, \quad {\hat Q}^{1 0} \, \omega_{10} + {\hat Q}^{11} \, \omega_{11} = 0, \quad {\hat Q}^{1 0} \, \omega_{20} + {\hat Q}^{11} \, \omega_{21} = 0, \nonumber\\
& & \hskip-1cm {\hat Q}^{1 0} \, \omega_{30} + {\hat Q}^{11} \, \omega_{31} = 0, \quad {\hat Q}^{1 0} \, \omega_{40} + {\hat Q}^{11} \, \omega_{41} = 0\, , \quad \hat\omega_1{}^{0} {\hat Q}^{11} - \hat\omega_1{}^{1} {\hat Q}^{10} =0 \,, \nonumber\\
& & \hskip-1cm H_0 R_1 - \omega_{10}\, Q^1{}_1 - \omega_{20} \, Q^2{}_1 - \omega_{30} Q^3{}_1 - \omega_{40} Q^4{}_1 \nonumber\\
& &  \hskip1cm = H_1 R_0 - \omega_{11}\, Q^1{}_0 - \omega_{21} \, Q^2{}_0 - \omega_{31} Q^3{}_0 - \omega_{41} Q^4{}_0 \,, \nonumber
\eea
while there are 16 constraints which are not the part of the second formulation. Twelve of these constraints are the followings which appear in both the Models B1 and B2,
\bea
\label{eq:ModelBHRwQExpand1}
& &  \hskip-0.3cm H_0 \, R_0 = \omega_{30}\, Q^3{}_0, \quad H_1 \, R_1 = \omega_{31}\, Q^3{}_1, \quad H_0 \, R_1 + H_1 \, R_0 = \omega_{30}\, Q^3{}_1+ \omega_{31}\, Q^3{}_0\,, \\
& &  \hskip-0.3cm \omega_{10} \, Q^1{}_0 = \omega_{20} \, Q^2{}_0, \quad \omega_{11} \, Q^1{}_1 = \omega_{21} \, Q^2{}_1, \quad \omega_{11} \, Q^1{}_0 + \omega_{10} \, Q^1{}_1 = \omega_{21} \, Q^2{}_0 + \omega_{20} \, Q^2{}_1, \nonumber\\
& &  \hskip-0.3cm \omega_{40} \, Q^1{}_0 + \omega_{20} \, Q^4{}_0 =0, \, \, \omega_{41} \, Q^1{}_1 + \omega_{21} \, Q^4{}_1=0, \, \, \omega_{41} \, Q^1{}_0 + \omega_{40} \, Q^1{}_1 + \omega_{21} \, Q^4{}_0 + \omega_{20} \, Q^4{}_1=0, \nonumber\\
& &  \hskip-0.3cm \omega_{40} \, Q^2{}_0 + \omega_{10} \, Q^4{}_0=0, \, \, \omega_{41} \, Q^2{}_1 + \omega_{11} \, Q^4{}_1=0, \, \, \omega_{41} \, Q^2{}_0 + \omega_{40} \, Q^2{}_1 + \omega_{11} \, Q^4{}_0 + \omega_{10} \, Q^4{}_1=0, \nonumber
\eea
while there are four more identities which apparently differ in the two constructions, and these are given as under,
\bea
\label{eq:ModelBHRwQExpand2}
& & \hskip-0.2cm {\rm Model \, \, B1:} \quad \quad \quad Q^1{}_0\, \hat{\omega}_1{}^0 - Q^1{}_1\, \hat{\omega}_1{}^1 =\omega_{21} \, \hat{Q}^{11} - \omega_{20} \, \hat{Q}^{10}\,, \nonumber\\
& &  \hskip-0cm  Q^2{}_0\, \hat{\omega}_1{}^0 - Q^2{}_1\, \hat{\omega}_1{}^1 = \omega_{11} \, \hat{Q}^{11}- \omega_{10} \, \hat{Q}^{10}\,, \quad Q^4{}_0\, \hat{\omega}_1{}^0 - Q^4{}_1\, \hat{\omega}_1{}^1 = \omega_{40} \, \hat{Q}^{10} - \omega_{41} \, \hat{Q}^{11}\,,\nonumber\\
& & \hskip-0cm H_0 R_1 + \, H_1 R_0 - \omega_{11}\, Q^1{}_0 - \omega_{10}\, Q^1{}_1 - \omega_{21} \, Q^2{}_0 - \omega_{20} \, Q^2{}_1 \nonumber\\
& & \hskip1.0cm + \, \omega_{31} Q^3{}_0  + \, \omega_{30} Q^3{}_1 + \omega_{41} Q^4{}_0 + \omega_{40} Q^4{}_1 + \hat{\omega}_1{}^0 \, \hat{Q}^{11} + \hat{\omega}_1{}^1 \, \hat{Q}^{10} = 0, \nonumber\\
& & \\
& & \hskip-0.2cm {\rm Model \, \, B2:} \quad \quad \quad Q^1{}_1\, \hat{\omega}_1{}^0 + Q^1{}_0\, \hat{\omega}_1{}^1 + \omega_{21} \, \hat{Q}^{10}\, + \omega_{20} \, \hat{Q}^{11} =0, \nonumber\\
& &  \hskip-0cm  Q^2{}_1\, \hat{\omega}_1{}^0 + Q^2{}_0\, \hat{\omega}_1{}^1 + \omega_{11} \, \hat{Q}^{10}\, + \omega_{10} \, \hat{Q}^{11} =0, \quad Q^4{}_1\, \hat{\omega}_1{}^0 + Q^4{}_0\, \hat{\omega}_1{}^1 + \omega_{41} \, \hat{Q}^{10}\, + \omega_{40} \, \hat{Q}^{11}=0, \nonumber\\
& & \hskip-0cm H_0 R_0 - \omega_{10}\, Q^1{}_0 - \omega_{20} \, Q^2{}_0 + \omega_{30} Q^3{}_0 + \omega_{40} Q^4{}_0 + \hat{\omega}_1{}^0 \, \hat{Q}^{10} \nonumber\\
& &  \hskip1cm = H_1 R_1 - \omega_{11}\, Q^1{}_1 - \omega_{21} \, Q^2{}_1 + \omega_{31} Q^3{}_1 + \omega_{41} Q^4{}_1 + \hat{\omega}_1{}^1 \, \hat{Q}^{11}.\nonumber
\eea

\subsubsection*{Class (IV) identities:}
In this case, there are 5 constraints which correspond to the second formulation, and the same are given as under,
\bea
\label{eq:ModelBRwQQExpand0}
& & R_0 \, \hat{\omega}_{1}{}^{0} + R_1 \, \hat{\omega}_{1}{}^{1} = 0, \quad \quad \quad Q^{1}{}_{0}\, \hat{Q}^{10} + Q^{1}{}_{1}\, \hat{Q}^{11}=0, \\
& & Q^{2}{}_{0}\, \hat{Q}^{10} + Q^{2}{}_{1}\, \hat{Q}^{11}=0, \quad \quad Q^{3}{}_{0}\, \hat{Q}^{10} + Q^{3}{}_{1}\, \hat{Q}^{11}=0, \quad \quad Q^{4}{}_{0}\,\hat{Q}^{10} + Q^{4}{}_{1}\, \hat{Q}^{11}=0, \nonumber
\eea
while there are 11 constraints which are not the part of the second formulation. Nine of these are the followings which appears in both the Models B1 and B2,
\bea
\label{eq:ModelBRwQQExpand1}
& & \hskip-0.2cm Q^{1}{}_{0} \, Q^{3}{}_{0} = R_0 \, \omega_{20}, \quad Q^{1}{}_{1} \, Q^{3}{}_{1} = R_1 \, \omega_{21}, \quad {Q}^{1}{}_{1} \, {Q}^{3}{}_{0} + {Q}^{1}{}_{0} \, {Q}^{3}{}_{1} = R_1 \, \omega_{20} + R_0 \, \omega_{21}\, , \nonumber\\
& & \hskip-0.2cm Q^{2}{}_{0} \, Q^{3}{}_{0} = R_0 \, \omega_{10}, \quad Q^{2}{}_{1} \, Q^{3}{}_{1} = R_1 \, \omega_{11}, \quad {Q}^{2}{}_{1} \, {Q}^{3}{}_{0} + {Q}^{2}{}_{0} \, {Q}^{3}{}_{1} = R_1 \, \omega_{10} + R_0 \, \omega_{11}\, ,\\
& & \hskip-0.2cm Q^{4}{}_{0} \, Q^{3}{}_{0} + R_0 \, \omega_{40}=0, \, \, Q^{4}{}_{1} \, Q^{3}{}_{1} + R_1 \, \omega_{41}=0, \, \, {Q}^{3}{}_{1} \, {Q}^{4}{}_{0} + {Q}^{3}{}_{0} \, {Q}^{4}{}_{1} + R_1 \, \omega_{40} + R_0 \, \omega_{41}=0, \nonumber
\eea
while there are two more identities which apparently differ in the two constructions, and these are given as under,
\bea
\label{eq:ModelBRwQQExpand2}
& & \hskip-0.50cm {\rm Model \, \, B1:} \quad {Q}^{3}{}_{0} \, {\hat Q}^{10} - {Q}^{3}{}_{1} \,{\hat Q}^{11} + R_0 \, {\hat \omega}_{1}{}^{0} - R_1 \, \hat{\omega}_{1}{}^1 =0\, , \nonumber\\
& & R_0 \, \omega_{31} + R_1 \, \omega_{30} - Q^{1}{}_{1} \, Q^{2}{}_{0} - Q^{1}{}_{0} \, Q^{2}{}_{1} + \, Q^4{}_0\, Q^4{}_1  + \, {\hat Q}^{10} \, {\hat Q}^{11} = 0\, . \nonumber\\
& & \\
& & \hskip-0.50cm {\rm Model \, \, B2:} \quad {Q}^{3}{}_{1} \,{\hat Q}^{10} + {Q}^{3}{}_{0} \, {\hat Q}^{11} + R_1 \, {\hat \omega}_{1}{}^{0} + R_0 \, \hat{\omega}_{1}{}^1 = 0 \, , \nonumber\\
& & \hskip-0.2cm R_0 \, \omega_{30} - Q^{1}{}_{0} \, Q^{2}{}_{0} + \frac{1}{2}\, {(Q^4{}_0)}^2  + \frac{1}{2}\, {({\hat Q}^{10})}^2 = R_1 \, \omega_{31} - \, Q^{1}{}_{1} \, Q^{2}{}_{1} + \frac{1}{2} \, {(Q^{4}{}_{1})}^2  + \frac{1}{2} \, {({\hat Q}^{11})}^2\, . \nonumber
\eea
%\subsubsection*{Invoking a model-independent form}
%Let us now discuss an apparent ambiguity seen from the collection of identities presented in eqns. (\ref{eq:ModelBHQwwExpand2}),  (\ref{eq:ModelBHRwQExpand2}) and (\ref{eq:ModelBRwQQExpand2}), where one observes that although most of the identities in the model B1 and B2 are the same, there are a couple of identities which naively appear to be not correlated in the cohomology version. This should not be the case as the global topological numbers such as Hodge numbers and the intersections numbers are the same in Model B1 and B2, and so one would naively expect to get the same, or at least correlated identities. However this ambiguity disappears after considering the tracelessness condition given in eqn. (\ref{eq:tracelessQw1}). To be specific, one finds that the first constraint in each of the Model B1 and B2 as given in eqn. (\ref{eq:ModelBHQwwExpand2}) reduces into the ones already contained in the collection (\ref{eq:ModelBHQwwExpand1}), and so this difference is trivialized. Similarly, one finds that the three out of four relations in each of the two sets of constructions given in eqn. (\ref{eq:ModelBHRwQExpand2}) reduce into the ones already contained in the collection (\ref{eq:ModelBHRwQExpand1}), and so this difference is trivialized. Finally, the first relation in each of the Model B1 and B2 as given in eqn. (\ref{eq:ModelBRwQQExpand2}) reduces into the ones already contained in the collection (\ref{eq:ModelBRwQQExpand1}), and so this difference is also trivialized.
On the lines of the discussion on the $(2,1)$-cohomology structure in section \ref{sec_BIsGeneric}, we find that the apparent different constraints in each of the three classes, namely {\bf (II)}, {\bf (III)} and {\bf (IV)}, are also correlated, and can be combined into the following relations, 
\bea
& & \hskip-1.5cm l_{0JK}^{-1} \,\biggl[H_{(\underline J} \, Q^3{}_{\underline K)} - \frac{1}{2}\, \kappa_{3bc}^{-1} \, \omega_{b \, (\underline J} \, \omega_{c \, \underline K)}\biggr]  = l_{0JK}\, \biggl[\frac{1}{2} \,\hat{\kappa}_{3\alpha\beta}^{-1} \, \hat{\omega}_{\alpha}{}^{(\underline J} \, \hat{\omega}_{\beta}{}^{\underline K)} \biggr] \qquad \forall \, J, \, K \,, \nonumber\\
& & \hskip-1.5cm l_{0JK}^{-1} \,\biggl[3\, H_{(\underline J} \, R_{\underline K)} - \, Q^{a}{}_{(\underline J} \, \omega_{a}{}_{\underline K)} \biggr] = l_{0JK}\, {\hat Q}^{\alpha (\underline J} \, \hat\omega_{\alpha}{}^{\underline K)}\qquad \forall \, J, \, K \,, \nonumber\\
& & \hskip-1.5cm l_{0JK}^{-1} \,\biggl[R_{(\underline J} \, \omega_{3 \, \underline K)} - \frac{1}{2} \, \kappa_{3bc} \, Q^{b}{}_{(\underline J} \, Q^{c}{}_{\underline K)} \biggr] = l_{0JK}\, \biggl[\frac{1}{2} \, \hat{\kappa}_{3\alpha\beta} \, \hat{Q}^{\alpha \, (\underline J} \, \hat{Q}^{\beta \, \underline K)} \biggr] \qquad \forall \, J, \, K \,.
\eea
%Therefore, there is no conceptual difference between the Bianchi identities in Model B1 and Model B2. This also illuminates the motivation of using two constructions for the model B. The resolution of the apparent ambiguity via using tracelessness condition hints that this has to be imposed as done most of the times in the toroidal orieintifold literature. However, if we impose this at the very beginning of the simplification, then we could have missed the beautiful contractions appearing with even/odd cohomology indices, and it would have been hard to make the educated guess about the generic form for (some of) the missing identities.
Finally, let us give some remarks on the four identities mentioned in eqn. (\ref{eq:bianchids1additional}). In this case it turns out that the fourth identity in eqn. (\ref{eq:bianchids1additional}) is trivially satisfied while the first and third one give a single constraint each which can be rewritten as
\bea
& & H_0 {\hat Q}^{10} + H_1 {\hat Q}^{11} = 0 \,, \qquad R_0 \, \hat\omega_1{}^0 +  R_1 \, \hat\omega_1{}^1 = 0 \,,
\eea
which are the same as those of the second formulation. In addition, the most complicated identity out of (\ref{eq:bianchids1additional}) turns out to be the second one which, with a little bit of reshuffling and after being accompanied with the second formulation constraints along with condition (\ref{eq:tracelessQw1}), can produce the following six `missing' constraints,
\bea
\label{eq:bianchids1additional3}
& & \hskip-0.75cm {\rm Model \, \, B1:} \quad \quad \quad  H_0 R_0 = \, \omega_{30} Q^3{}_0, \quad H_1 R_1 = \, \omega_{31} Q^3{}_1, \nonumber\\
& & \hskip-0.1cm \qquad H_0 \, R_1 + H_1 \, R_0 = \omega_{10}\, Q^1{}_1+ \omega_{11}\, Q^1{}_0 = \omega_{20}\, Q^2{}_1+ \omega_{21}\, Q^2{}_0, \nonumber\\
& & \hskip-0.1cm \omega_{40}\,Q^2{}_0 +\omega_{11}\,Q^4{}_1 = \omega_{41}\,Q^2{}_1 + \omega_{10}\,Q^4{}_0, \quad \omega_{40}\,Q^1{}_0 + \omega_{21}\,Q^4{}_1 = \omega_{41}\,Q^1{}_1 + \omega_{20}\,Q^4{}_0,  \nonumber\\
& & \\
& & \hskip-0.75cm {\rm Model \, \, B2:} \quad \quad \quad 2 H_0 R_0 -  \, \omega_{10} Q^1{}_0 -  \omega_{20} Q^2{}_0  = 2 H_1 R_1 -  \, \omega_{11} Q^1{}_1 -  \omega_{21} Q^2{}_1, \nonumber\\
& & \hskip-0.5cm H_0 \, R_0 + H_1 \, R_1 = \omega_{30}\, Q^3{}_0 + \omega_{31}\, Q^3{}_1, \quad H_0 \, R_1 + H_1 \, R_0 = \omega_{30}\, Q^3{}_1+ \omega_{31}\, Q^3{}_0\, , \nonumber\\
& & \hskip-0.5cm \omega_{40}\,Q^1{}_0 + \omega_{20}\,Q^4{}_0 = \omega_{41}\,Q^1{}_1 + \omega_{21}\,Q^4{}_1, \quad \omega_{40}\,Q^2{}_0 + \omega_{10}\,Q^4{}_0 = \omega_{41}\,Q^2{}_1 +\omega_{11}\,Q^4{}_1,  \nonumber\\
& & \hskip-0.5cm \omega_{10} Q^1{}_0 + \omega_{21}\,Q^2{}_1 = \omega_{11}\,Q^1{}_1 + \omega_{20} Q^2{}_0, \nonumber
\eea
which is indeed a subset of the identities arising from their respective first formulation. 

We summarize the correlation of the weaker first formulation identities for the Model A and Model B in Table \ref{tab_BIsNew}.
\begin{table}[H]
  \centering
 \begin{tabular}{|c||c|c|}
\hline
%& & \\
 BIs & First formulation & Second formulation  \\
% & & \\
 \hline
 & & \\
 {\bf (1)} & $2\, H_{klm}\, R^{klm} = 3\, \omega_{kl}{}^k\, Q_m{}^{lm}$ (Trivial) & Model A: Trivial  \\
& & Model B: Trivial \\
& & \\
{\bf (2)} & $H_{kl[\ov i} \,\, \, Q_{\ov j]}{}^{kl}{} -\frac{1}{2}\, Q_k{}^{kl}\,H_{lij} = \,\frac{1}{2}\, \omega_{kl}{}^k\, \omega_{ij}{}^l$ & Model A: Trivial\,  \\
& & Model B:  \, \, $H_K\, \hat{Q}^{\alpha K} = 0$ \\
& & \\
{\bf (3)} & $ \omega_{kl}{}^{[\ov i} \, R^{kl \ov j]}  +\frac{1}{2}\,\omega_{kl}{}^k\, R^{lij} + \frac{1}{2}\,Q_k{}^{kl}{}\,Q_l{}^{ij}{} = 0$ & Model A: Trivial\,  \\
& & Model B:  \, \, $R_K \, \hat{\omega}_\alpha{}^K = 0$ \\
& & \\
{\bf (4)} & $H_{kli}\, R^{klj} = Q_i{}^{kl}{}\, \omega_{kl}^j  + \omega_{kl}{}^k\, Q_i{}^{lj}{} + \, Q_k{}^{kl}\,\omega_{li}{}^j$ & Model A: eqn. (\ref{eq:bianchids1additional2}) \\
&&Model B: eqn. (\ref{eq:bianchids1additional3})\\
%&& \\
 \hline
  \end{tabular}
  \caption{A correlation of the weaker first formulation identities in Model A and Model B.}
  \label{tab_BIsNew}
 \end{table}

%\newpage
%\bibliographystyle{JHEP}
%\bibliography{reference}

\newpage
\bibliographystyle{utphys}
\bibliography{reference}

\end{document}